\documentclass[fleqn,usenatbib]{mnras}

\usepackage{newtxtext,newtxmath}

\usepackage[T1]{fontenc}

\usepackage{float}
\usepackage{fleqn}
\usepackage{graphicx}
\usepackage{color}
\usepackage{hyperref}
\usepackage{epstopdf}
\usepackage{url}
\usepackage{subfloat}
\usepackage{caption}
\usepackage{gensymb}
\usepackage{multirow}
\usepackage{makecell}
\usepackage{esvect}
\usepackage{xcolor}
\usepackage{lscape}
\usepackage{pdflscape}
\usepackage{ae,aecompl}
\usepackage{wrapfig}
\usepackage{enumitem}

\usepackage{capt-of}

\DeclareRobustCommand{\VAN}[3]{#2}
\let\VANthebibliography\thebibliography
\def\thebibliography{\DeclareRobustCommand{\VAN}[3]{##3}\VANthebibliography}

\usepackage{graphicx}	
\usepackage{amsmath}	

\title[PolHEx (Polarisation of Hot Exoplanets)]{Modelling reflected polarised light from close-in giant exoplanet WASP-96b using PolHEx (Polarisation of Hot Exoplanets)}

\author[K. L. Chubb et al.]{
	Katy L. Chubb,$^{1,2}$\thanks{E-mail: klc20@st-andrews.ac.uk}
	Daphne M. Stam,$^{3}$
	Christiane Helling,$^{2,4}$
	Dominic Samra,$^{2}$
\newauthor and Ludmila Carone$^{2}$
	\\
	$^{1}$Centre for Exoplanet Science, University of St Andrews, North Haugh, St Andrews, UK\\
	$^{2}$Space Research Institute, Austrian Academy of Sciences, Schmiedlstr. 6, A-8042, Graz, Austria \\
 	$^{3}$Leiden Observatory, Niels Bohrweg 2, 2333 CA Leiden, Netherlands \\
	$^{4}$Fakultät für Mathematik, Physik und Geodäsie, TU Graz, Petersgasse 16, Graz, A-8010, Austria
}

\date{Accepted XXX. Received YYY; in original form ZZZ}

\pubyear{2023}

\begin{document}
	\label{firstpage}
	\pagerange{\pageref{firstpage}--\pageref{lastpage}}
	\maketitle
	
	\begin{abstract}
			We present the Polarisation of Hot Exoplanets (PolHEx) code for modelling the total flux ($F$) and degree of linear polarisation ($P$) of light spectra reflected by close-in, tidally locked exoplanets. We use the output from a global climate model (GCM) combined with a kinetic cloud model of hot Jupiter WASP-96b as a base to investigate effects of atmospheric longitudinal-latitudinal inhomogeneities on these spectra. 
			We model $F$ and $P$-spectra as functions of wavelength and planet orbital phase for various model atmospheres. We find different materials and sizes of cloud particles to impact the reflected flux $F$, and particularly the linear polarisation state $P$. A range of materials are used to form inhomogeneous mixed-material cloud particles (Al$_2$O$_3$, Fe$_2$O$_3$, Fe$_2$SiO$_4$, FeO, Fe, Mg$_2$SiO$_4$, MgO, MgSiO$_3$, SiO$_2$, SiO, TiO$_2$), with Fe$_2$O$_3$, Fe, and FeO the most strongly absorbing species. The cloud particles near the relatively cool morning terminator are expected to have smaller average sizes and a narrower size distribution than those near the warmer evening terminator, which leads to different reflected spectra at the respective orbital phases.
            We also find differences in the spectra of $F$ and $P$ as functions of orbital phase for irregularly or spherically shaped cloud particles.
			This work highlights the importance of including polarisation in models and future observations of the reflection spectra of exoplanets.
	\end{abstract}
	
	
	
	
	\section{Introduction}
	
	The theoretical groundwork for the scattering properties of atmospheric particles derives from prominent works such as \cite{08Gustav,18Rayleigh,50Chandrasekhar}. It was the inclusion of the polarization state in the modelling of scattered light, however, that was crucial in enabling the identification of cloud types on Venus during the 1970s by \cite{74HaHo}. Via polarimetry they were able to deduce the clouds on Venus were most likely formed from sulphuric acid, with a narrow distribution of particle size and mean radius of  $\sim$1~$\mu$m. The method was later utilised further for Venus~\citep[e.g.][]{15RoMaMo} and other solar system planets, such as Mars~\citep{08Sc}, Jupiter~\citep{91WeSm,17McStBa}, Neptune and Uranus~\citep{07JoSc}, and Saturn~\citep{84ToDo}.
	
			Measuring reflection spectra, particularly when including polarisation, is a highly powerful and complementary observation technique to transmission and emission spectroscopy (i.e.\ observing the light from a star as a function of wavelength during the primary and/or secondary eclipse of a transiting planet) for revealing additional information about transiting exoplanet atmospheres~\citep{18Munoz,18MiBl,19FoRoSt}. There have been a range of theoretical studies and models of the polarised flux of exoplanet systems, including \cite{00SeWhSa,18BaKeBo} for close-in giant planets, and \cite{08Stam,12KaSt,17FaRoSt,17RoSt,20GrRoTr,22TrSt,22WeDuHu} for Earth-like or habitable-zone exoplanets. Observations of polarisation have been proposed in the context of searching for liquid water and biosignatures on Earth-like planets~\citep{23Vaughan,16BeKuHa,19TreesStam,21SpPaBl}. 
			
				By following the orbital phase of a transiting exoplanet, information on the scattered (reflected) light can be determined via the secondary eclipse (when the planet is hidden behind the star) and phase curve mapping~\citep{21WoKiSh,21HeMoKi}. Although this technique is the same as used for measuring emission spectra, emission and reflection spectra can be disentangled from one another due to the fact they typically dominate across different wavelength regions to one another, with reflection spectroscopy in the visible/near-IR and emission spectroscopy in the IR. There are a number of studies and tools for modelling reflection spectra~\citep{14BaAiIr,18MaMaFo,19BaMaLe,19KaRu}, often with a focus on the overall flux and not considering the polarisation state. Transmission spectra, which can be observed across the whole wavelength region from visible to IR (see, for example, \cite{23AlWaAl,23FeRaWe,23AhStMa,23RuSiMu}), do allow some information of spectral cloud features to be inferred~\citep{15WaSi,17MoBoBo,18OrMi.arcis,19PoLoKr,20SaHeMi,22LoSiRu}, in particular at longer wavelengths, around 10$\mu$m, where there are typically signatures from vibrational modes~\citep{18OrMi.arcis,23BoKeGr}. However, the extent of information to be inferred from these observations is limited. Including the polarisation state in reflection spectroscopy, on the other hand, makes the technique particularly sensitive to microphysical cloud properties, such as material and size distribution. Being sensitive to different components of an atmosphere  makes polarised reflection spectroscopy a very complementary technique to transmission and emission spectroscopy. 
				
				There are a number of  ground-based telescopes that can measure the state of polarisation of light, such as HARPSpol~\citep{11PiSnDo}, CRIRES+/VLT~\citep{23DoBrSm}, SPHERE/VLT~\citep{20BoLaHo}, ZIMPOL/VLT~\citep{04GiScTh}, ESPaDOnS~\citep{06DoCaLa}, WIRC+Pol~\citep{19TiMiJo}, PEPSI~\citep{15StIlJa}, HIPPI-2~\citep{20BaCoKe}. Some of these have been pointed towards exoplanets~\citep{08BeBeFl,11BeBeFl,18BoBaCo,21BaBoCo} and Brown Dwarfs~\citep{20MiGiKa}, however there is still some discussion over the reliability and interpretation of these exoplanet observations~\citep{16BoBaKe,18BoBaCo}. There are  already some plans to include polarimeters on future space-based instruments~\citep{17TaMaIt}, such as LUVOIR/POLLUX~\citep{18BoNeGo}, and the Nancy Grace Roman Space Telescope~\citep{21GrZiSu}. In order to further motivate the implementation and to facilitate the design of such instruments it is important to have detailed and accurate theoretical models of real systems that are likely to be observed by such instruments, which is the motivation behind the present study. 
				
					In this work we present PolHex, a numerical code for modelling the polarised reflected flux of close-in transiting exoplanet atmospheres. PolHEx is based on the adding-doubling radiative transfer algorithm of \cite{87HaBoHo}, which has been built upon over the years for application to exoplanet atmospheres~\citep{06StRoCo, 08Stam}. Versions of the code have been used for modelling for exoplanet atmospheres by many studies, such as \cite{99StHaHo,00StHaHo,04StHoWa,06StRoCo,08Stam,11KoHeSt,12KaStHo,12KaSt,13KaStGu, 17FaRoSt,17RoSt,17McStBa,17PalmerEPSC,19TrSt,20GrRoTr,22MeStVi,22TrSt}. There is a publicly available code written in python and fortran called PyMieDap\footnote{\url{https://gitlab.com/loic.cg.rossi/pymiedap.git}}~\citep{18RoBeSt}, which shares much of the functionality and origins with the code in the present study. PolHEx has been specifically tailored for modelling close-in hot exoplanets, which are assumed to be tidally locked. This allows us to directly link longitude/latitude atmospheric variation to orbital phase, and gives a simple way of specifying inhomogeneities in the atmosphere. 
				We use atmospheric climate and kinetic cloud models of hot gas giant exoplanet WASP-96b from \cite{22SaHeCh} as a base atmosphere to study the impact of inhomogeneous atmospheric composition on reflected flux ($F$) and degree of linear polarisation ($P$) using PolHEx. These kinetic cloud models build on a global climate model (GCM) of WASP-96b which was produced using expeRT/MITgcm~\citep{22ScCaDe,20CaBaMo,21BaDeCa}. 
				
					This paper is structured as follows. Section~\ref{sec:theory} outlines the relevant theory behind our modelling techniques, with details on the PolHEx code itself given in Section~\ref{sec:PolHEx}. Section~\ref{sec:atmosphere} summarises the atmospheric properties of hot gas giant exoplanet WASP-96b which are used in this work, based on the outputs of a GCM and kinetic cloud models.  This includes the molecular and cloud composition, pressure-temperature profiles and longitude-latitude grid. Section~\ref{sec:setups} then details the different theoretical models we have set up based on these atmospheric properties. Here, a number of inhomogeneous (i.e. varying longitude and latitude) and homogeneous (no variation in longitude or latitude) atmospheric models are considered. The results of these models are presented in Section~\ref{sec:results}, followed by a discussion in Section~\ref{sec:discussion}. We present our conclusions in Section~\ref{sec:conclusion}. 
				
	\section{Theory}\label{sec:theory}

\subsection{Atmospheric scattering}

When an oscillating plane electromagnetic (EM) wave emitted from a host star encounters particles (molecular, atomic, cloud) in an orbiting exoplanet's atmosphere, the wave will interact with the charged components of the atmospheric particles. This interaction causes the charged components such as electrons to oscillate with the same frequency as the incident wave, which in turn induces new EM waves that propagate out in all directions from the particle. If a new wave which is propagating in the same direction as the incident beam is out of phase with the incident beam then they will interfere with one another, leading to a change of direction of the incident wave. This process is known as scattering~\citep{02MiTrLa,04HoMeDo}.

In general, the polarisation state of the stellar EM wave will change during a scattering process. The refractive index $m$~=~$n$~+~$ik$ of the cloud particles holds information on their scattering properties. This is combined with the wavelength and particle size and shape in order to determine how light is scattered. The real part $n$ of the refractive index represents the phase velocity (rate of propagation) in the material, and the imaginary part $k$ the absorption of incoming radiation by the material. 

Each of the atmospheric particles which produce new secondary EM waves due to interaction with the incoming stellar wave will have an impact on the other particles around it. If the number of particles is small the secondary wave contribution can be neglected which leads to the single scattering approximation. 
If, however, the atmosphere contains many particles, then the scattering of light that has already been scattered by another particle needs to be taken into account. This is known as multiple scattering. PolHEx fully includes multiple scattering effects in it's adding-doubling radiative transfer algorithm~\citep{87HaBoHo,06StRoCo}. In wavelength regions where absorption is high (for example due to a strongly absorbing molecular or atomic transition), multiple scattering effects are reduced. As multiple scattering is a process which tends to reduce the degree of linear polarisation, $P$, of light in comparison to single scattering, $P$ will generally tend to be higher within strong absorption bands \citep[see][and references therein]{17FaRoSt,06StRoCo,1999Stam}.

\subsection{Flux and polarisation state}

To describe the flux and polarisation state of the stellar radiation which is scattered towards us, the observer, by the exoplanet's atmosphere, we use a Stokes vector~\citep{50Chandrasekhar,83HoMe,04HoMeDo,06StRoCo}, as a function of wavelength $\lambda$ and orbital phase $\alpha$ of the planet orbiting a star (see Fig.~\ref{fig:phase_angles}): 
\begin{equation}\label{eq:flux_vector}
	\pi \textbf{F} (\lambda, \alpha) = \pi \begin{pmatrix}
		F (\lambda, \alpha) \\
		Q (\lambda, \alpha) \\
		U (\lambda, \alpha) \\
		V (\lambda, \alpha)
	\end{pmatrix}.
\end{equation}
The four Stokes parameters forming the vector are defined as follows: $F$ is the total flux, $Q$ and $U$ the linearly polarised fluxes (defined with respect to a reference plane), and $V$ the circularly polarised flux \citep[see][]{74HaTr}. The units of the Stokes parameters are W~m$^{-2}$~Hz$^{-1}$. In our simulations, we will ignore the circular polarisation as the signal is very small compared to the linear polarisation while including it significantly increases computation times \citep[][]{2018RS}.

The degree of linear polarisation of the radiation is defined as: 
\begin{equation}
	P(\alpha, \lambda) = \frac{\sqrt{Q(\alpha, \lambda)^2 + U(\alpha, \lambda)^2}}{F(\alpha, \lambda)}. 
\end{equation}
We define $Q$ and $U$ when integrated over the planetary disk with respect to the the planetary scattering plane , which is the plane through the centres of the star, the planet, and the observer. We assume that the rotation axis of a transiting planet is perpendicular to this plane (they are tidally locked), and that the observer is facing the system edge-on, i.e.\ so the planet's orbital inclination is 90$\degree$. 
We further assume that the planet is mirror-symmetric about the equator, which means that $U$~=~0 when integrated over the planetary disk. This allows the use of an alternative definition of the degree of polarisation $P$ that includes information about the direction of polarisation:
\begin{equation}\label{eq:linear_P}
	P(\alpha, \lambda)= -\frac{Q(\alpha, \lambda)}{F(\alpha, \lambda)}. 
\end{equation}
For positive values of $P$, the light is polarised perpendicular to the planetary scattering plane (thus perpendicular to the line between the planet and the star), and for negative values of $P$, the light is polarised parallel to the planetary scattering plane. We chose this convention to ensure $P$ is positive for a clear atmosphere (i.e. only scattering from gaseous species).

In our case, $F$ is the observed flux from the planet, which is composed both of stellar flux reflected by the planet's atmosphere (and surface if there were one), and also of thermal flux from the planet, i.e: 
\begin{equation}
	F (\alpha, \lambda)  = {F_{\rm reflected} (\alpha, \lambda) + F_{\rm thermal} (\alpha, \lambda)}.
\end{equation}
In the IR wavelength region, the planetary flux is expected to be dominated by emission, whereas in the visible it's expected to be dominated by reflected light. We only consider the degree of polarisation of reflected flux in this study and assume thermal flux is negligible, as we focus on the wavelength region 0.5~-~1~$\mu$m where the reflected flux will dominate. Therefore, in our case, $F$ in Eqs.~\ref{eq:flux_vector}~-~\ref{eq:linear_P} is really just $F_{\rm reflected}$. 

\subsection{Scattering by spherical particles}\label{sec:spherical_scattering}

In the models where we assume spherical cloud particles we use Mie theory~\citep{08Gustav} to compute how radiation is scattered as a function of scattering angle $\Theta$ (for forward scattered light, $\Theta = 0^\circ$), using a scattering matrix of the form (see, for example, \cite{70Hovenier,04HoMeDo}): 
\begin{equation}\label{eq:scat_mat}
	\textbf{F} (\Theta)  = \begin{pmatrix}
		\alpha_1 (\Theta) & \beta_1 (\Theta) & 0 & 0 \\
		\beta_1 (\Theta) & \alpha_2 (\Theta) & 0 & 0 \\
		0 & 0 & \alpha_3 (\Theta) & \beta_2 (\Theta)  \\
		0 & 0 & - \beta_2 (\Theta) & \alpha_4 (\Theta)  
	\end{pmatrix}.
\end{equation}
The first matrix element, $\alpha_1$ ($\Theta$), is known as the phase function or scattering function. It would be the only element of \textbf{$F$}($\Theta$) needed if polarisation were to be ignored.
The degree of linear polarisation $P$ of light that is singly scattered by the cloud particles is related to the scattering matrix elements of Eq.~\ref{eq:scat_mat} by $P$~=~$\frac{-\beta_1(\Theta)}{\alpha_1(\Theta)}$.

With PolHEx, a single scattering matrix can be computed for a given particle size distribution. However, we do not use the matrix elements directly in the code's radiative transfer part. Instead we expand them into generalised spherical functions~\citep{84RoSt,59KuRi} and use the coefficients of this expansion.
Full details on the expansion of Mie scattering matrices into spherical functions as used in PolHEx can be found in \cite{84RoSt}.

\subsection{Scattering by non-spherical particles}\label{sec:nonspherical_scattering}

Scattering matrices (equivalent to Eq.~\ref{eq:scat_mat}) for light that is scattered by non-spherical (e.g.\ irregularly shaped) particles can be obtained using various methods, such as the Discrete Dipole Approximation (DDA)~\citep{11YuHo} or the T-matrix method~\citep{02MiTrLa,17MiZaKh}. Although accurate, these methods can take a considerable amount of computational time, especially for particles with a large size parameter $x$, as defined by
\begin{equation}\label{eq:size_par}
	x = \frac{2\pi r}{\lambda},
\end{equation}
where $r$ is the (equivalent) radius of the particles.

Other, more efficient methods have been developed, such as the Distribution of Hollow Spheres (DHS)~\citep{03MiHoKo,05MiHoKo}. In this method, which has been employed by various studies such as \cite{20SaHeMi}, the optical properties of a collection of non-spherical particles with random orientations are approximated by the optical properties of a collection of basic shapes, i.e.\ spherical particles with varying amounts of vacuum inside. It has been shown by \cite{03MiHoKo} to recreate the measured absorption cross-sections of small crystalline forsterite particles well. We use the publicly available code optool\footnote{\url{https://github.com/cdominik/optool}}~\citep{21DoMiTa} to compute scattering matrices of irregularly shaped cloud particles. It is derived from codes by \cite{05MiHoKo} (DHS model for irregular grains) and \cite{18TaTa} (scattering by fractal dust aggregates). Optool allows a material to be defined, with specified refractive indices, size distribution, wavelength, and degree of irregularity. This degree of irregularity is defined using a parameter called $f_{\rm max}$, which ranges from 0 for spherical particles to close to 1 for very irregular particles (computationally $f_{\rm max}$ should stay just below 1, e.g.\ 0.98). 

\subsection{Rayleigh scattering by molecules}\label{sec:rayleigh}

Rayleigh scattering is essentially Mie scattering in the limit of a very small size parameter $x$ (see Eq.~\ref{eq:size_par}).
Rayleigh scattering will occur due to the small gaseous (molecular and atomic) atmospheric particles. Incident radiation (in this case from the planet's host star) induces a dipole moment in the particle, which is proportional to the incident electric field, with a proportionality constant known as the polarisability. This polarisability can be isotropic or non-isotropic. If a particle with $x$~$<<$~1 (Eq.~\ref{eq:size_par}) has isotropic polarisability then Rayleigh scattering without depolarisation, or isotropic Rayleigh scattering, occurs~\citep{04HoMeDo}. If the particle has anisotropic polarisability, as is the case for H$_2$~\citep{04KoWo} for example, then Rayleigh scattering with depolarisation, or anisotropic Rayleigh scattering, occurs. This can be quantified using the depolarisation factor $\delta$, which appears in the equation for  the scattering matrix for Rayleigh scattering~\citep{18Rayleigh,50Chandrasekhar,74HaTr}: 
\begin{multline}
	\small{\textbf{$P_m$} (\Theta) = \Delta \begin{pmatrix}
		\frac{3}{4} (1+\cos^2\Theta) & -\frac{3}{4}(\sin^2\Theta) & 0 & 0 \\
		-\frac{3}{4}(\sin^2\Theta) &\frac{3}{4} (1+\cos^2\Theta) & 0 & 0 \\
		0 & 0 & \frac{3}{2}\cos\Theta & 0 \\
		0 & 0 & 0 & \Delta^{'} \frac{3}{2}\cos\Theta 
	\end{pmatrix}} \\\\
        \small{ +  ( 1  -  \Delta ) \begin{pmatrix}
		1 & 0 & 0 & 0 \\
		0 & 0 & 0 & 0 \\
		0 & 0 & 0 & 0 \\
		0 & 0 & 0 & 0
	\end{pmatrix}}  , 
\end{multline}
where: 
\begin{equation}
	\Delta = \frac{1-\delta}{1+ \frac{\delta}{2}}, 
\end{equation}
\begin{equation}
	\Delta^{'} = \frac{1-2\delta}{1- \delta}.
\end{equation}
The average molecular scattering cross-section $\sigma_m$ per particle (over all scattering angles $\Theta$) for Rayleigh molecules~\cite{74HaTr} is:
\begin{equation}\label{eq:mol_scat}
	\sigma_m = \frac{8\pi^3}{3}\frac{(n^2 - 1)^2}{\lambda^4 N^2} \frac{6+3\delta}{6-7\delta}, 
\end{equation}
with  $\sigma_m$~$\propto$~$\lambda^{-4}$ indicating that molecular scattering decreases with increasing wavelength. Here, $n$ is the real part of the refractive index of the gas, and $N$ is the number of molecules per unit volume (which depends on the gas temperature, $T_{\rm gas}$). We use a (wavelength independent) $\delta$ of 0.02 for H$_2$~\citep{57Penndorf,74HaTr}, as this is the main gaseous component of our model WASP-96b atmosphere. 

\subsection{Integrated reflection spectra across the planet}\label{sec:ref}

For a given orbital phase, the flux vectors from various positions on the planet
are integrated over the illuminated part of the planetary disk that is visible by the observer. 
The Stokes vector of the light that is reflected by the planet and that arrives at the observer can be described by:
\begin{equation}\label{eq:ref_flux}
	\textbf{F} (\lambda, \alpha) =  \frac{r^2}{d^2} \frac{R^2}{D^2}\frac{1}{4} \textbf{S}(\lambda, \alpha) \pi \textbf{B}_0 (\lambda),
\end{equation}
where $R$ is the stellar radius, $D$ the star-planet distance, $d$ is the planet-Earth distance, $r$ is the planet radius, and $S$ the planetary scattering matrix (this describes the light that is reflected by the planet towards the observer)~\citep{04StHoWa}. We use c.g.s.\ units for the distances, and the scattering matrices are unit-less. $B_0$ is the Stokes column vector:
\begin{equation}
	\begin{pmatrix}
		B_0 \\
		0\\
		0\\
		0
	\end{pmatrix},
\end{equation}
with $\pi B_0$ the stellar surface flux (units of erg~s$^{-1}$~cm$^{2}$). The stellar surface flux is assumed to be unpolarised when integrated over the stellar disk~\citep{87KeHeSt}. Although PolHEx could be adapted for polarised incoming stellar radiation if necessary, we do not in this case because it has been found that most FGK stars have negligible intrinsic polarization as long as they are inactive~\citep{17CoMaBa}, and WASP-96 is not known to be particularly active.

In this work we only compute the planetary scattering matrix, \textbf{S}($\lambda$,$\alpha$)~\citep{04StHoWa}, of our model planets: 
\begin{equation}\label{eq:planetary_scat_mat}
	\textbf{S} (\lambda,\alpha)  = \begin{pmatrix}
		a_1 (\lambda,\alpha) & b_1 (\lambda,\alpha) & 0 & 0 \\
		b_1 (\lambda,\alpha) & a_2 (\lambda,\alpha) & 0 & 0 \\
		0 & 0 & a_3 (\lambda,\alpha) & b_2 (\lambda,\alpha)  \\
		0 & 0 & - b_2 (\lambda,\alpha) & a_4 (\lambda,\alpha)  
	\end{pmatrix}.
\end{equation}
We set all other terms in Equation~\ref{eq:ref_flux} to 1, and normalise such that $F$ at $\alpha$~=~0$\degree$ is equal to the geometric albedo; see Section~\ref{sec:geo_albedo}. Thus our outputs labelled $F$ are really the $a_1$($\lambda$, $\alpha$) element of the planetary scattering matrix. 
These values can be scaled given the parameters of the system; we do not choose to do so in the figures presented here and just look at comparative values, as our main aim is to assess differences between different model atmospheres. 

Degree of polarisation $P$ is a relative measure and so does not need to be scaled. We do, however, give some typical scaled values using the WASP-96b system parameters in Appendix~\ref{sec:app1}, which we use to comment on the potential detectability of our model planets. In summary, we find a typical range of $\frac{P}{F_{\rm star}}$ (i.e. polarisation as a fraction of the observed stellar flux) of 0.1~-~30~ppm. For reference, the HIPPI-2 instrument can measure a polarised signal with a precision of around 3.5~ppm~\citep{20BaCoKe}, with higher precisions expected from potential future instruments. This places some of the polarised signals modelled in this work on the edge of the current detectability limits, which may provide guidance for the design of future polarimeters.

Further details on the computation of the planetary scattering matrix can be found in \cite{18RoBeSt}, including the expansion as a Fourier series and the choice of Gaussian abscissae for the integration across the planetary disk, which are also relevant to PolHEx. We find 80~Gaussian abscissae to be sufficiently accurate for our computations. 

\subsection{Bond and geometric albedo}\label{sec:geo_albedo}

The Bond albedo, ${\rm A_B}$, of a planet is essentially the efficiency with which the planet reflects incoming stellar radiation into all directions. It therefore determines how much energy from stellar radiation is absorbed and available for transport round the planet (see, for example, \cite{22ChMi}). The geometric albedo, ${\rm A_G}$, is defined as the ratio of the reflected flux at $\alpha$~=~0$\degree$ (i.e. scattering angle $\Theta$~=~180$\degree$) compared to a Lambertian (isotropically reflecting) flat disk of the same cross-sectional area that comprises the same solid angle in the sky. Unlike $A_B$, $A_G$ can be larger than 1. For a transiting exoplanet with known radius, $A_G$ can be measured just before the secondary transit. Although both the Bond and the geometric albedo are generally wavelength-dependent, they are typically measured and averaged over a band pass (see, for example, \cite{23KrLePa}). 

Geometric albedo $A_G$ can be found from the planetary scattering matrix element $a_1$ as an output of PolHex via: 
\begin{equation}\label{eq:ag}
    A_G = a_1 (\Theta = 180\degree) = a_1 (\alpha = 0\degree).
\end{equation}
For a model planet with no atmosphere and a Lambertian reflecting surface with an albedo of 1, the planetary scattering matrix element $a_1$ is given by 
\begin{equation}
	a_1 (\Theta) = \frac{2\pi}{3}(\sin{\Theta} - \Theta\cos{\Theta}),
\end{equation}
where $\Theta = 180^\circ - \alpha$. There is a factor of 4 difference compared to e.g. \cite{06StRoCo}, due to our normalisation.
For a white, Lambertian reflecting planet, at $\Theta$~=~180$\degree$ 
or $\alpha$~=~0$\degree$, $a_1$ = $A_G = \frac{2}{3}$. 

\section{The Polarisation of Hot Exoplanets (PolHEx) code}\label{sec:PolHEx}

The Polarisation of Hot Exoplanets (PolHEx) code is used throughout this study. This code is based on the adding-doubling radiative transfer algorithm of \cite{87HaBoHo}, adapted for and used to model polarised fluxes of light reflected by exoplanets by \cite{06StRoCo, 08Stam}. Variations of the code have been used in many studies such as \cite{99StHaHo,00StHaHo,04StHoWa,06StRoCo,08Stam,11KoHeSt,12KaStHo,12KaSt,13KaStGu, 17FaRoSt,17RoSt,17McStBa,17PalmerEPSC,19TrSt,20GrRoTr,22MeStVi,22TrSt}. A version of the code was used in a benchmark against the Monte Carlo based radiative transfer code ARTES in \cite{12KaStHo}. There is a publicly available code written in a combination of python and fortran called PyMieDap\footnote{\url{https://gitlab.com/loic.cg.rossi/pymiedap.git}}~\citep{18RoBeSt}, which shares much of the functionality and origins with PolHEx. 

\begin{figure*}
	\centering
	\includegraphics[width=0.99\textwidth]{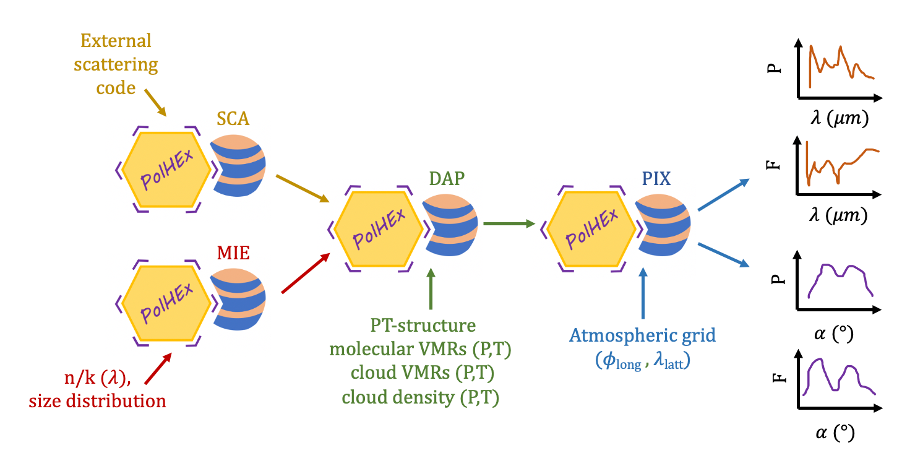}
	\caption{A simplified summary of the PolHEx code setup. Either the MIE routine within PolHEx or an external code can be used for computing the scattering matrices of cloud or aerosol particles. Use of an external code output requires an extra step (SCA) to expand these matrix elements (as functions of scattering angle) into spherical functions, and to format the output for input into the adding-doubling radiative transfer routine (DAP) part of PolHEx. The pressure-dependent atmospheric parameters are setup in DAP for each different atmospheric region on the planet, before combining different regions of the planet together in the PIX component of the code. 
		The outputs from PIX are either $F$ and $P$ as functions of wavelength $\lambda$ for a fixed orbital phase $\alpha$, or $F$ and $P$ as functions of $\alpha$ for a fixed $\lambda$.}\label{fig:code_diagram}
\end{figure*}

Figure~\ref{fig:code_diagram} gives a summary of the structure of PolHEx. 
Either the MIE routine within PolHEx or an external code can be used to compute single scattering matrices (as in Eq.~\ref{eq:scat_mat}) for  particle size distributions of cloud/aerosol particles, resulting in 6 independent matrix elements as functions of the single scattering angle.
The scattering matrix elements are expanded in general spherical functions (see Section~\ref{sec:spherical_scattering}) before being passed on to the adding-doubling radiative transfer (DAP) part of PolHEx. 
If an external code is used to compute the single scattering matrix elements of the cloud/aerosol particles, then an extra step is required for their expansion (this is done by the SCA component of the code, as labelled in Figure~\ref{fig:code_diagram}). 
The atmosphere is also built as input to DAP, as described in Section~\ref{sec:atmosphere} for our WASP-96b model atmospheres. Molecular absorption and scattering are both included here.
The pressure-dependent atmospheric parameters are set up in DAP for a number of atmospheric layers. The geometry of the system is set up in this part of the code, including the atmospheric composition as a function of longitude and latitude.

The output of DAP entails the reflection of unpolarised incident light for a range of local illumination and viewing angles for a given planetary model atmosphere. 
The matrix elements which describe this local reflection of the planet (in our case due to the atmosphere only) are expanded as a Fourier series, the coefficients of which can be read into the PIX part of PolHEx, for integrating the locally reflected light across the illuminated and visible part of the planetary disk for a given phase angle $\alpha$. 

The planetary disk can be horizontally inhomogeneous with different local atmospheres on different parts of the disk. In order to perform the integration across the inhomogeneous planetary disk,
different Fourier coefficients are computed for the different regions on the planet and these different coefficients are then read and used for the integration.
With an increasing number of different regions, the computation time also increases, so some compromise needs to be made between the degree of complexity regarding the inhomogeneities and computational speed. 

PolHEx computes planet reflection for $\alpha$ from 0$\degree$ to 360$\degree$ in steps of 10$\degree$, with 0$\degree$ being when the planet is directly behind the star, 90$\degree$ when the morning terminator is turned towards the observer, 180$\degree$ when the planet is in front of the star (mid transit), 270$\degree$ when the evening terminator is turned towards the observer, and 360$\degree$ when the planet is again behind the star.  
The outputs of PolHEx, as labelled in Figure~\ref{fig:code_diagram}, are either $F$ and $P$ as functions of wavelength $\lambda$ for a fixed value of $\alpha$, or $F$ and $P$ as functions of $\alpha$ for a fixed value of $\lambda$.

\section{The atmosphere of WASP-96b}\label{sec:atmosphere}

WASP-96b is a hot gaseous exoplanet with a mass of 0.48~$\pm$~0.03~M$_{\rm Jup}$, and a mean radius of 1.2~$\pm$~0.06~R$_{\rm Jup}$.  It orbits close to its host star, with a semi-major axis of 0.045~AU and an orbital period of 3.4~days~\citep{14HeAnCa}. 
The transmission spectra of WASP-96b have been observed and its atmosphere characterised using the Very Large Telescope (VLT), Hubble Space Telescope (HST) and the Spitzer space telescope~\citep{18NiSiFo,22NiSiSp}, and more recently with JWST's NIRISS/SOSS instrument~\citep{23RaWeEs,23TaRaWe}.

\begin{figure*}
	\centering
	\includegraphics[width=0.7\textwidth]{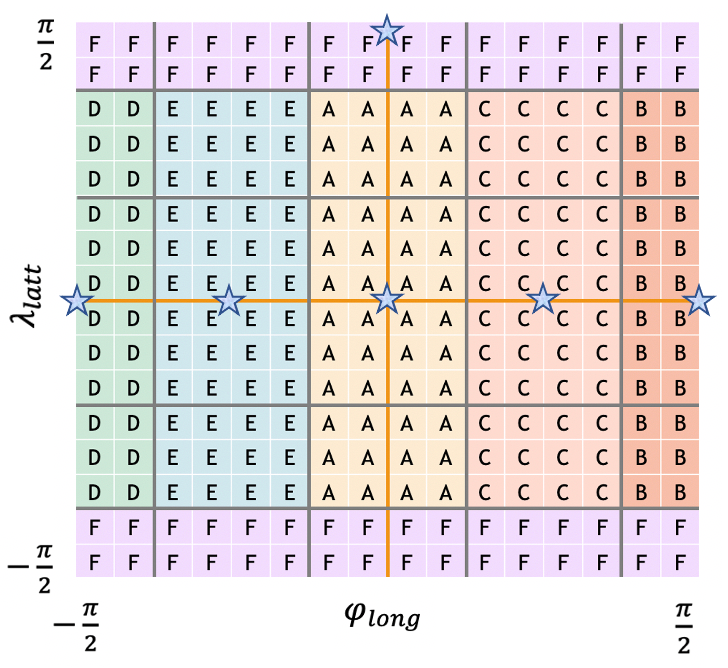}
	\caption{An illustration of the division of the dayside of WASP-96b into different local model atmospheres. The letter denoting the composition is linked to the output of StaticWeather~\citep{22SaHeCh} for WASP-96b for the following longitude ($\phi_{\rm long}$)~/~latitude ($\lambda_{\rm latt}$) points: A~=~0$\degree$~/~0$\degree$, B~=~90$\degree$~/~0$\degree$, C~=~45$\degree$~/~0$\degree$, D~=~-90$\degree$~/~0$\degree$, E~=~-45$\degree$~/~0$\degree$, F~=~0$\degree$~/~86$\degree$. The blue stars indicate the locations of these longitude~/~latitude points. A larger grid (64~$\times$~64) is used in this work but with the same proportions covered by each atmosphere type.}\label{fig:wasp96b_atm_grid}
\end{figure*}

Our base planetary atmosphere setup is derived from the output of a global circulation model (GCM) of WASP-96b which was produced using expeRT/MITgcm~\citep{22ScCaDe,20CaBaMo,21BaDeCa} and combined with a kinetic cloud modelling routine~\citep{19HeGoWo,21HeLeSa} in \cite{22SaHeCh}. We explore how varying certain parameters, such as the materials forming cloud particles, impacts the reflected flux and polarisation of the planet.
We utilise the output provided as a result of \cite{22SaHeCh} for 6 different longitude and latitude dependent atmospheric regions, which we label A$\dots$F, as shown in Figure \ref{fig:wasp96b_atm_grid}. 

\begin{figure*}
	\centering
 \includegraphics[width=0.49\textwidth]{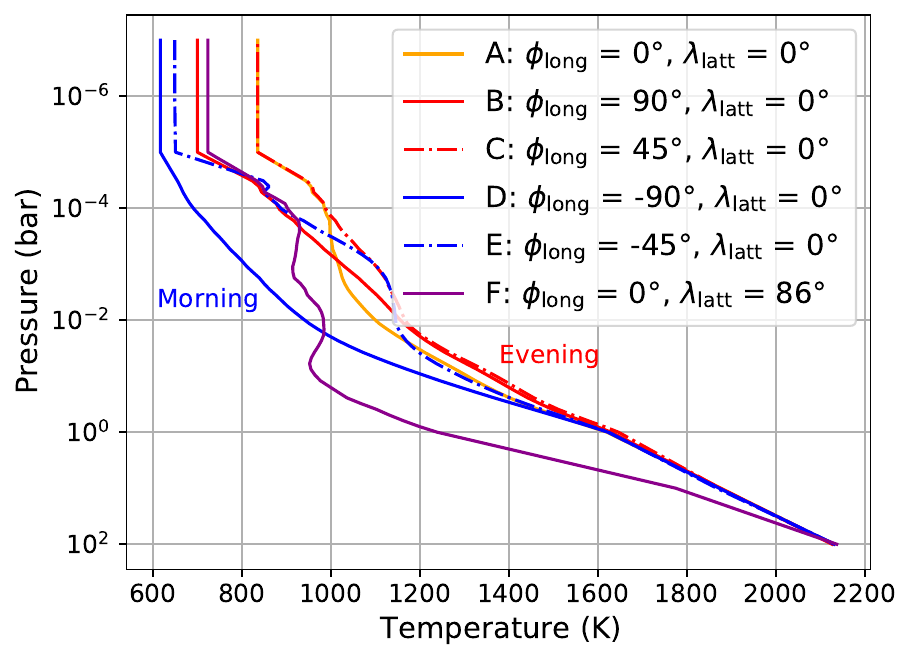}
 	\includegraphics[width=0.49\textwidth]{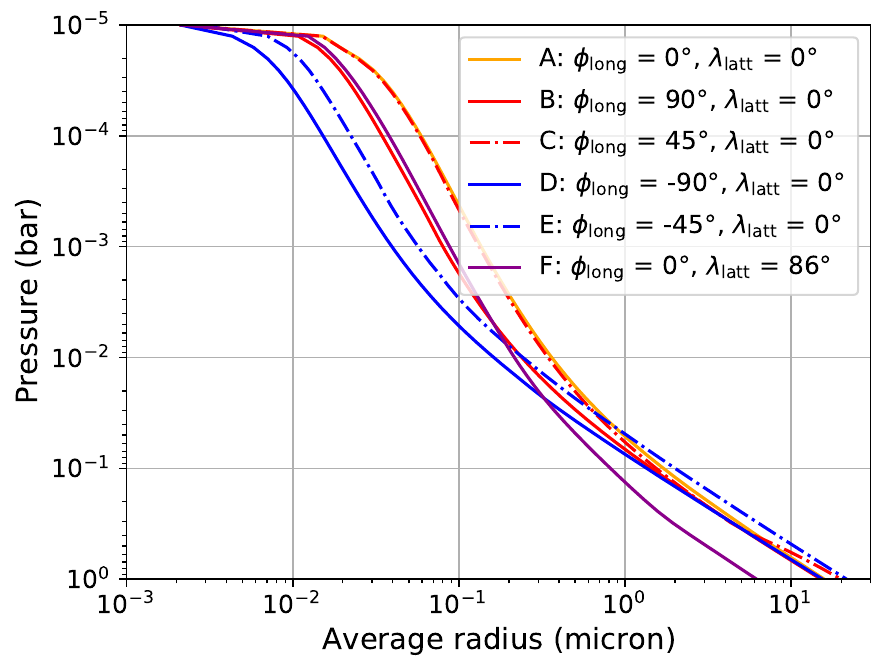}
	\caption{Left: pressure-temperature profiles for the atmospheric regions A to F shown in Fig.~\ref{fig:wasp96b_atm_grid}. Right: Average particle radii for the same regions.}\label{fig:PT_all6}
\end{figure*}

For each atmospheric region A$\dots$F, we build an atmosphere with 44 plane-parallel atmospheric layers, with the pressure and temperatures for each taken from \cite{22SaHeCh} and shown in Figure~\ref{fig:PT_all6}. We assign the volume mixing ratio (VMR) of molecules, material volume fractions of the clouds, and optical depth of clouds for each region as described below. 
There is no clearly defined surface on hot gaseous exoplanets, so we use atmospheric layers down to a pressure of 6~bar, and describe the deeper layers as a black surface, i.e.\ all the light emerging from the bottom of the lowest layer (at 6~bar) is absorbed. 

\subsection{Gaseous composition}\label{sec:gas}

\begin{figure*}
	\centering
	\includegraphics[width=0.45\textwidth]{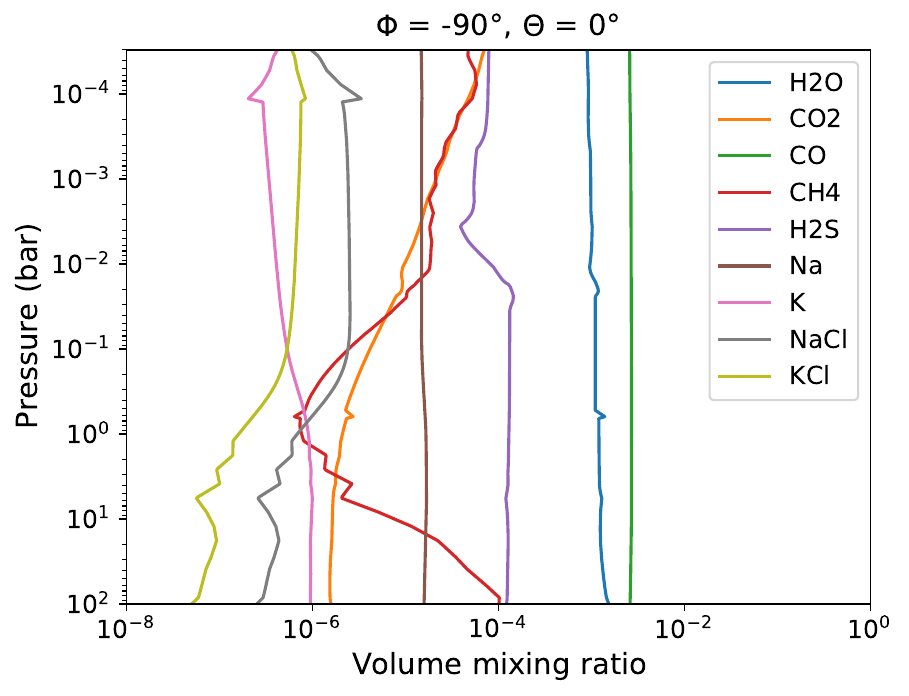}
	\includegraphics[width=0.45\textwidth]{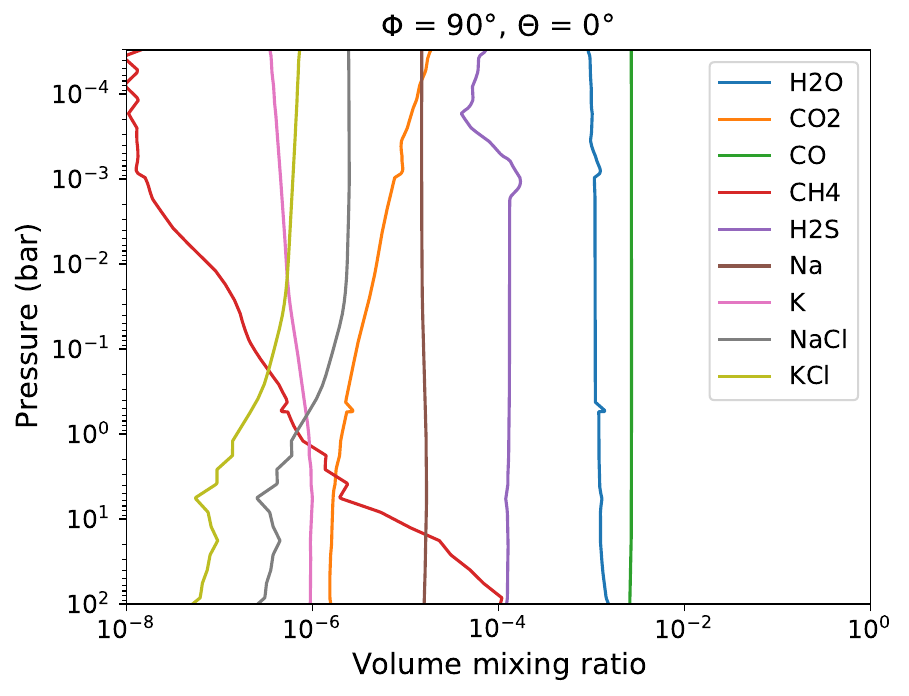}
	\caption{Molecular and atomic concentrations ${n_i}/{n_{\rm tot}}$ for WASP-96b atmospheric models. Left: 
	at $\phi_{\rm long}$~=~-90$\degree$, $\lambda_{\rm latt}$~=~0$\degree$ (cooler morning terminator), and Right: $\phi_{\rm long}$~=~90$\degree$, $\lambda_{\rm latt}$~=~0$\degree$ (warmer morning terminator).}\label{fig:mixing_ratios_plus90}
\end{figure*}

The molecular volume concentrations $\frac{n_i}{n_{\rm tot}}$ (number of molecules of a given species per unit volume divided by the total number of molecules in that volume) at each of the 6~atmospheric regions A$\dots$F in Figure~\ref{fig:wasp96b_atm_grid} are shown in Figure~\ref{fig:mixing_part1}. For simplicity, and because most of the molecular VMRS are very similar between different regions, we only use the two terminator regions (B and D) for modelling the molecular composition, see Figure~\ref{fig:mixing_ratios_plus90}. The model atmosphere contains only the most abundant species (H$_2$O~\citep{jt734}, CO~\citep{15LiGoRo.CO}, H$_2$S~\citep{jt640}, CH$_4$~\citep{17YuAmTe.CH4},  Na~\citep{NISTWebsite,19AlSpLe.broad}, and K~\citep{NISTWebsite,16AlSpKi.broad}). The remainder of the atmosphere is comprised of H$_2$ and He, in solar abundances. 

We use absorption cross-sections (in ${\rm cm^2}/{\rm molecule}$) computed using ExoCross~\citep{ExoCross} as part of the ExoMolOP database~\citep{20ChRoAl.exo}, with the line list for each as specified above. In general we do not expect to see spectral features from these species at abundances below around 1$\times$10$^{-6}$~\citep{22GaMiCh}, however we include the atoms Na and K because of their very strong resonance doublet features. Such features have been observed in the atmosphere of WASP-96b using the Very Large Telescope (VLT)~\citep{18NiSiFo}. We bin the combined cross-sections which include all these species down to a small number of wavelengths (58), ensuring  sufficient sampling around the prominent spectral features. These absorption features are most apparent in our clear (i.e. cloud-free) atmospheric models, but are also important in our cloudy models in order to explore the scattering behaviour within and outside the regions of the absorption features. 

\subsection{Cloud composition}\label{sec:cloud}

\cite{22SaHeCh} predict a variety of different species to form clouds in WASP-96b-like atmospheres. 
Building an inhomogeneous atmosphere of WASP-96b allows us to investigate: 
\begin{itemize}
	\item different cloud compositions
	\item different cloud particle sizes and distributions
	\item clouds at different layers of the atmosphere
\end{itemize}
We then explore these further using model homogeneous atmospheres, largely based on either atmospheric region B (around the evening terminator) or D (around the morning terminator). 
We set up an atmosphere using PolHEx, which allows for a user-defined number of atmospheric layers, each with its own pressure, temperature, gaseous abundance, and cloud layer. For the cloud layer at each pressure level, expansion coefficients of the respective single scattering matrix are read in, with clouds composed of different materials already pre-mixed, as explained in Section~\ref{sec:eff_med_theory}. The scattering properties of the cloud particles can either be computed using the internal PolHEx-MIE computation, which uses Mie theory and therefore assumes spherical particles, or from an external source. We use the optool code\footnote{\url{https://github.com/cdominik/optool}}~\citep{21DoMiTa} (see Section~\ref{sec:nonspherical_scattering}) for exploring the impact of irregularly shaped rather than spherical particles in the atmosphere. We check for consistency with the MIE computations of PolHEx for spherical particles for each set of refractive indices and size distributions, and find identical results. 

\subsubsection{Materials used for clouds}\label{sec:cloud_materials}

\begin{figure*}
	\centering
	\includegraphics[width=0.43\textwidth]{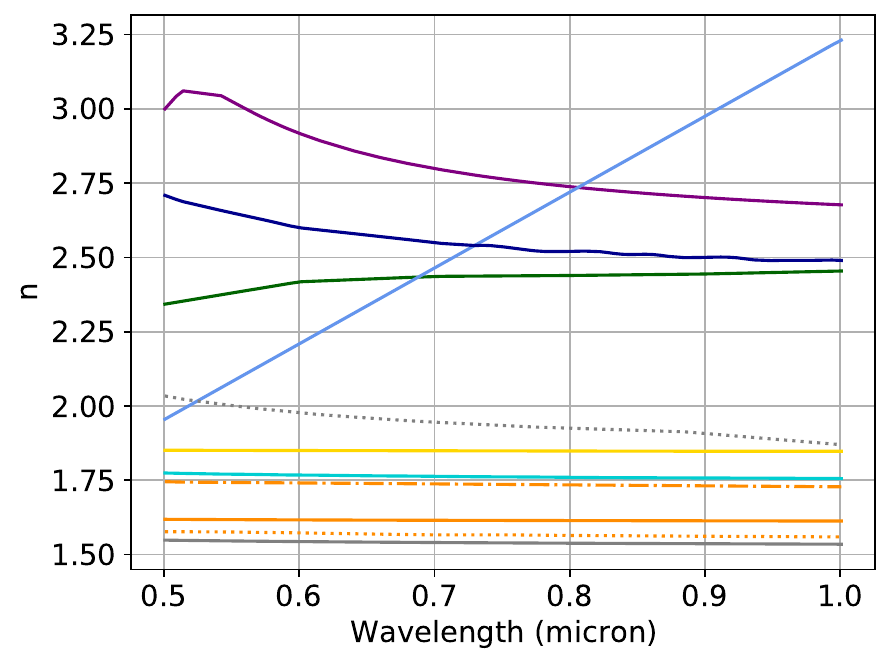}
	\includegraphics[width=0.55\textwidth]{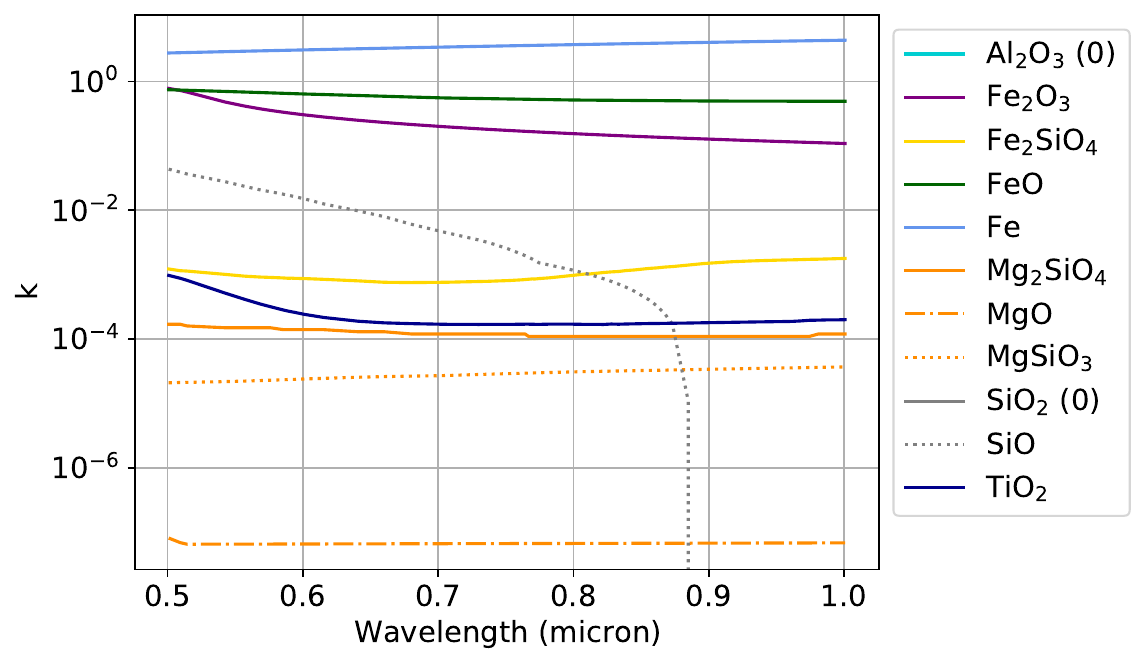}
	\caption{The real (left) and imaginary (right) parts $n$ and $k$ of the refractive index for species used in this work (for references, see Table~\ref{t:ref_index}). The real part of the refractive index indicates the phase velocity (rate of propagation), which relates to scattering, whereas the imaginary part relates to the material's absorption properties. The (0) in the legend refers to the imaginary part $k$ being zero across all wavelengths shown for that species. }
	\label{fig:n_k}
\end{figure*}

Table~\ref{t:ref_index} gives a summary of the different materials used to form the clouds in this work, along with the references for their refractive indices as a function of wavelength. 
The real $n$ and imaginary $k$ parts of the refractive indices (sometimes known as optical constants) as functions of wavelength can be seen in Figure~\ref{fig:n_k}. It is worth noting that the Fe-bearing species Fe, FeO, Fe$_2$O$_3$ have the highest values of $k$, which signifies they are highly absorbing. 
The same species, along with TiO$_2$ have relatively high values of $n$, which indicates a high phase velocity (rate of propagation) which impacts scattering. We note that some typical spectral features that are driven by the imaginary component of the refractive index occur at longer wavelengths than those shown here. 

There are a number of databases which can be used to search for the wavelength-dependent refractive indices of materials which are used to form the clouds. For example, the Database of Optical Constants for Cosmic Dust\footnote{\url{https://www.astro.uni-jena.de/Laboratory/OCDB/}}, and the Aerosol Refractive Index Archive (ARIA)~\footnote{\url{http://eodg.atm.ox.ac.uk/ARIA/}} were both used in the present work (see Table~\ref{t:ref_index}).
The optical properties of potential condensates in exoplanetary atmospheres, including many of those included in this work, have also been compiled and made publicly available in the github associated with various works, such as \cite{17KiHe}.

\begin{table*}
	\centering
	\begin{tabular}{lllll} 
		\hline
		\rule{0pt}{3ex}Species & Name & $n$  &  $k$ & Source \\
		\hline
		\rule{0pt}{3ex}Al$_2$O$_3$ (crystalline) & Corundum & 1.76 & 0 & \cite{12Palik} \\
		\rule{0pt}{3ex}Fe$_2$O$_3$ (solid) & Hematite & 2.79 & 0.22 & \cite{05Triaud}$^a$\\
		\rule{0pt}{3ex}Fe$_2$SiO$_4$ (crystalline) & Fayalite & 1.85 & 1.16~$\times$~10$^{-3}$ &  Unpublished$^b$\\
		\rule{0pt}{3ex}FeO (amorphous) & Wustite &  2.43 & 0.55 & \cite{95HeBeMu}\\
		\rule{0pt}{3ex}Fe (metallic) & Iron & 2.66 & 3.64 & \cite{12Palik} \\
		\rule{0pt}{3ex}Mg$_2$SiO$_4$ (amorphous) & Forsterite & 1.61 & 1.22~$\times$~10$^{-4}$ & \cite{03JaDoMu}\\ 
		\rule{0pt}{3ex}MgO (cubic) & Magnesium oxide & 1.74 & 6.76~$\times$~10$^{-8}$ & \cite{12Palik} \\
		\rule{0pt}{3ex}MgSiO$_3$ (amorphous) & Enstatite  & 1.57 & 2.99~$\times$~10$^{-5}$ & \cite{95DoBeHe}\\
		\rule{0pt}{3ex}SiO$_2$ (crystalline) & Quartz &  1.54 & 0  & \cite{12Palik} \\
		\rule{0pt}{3ex}SiO (non-crystalline) & Silicon oxide & 1.93 & 6.61~$\times$~10$^{-3}$ & \cite{12Palik} \\
		\rule{0pt}{3ex}TiO$_2$ (rutile) & Rutile & 2.54  & 2.40~$\times$~10$^{-4}$ & \cite{11ZePoMu}\\
	\end{tabular}
	\caption{Sources used for the real ($n$) and imaginary ($k$) parts of the refractive indices of the species that form cloud particles in this work. Average values across the wavelength range we consider (0.5~-~1~$\mu$m) are given. All species are in solid phase.}
	\label{t:ref_index}
	\rule{0pt}{0.2ex}
	\flushleft{\textit{$^a$: Downloaded via the 
			Aerosol Refractive Index Archive (ARIA) at \url{http://eodg.atm.ox.ac.uk/ARIA/}}}\\
	\flushleft{\textit{$^b$: Accessed via the Database of Optical Constants for Cosmic Dust at  \url{https://www.astro.uni-jena.de/Laboratory/OCDB/crsilicates.html}}} \\
\end{table*}

\subsubsection{Mixing materials to form clouds using effective medium theory}\label{sec:eff_med_theory}

\begin{figure*}
	\centering
	\includegraphics[width=0.49\textwidth]{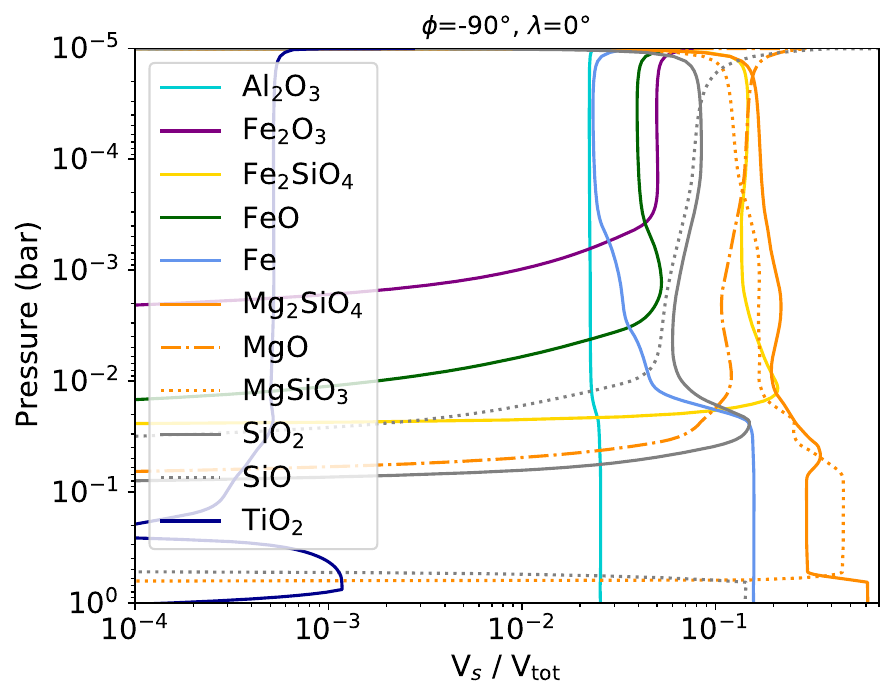}
	\includegraphics[width=0.49\textwidth]{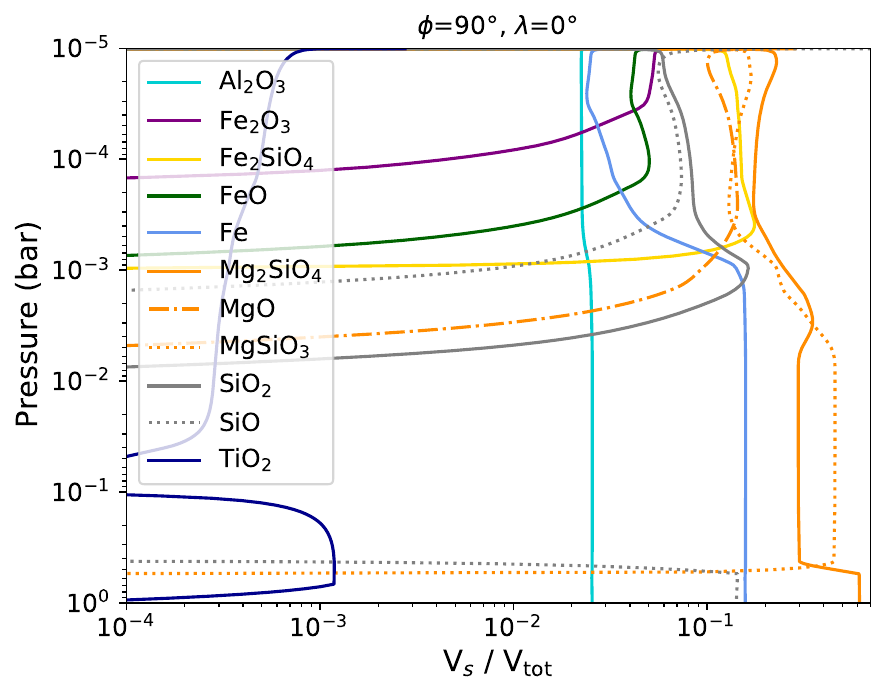}
	\includegraphics[width=0.49\textwidth]{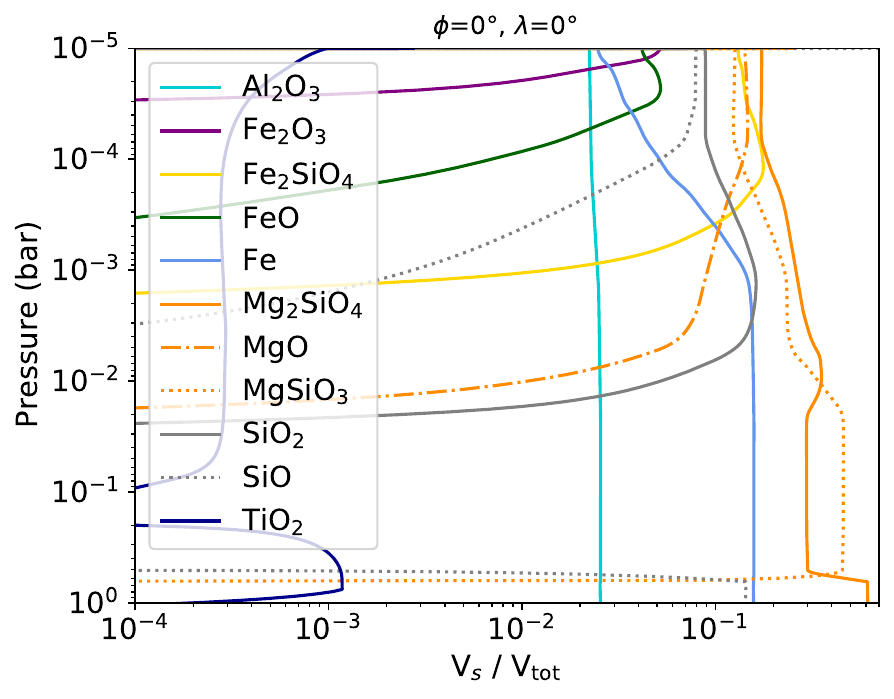}
	\caption{Material volume fractions V$_{s}$~/~V$_{\rm tot}$ of the different materials forming the mixed-material cloud particles. Upper Left: for $\phi_{\rm long}$~=~-90$\degree$ , $\lambda_{\rm latt}$~=~0$\degree$ (cooler morning terminator); Upper Right: $\phi_{\rm long}$~=~90$\degree$ , $\lambda_{\rm latt}$~=~0$\degree$ (warmer evening terminator); and Lower Center: $\phi_{\rm long}$~=~0$\degree$, $\lambda_{\rm latt}$~=~0$\degree$ (the sub-stellar point).}
	\label{fig:cloud_abundance_plus_minus90}
\end{figure*}

In realistic scenarios, and as demonstrated in \cite{22SaHeCh} for WASP-96b, we expect clouds to be formed not of just one single material, but of several different species. In order to model clouds formed of different materials, we use effective medium theory to compute the complex refractive indices which result from the different materials combined~\citep{16MiDiLi}. Each material has it's own real $n$ and imaginary $k$ part of the refractive index, which varies by wavelength, as illustrated by Figure~\ref{fig:n_k}. 
We take into account the material volume fractions of different cloud materials for different longitude, latitude and pressure layer (as illustrated in Figure~\ref{fig:cloud_abundance_plus_minus90}; here, the volume fractions are only as proportions of the total cloud composition and do not take molecular abundances into account), along with the refractive index for each material as a function of wavelength, in order to get a mixed refractive index as a function of wavelength and pressure layer for each atmospheric region (Figure~\ref{fig:nk_combined}). We mix all species given in Figure~\ref{fig:cloud_abundance_plus_minus90} together, taking their relative material volume fractions into account. An example of how these species contribute to the mixed-material refractive indices of Figure~\ref{fig:nk_combined} is illustrated in Figure~\ref{fig:nk_combined_1e-1_1e-4}, which shows the real part of the mixed-material refractive index $n$ when including selected materials for atmospheric region B (evening terminator) at 0.1~bar and 1~$\times$~10$^{-4}$~bar. It can be seen that $n$ when including all materials is identical to $n$ with only the four most abundant species in the left panel (0.1~bar), with Al$_2$O$_3$ only acting to reduce the value of $n$ at the higher wavelengths. The right panel (1~$\times$~10$^{-4}$~bar), however, demonstrates that a larger number of materials are contributing to $n$ in the upper atmosphere. The number of species which contribute to the mixed-material refractive index is therefore dependent on the atmospheric layer and region, and only including a small number of the most abundant species rather than all materials will have an impact on the resulting modelled spectra. The materials used to form all cloud particles are based on the assumed elements available in the exoplanet's atmosphere, along with the local thermodynamic conditions. 132 gas-surface growth reactions are taken into account for the mixed-material cloud particle formation~\citep{23HeSaLe,22Helling}. 

We use the Bruggeman mixing rule~\citep{35Bruggeman}, as was used, for example, in recent works such as \cite{22SaHeBi,SamraThesis}: 
\begin{equation}\label{eq:eff_med}
	\sum_{s} \frac{V_s}{V_{\rm tot}} \frac{\epsilon_s - \epsilon_{\rm eff}}{\epsilon_s + 2\epsilon_{\rm eff}} = 0 
\end{equation}
Here, $\epsilon_s$ is the dielectric constant of each individual condensate material which makes up the inhomogeneous cloud particles. $\epsilon_s$  is related to the refractive index by: 
$\epsilon_s$~=~($n$~+~$ik$)$^2$. We solve Eq.~\ref{eq:eff_med} iteratively using Mathematica~\citep{Mathematica} to get the combined (or effective) dielectric constant $\epsilon_{\rm eff}$, which can be split into real $n_{\rm eff}$ and imaginary $k_{\rm eff}$ parts. We compute these effective refractive indices  as a function of longitude, latitude, pressure layer, and wavelength, and use them as input into our PolHex WASP-96b models. 

\subsection{Optical depth of clouds}\label{sec:optical_depth}

Strong spectral features of Na and K have been observed in the transmission spectra of WASP-96b using the Very Large Telescope (VLT), Hubble Space Telescope (HST) and the Spitzer space telescope~\citep{18NiSiFo,22NiSiSp}.  
The conclusion of \cite{18NiSiFo} was that the atmosphere must be cloud-free in order for the line-wings of the atomic absorption features to be visible. The GCM and kinetic cloud models of \cite{22SaHeCh}, however, find that it would be very unlikely for WASP-96b to be cloud free. \cite{22SaHeCh} therefore explored how their models could better match up with the observations, with one of the processes being a reduced atmospheric vertical mixing, which would cause clouds to settle deeper in the atmosphere than originally predicted. In this work we choose the lower altitude cloud layer in our inhomogeneous model atmosphere models, in order to be more consistent with the observations of  \cite{18NiSiFo}. We do also demonstrate an inhomogeneous model where we place the cloud layer higher up in the atmosphere (to become optically thick at 1~$\times$~10$^{-4}$~bar), as a comparison (see Table~\ref{t:setups} for a summary of the different models computed in this study). 

We assume that within each atmospheric layer the number density $N_{\rm cloud}$ of the materials (both molecules/atoms and clouds) within remains constant. This means the optical depth $\tau$ of a given layer of length $l$ due to clouds composed of a variety of materials with combined extinction coefficient $k_{\rm ext}$ 
can be deduced by:
\begin{equation}
	\tau =k_{\rm ext} N_{\rm cloud} l.
\end{equation}

We use cgs units in our code, with $l$ in cm, $N_{\rm cloud}$ in $\frac{g}{cm^3}$, and $k_{\rm ext}$ in $\frac{cm^2}{g}$. Optical depth $\tau$ is unitless. Extinction coefficient $k_{\rm ext}$ is sometimes called attenuation cross section  $\sigma_{\rm cloud}$. $k_{\rm ext}$ is the sum of the scattering $k_{\rm scat}$ and absorption  $k_{\rm abs}$ cross sections: 
\begin{equation}
	k_{\rm ext} = k_{\rm scat} + k_{\rm abs}
\end{equation}
The single scattering albedo ($\rm ssa$) can be found from these: 
\begin{equation}\label{eq:ssa}
	{\rm ssa} = \frac{k_{\rm scat}}{k_{\rm ext}}
\end{equation}

\subsubsection{Particle Size distribution}\label{sec:size_dist}

Although various size distributions can be used in the Mie computations within PolHex~\citep{74HaTr,84RoSt}, we choose to use a simple Gaussian distribution~\citep{20SaHeMi} for the local size distribution of cloud particles in each atmospheric layer and region. See Section~\ref{sec:size_dist_discussion} for a discussion on different particle size distributions. The average particle size $r_g$ as a function of pressure for each atmospheric region A$\dots$F are illustrated in Figure~\ref{fig:PT_all6}. We assume a Gaussian distribution standard deviation around each of these average particle sizes which is an order of magnitude less than each of the average particle sizes assumed. It can be seen from Figure~\ref{fig:PT_all6} that the evening terminator region (B) is warmer than the morning terminator region (D). This generally implies a smaller average particle size for the cooler morning terminator than for the warmer evening terminator.

\subsection{Geometry of WASP-96b's transit}

\begin{figure*}
	\centering
	\includegraphics[width=0.8\textwidth]{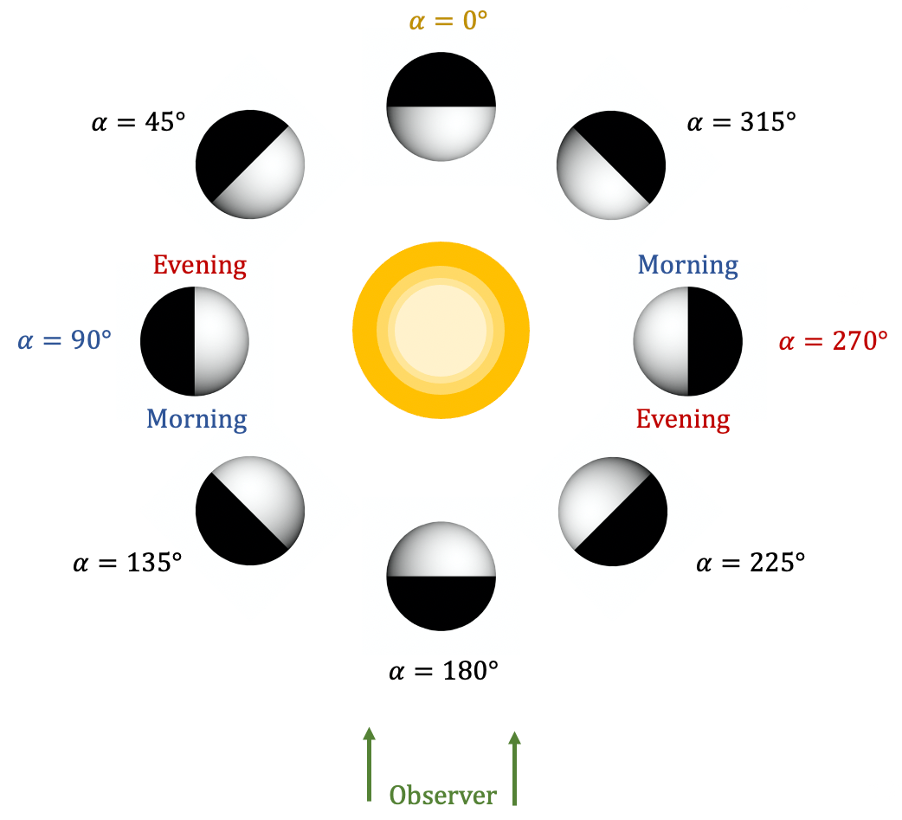}
	\caption{An illustration of how we define phase angle $\alpha$, with a face-down view of the planet-star system. We assume tidally locked planets, so the rotation period of the planet on its axis is the same as the orbital period, both in the anti-clockwise direction in this diagram. The direction of observation is indicated, with the cooler morning terminator in view at $\alpha$~=~90$\degree$ and the warmer evening terminator in view at $\alpha$~=~270$\degree$. }\label{fig:phase_angles}
\end{figure*}

\begin{figure*}
	\centering
	\includegraphics[width=0.8\textwidth]{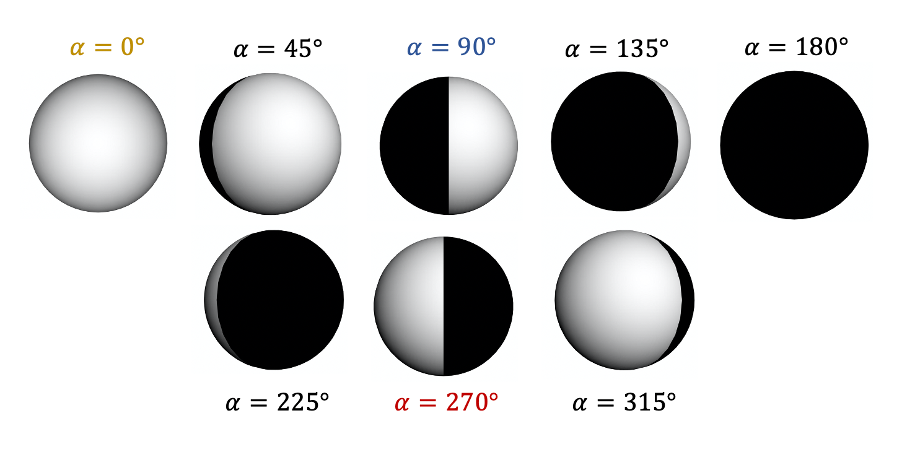}
	\caption{An illustration of the planetary disk as seen by the observer (if it could be resolved) as a function of $\alpha$. 
    The cooler morning terminator at $\alpha$~=~90$\degree$ and the warmer evening terminator at $\alpha$~=~270$\degree$ are highlighted as the phases we focus on in this paper.}\label{fig:phase_angles_face_on}
\end{figure*}

PolHEx is setup with the hot exoplanets assumed to be tidally locked to their host star. This means that the same face of the planet is always facing the star. This simplifies the part of the planet visible to the observer, with the convention of phase~=~0$\degree$ for the dayside of the planet facing the observer and phase~=~180$\degree$ for the nightside of the planet facing the observer. Phases of 90$\degree$ and 270$\degree$ correspond to the morning and evening terminators directly facing the observer, respectively. An illustration of this and a definition of how the phase angles are defined is given in Figure~\ref{fig:phase_angles}, which gives a face-down perspective of the geometry. Only the dayside part of the planet ($-\frac{\pi}{2}$ $\geq$ $\phi_{\rm long}$ $\leq$ $\frac{\pi}{2}$ , $-\frac{\pi}{2}$ $\geq$ $\lambda_{\rm latt}$ $\leq$ $\frac{\pi}{2}$) will give non-zero reflected stokes vectors. Under our assumption of a tidally locked planet we can therefore assign atmospheric types based on regions of longitude and latitude, and these definitions will hold for all phase angles. The illuminated part of the planet which is visible to an observer of course changes as a function of orbital phase, as illustrated in Figure~\ref{fig:phase_angles_face_on}. 

\section{Different atmospheric setups}\label{sec:setups}

The atmosphere is divided into \textit{nlatt}~=~64 longitude $\phi_{\rm long}$  and \textit{nlong}~=~64 latitude $\lambda_{\rm latt}$ points. Each of these grid points is assigned an atmospheric type, computed from the radiative transfer adding-doubling part of the code. For simplicity we have divided WASP-96b up into six atmospheric regions A$\dots$F, as described in Section~\ref{sec:atmosphere} and illustrated in Figure~\ref{fig:wasp96b_atm_grid}.
The locally reflected Stokes vectors are computed for each longitude-latitude grid point and then integrated over the visible and illuminated part of the planetary disk, in order to get the total reflected stokes parameters for a given orbital phase (as in Figures~\ref{fig:phase_angles} and~\ref{fig:phase_angles_face_on}).  
Enough latitude and longitude grid points need to be used such that the stokes vector does not vary significantly between adjacent grid points. The method of assigning grid points ensures that the planet is well sampled around the terminator and polar regions.  An adequate number of grid points to converge was found to be 64~$\times$~64.  

Each model atmosphere consists of 44 layers. The following parameters (see Section~\ref{sec:atmosphere} for details) are varied for each model atmosphere type (region in longitude and latitude space) and atmospheric layer (altitude or pressure layers): 
\begin{itemize}
	\item \textbf{gas temperature} (K) as a function of gas pressure (bar)
	\item \textbf{molecular/atomic number densities (cm$^{-3}$)} as a function of pressure 
	\item \textbf{size distribution} of cloud particles for that given pressure layer$^*$
	\item \textbf{complex refractive index} as a function of pressure, computed using effective medium theory to mix different materials together$^*$
	\item the wavelength-dependent \textbf{optical depth} of clouds as a function of pressure layer 
\end{itemize}
$^*$For these cloud parameters we do not compute a different set for every one of the 44 layers, but group them together into sub-groups of similar composition and size distribution (cloud layers). This is largely due to the computational load of including separate scattering matrices into every layer for the radiative transfer part of the code. We use five different cloud layers (distributed evenly in log-pressure) to capture the variations in size distributions and compositions throughout the atmospheres. We note that the total number of atmospheric layers does not significantly affect our model results, but the reflected flux and polarisation are sensitive to the minimum and maximum pressures for each cloud layer because those pressures determine the ratio of gas molecules versus cloud particles in each layer.

The different model setups that we compute in this work are summarised in Table~\ref{t:setups}, along with the figure(s) which demonstrate the results of these models. In general we produce figures for the total flux $F$($\lambda$,$\alpha$) and degree of linear polarisation $P$($\lambda$,$\alpha$) for model atmospheres as either a function of wavelength (between 0.5~-~1~$\mu$m) for $\alpha$~=~90$\degree$ or $\alpha$~=~270$\degree$, or as a function of orbital phase (between 0~-~360$\degree$) for a selection of wavelengths. Here, we use the term homogeneous in terms of longitude and latitude; there is variation with altitude in all models we call homogeneous. 

\begin{table*}
	\centering  
	\begin{tabular}{ll}
		\hline
		\hline
		\rule{0pt}{3ex}Description & Associated figure \\
		\hline
		\multicolumn{2}{l}{\rule{0pt}{3ex}\textbf{Inhomogeneous model atmospheres: $F$ and $P$ as a function of wavelength for $\alpha$~=~90$\degree$ and $\alpha$~=~270$\degree$}}\\
		\rule{0pt}{3ex} WASP-96b setup including atmosphere types A$\dots$F, with and without clouds, and varying optical depth&  \makecell{\rule{0pt}{3ex} Fig.~\ref{fig:FP_all6_combined_90_270}}\\
		\rule{0pt}{3ex} WASP-96b setup including atmosphere types A$\dots$F, for irregularly shaped particles above 0.01~bar&  \makecell{\rule{0pt}{3ex} Fig.~\ref{fig:FP_all6_combined_90_270}}\\
		\Xhline{2\arrayrulewidth}
		\multicolumn{2}{l}{\rule{0pt}{3ex}\textbf{Inhomogeneous model atmospheres: $F$ and $P$ as a function of orbital phase for selected wavelengths}} \\
		\rule{0pt}{3ex}WASP-96b setup including atmosphere types A$\dots$F with clouds of mixed composition &  \makecell{\rule{0pt}{3ex} Fig.~\ref{fig:AF_phase_plot}}\\
		\rule{0pt}{3ex}WASP-96b setup including atmosphere types A$\dots$F with no clouds (clear) &  \makecell{\rule{0pt}{3ex} Fig.~\ref{fig:AF_phase_plot_clear}}\\
		\rule{0pt}{3ex} WASP-96b setup including atmosphere types A$\dots$F, for irregularly shaped particles above 0.01~bar&  \makecell{\rule{0pt}{3ex} Fig.~\ref{fig:AF_phase_plot_irregular}}\\
		\Xhline{2\arrayrulewidth}
		\multicolumn{2}{l}{\rule{0pt}{3ex}\textbf{Homogeneous model atmospheres: $F$ and $P$ as a function of wavelength for $\alpha$~=~90$\degree$}}\\
		\rule{0pt}{3ex}A series of 6 model atmospheres, of types A$\dots$F, each with mixed species used to form clouds &  \makecell{\rule{0pt}{3ex} Fig.~\ref{fig:FP_all6_individual}}\\
		\rule{0pt}{3ex}A series of models of atmosphere type B, each with one single species used to form the clouds &  \makecell{\rule{0pt}{3ex} Fig.~\ref{fig:B_species}}\\
		\rule{0pt}{3ex}A series of models of atmosphere type D, each with one single species used to form the clouds &  \makecell{\rule{0pt}{3ex} Fig.~\ref{fig:D_species}}\\
		\rule{0pt}{3ex}A series of 5 model atmospheres of type B, with varying irregularity of clouds particles &  \makecell{\rule{0pt}{3ex} Fig.~\ref{fig:FP_B_irregular}}\\
		\rule{0pt}{3ex}11 models of  type B, with one single species used to form clouds and 0.25/2.5~$\times$~10$^{-4}$~$\mu$m size distribution &  \makecell{\rule{0pt}{3ex} Fig.~\ref{fig:single_species_FP}}\\
		\rule{0pt}{3ex}11 models of  type B, with one single species used to form clouds and 0.1/0.01~$\mu$m size distribution &  \makecell{\rule{0pt}{3ex} Fig.~\ref{fig:single_species_FP_01}}\\
		\rule{0pt}{3ex}Models of  type B, each with Mg$_2$SiO$_4$ only used to form clouds and various Gaussian size distributions &  \makecell{\rule{0pt}{3ex} Fig.~\ref{fig:Mg2_size_dist_FP}}\\
		\Xhline{2\arrayrulewidth}
		\multicolumn{2}{l}{\rule{0pt}{3ex}\textbf{Homogeneous model atmospheres: $F$ and $P$ as a function of orbital phase for selected wavelengths}}\\
		\rule{0pt}{3ex}Mixed cloud composition type B, compared to models with one single species used to form the clouds &  \makecell{\rule{0pt}{3ex} Fig.~\ref{fig:phase_curves_single_species_F}/\ref{fig:phase_curves_single_species_P}}\\
		\hline
		\hline
	\end{tabular}
	\caption{A summary of the different model atmospheres computed in this study. Note that $\alpha$~=~90$\degree$ and $\alpha$~=~270$\degree$ are identical in the cases of (longitudinally/latitudinally) homogeneous atmospheres, so only $\alpha$~=~90$\degree$ is shown in those cases.}
	\label{t:setups} 
	\rule{0pt}{0.2ex}
\end{table*}

\section{Results for WASP-96b}\label{sec:results}

\subsection{Inhomogeneous model atmospheres: $F$ and $P$ as a function of wavelength for $\alpha$~=~90$\degree$ and $\alpha$~=~270$\degree$}\label{sec:FP_figures}

\begin{figure*}
	\centering
 	\includegraphics[width=0.49\textwidth]{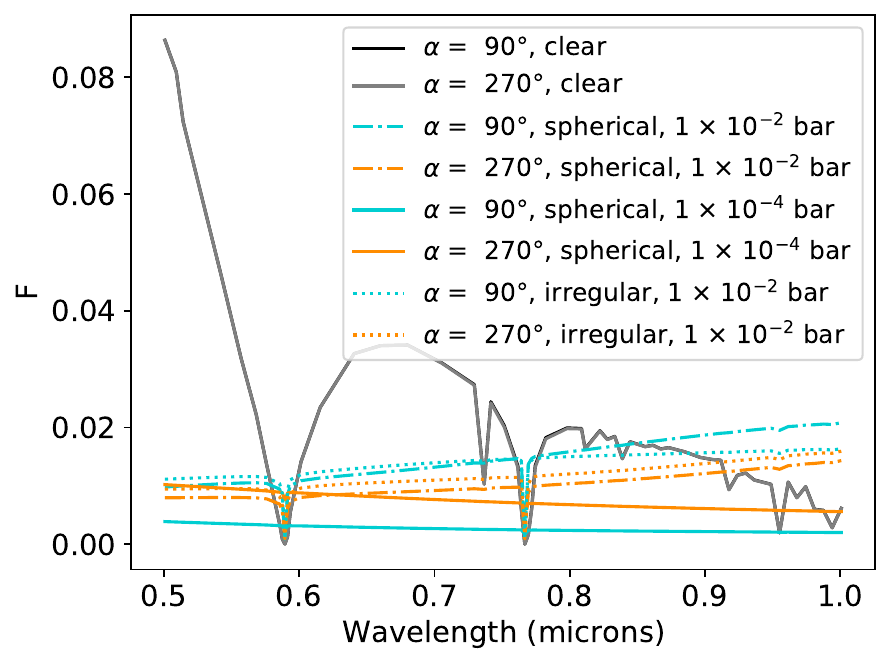}
	\includegraphics[width=0.49\textwidth]{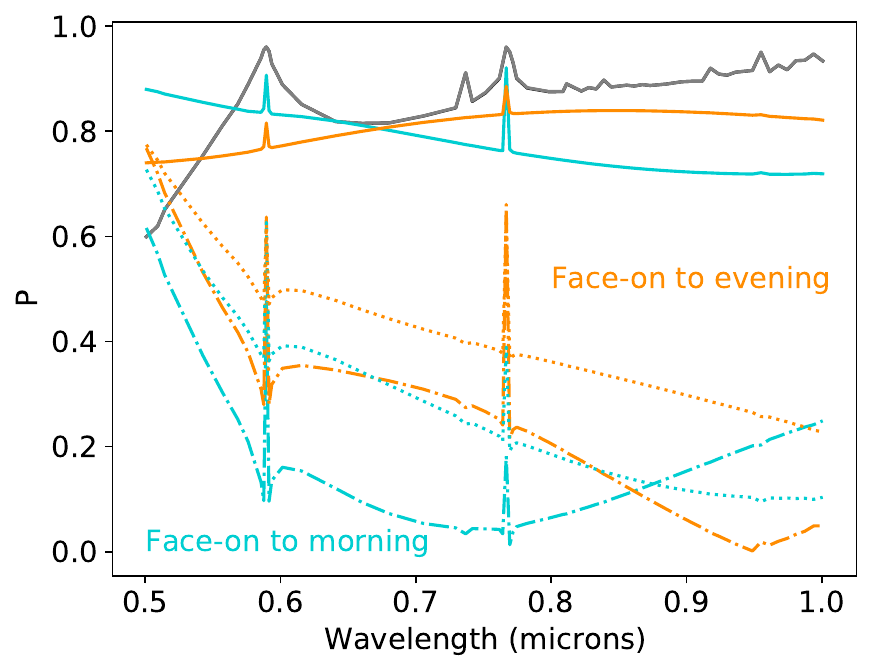}
	\caption{Reflected flux $F$ (left) and degree of linear polarisation $P$ (right) for our model \textbf{inhomogeneous} (i.e. varying as a function of longitude and latitude) WASP-96b atmosphere, assuming different properties for atmospheric regions $A$ to $F$, split as illustrated by Figure~\ref{fig:wasp96b_atm_grid}. The optical depth reaches unity due to clouds at the pressure level specified in the legend, i.e. 1~$\times$~10$^{-2}$~bar (dash-dotted orange and turquoise lines) and 1~$\times$~10$^{-4}$~bar (solid orange and turquoise lines).
 A comparison to the same model setup (same pressure-temperature profiles and molecular compositions) but with completely clear atmospheres is shown. The clear atmospheres are nearly identical so cannot easily be distinguished here. The dotted lines are for model atmospheres which reach optical depth at 1~$\times$~10$^{-2}$~bar but with irregularly shaped particles ($f_{\rm max}$~=~0.8) from 0.01~bar and above. We consider orbital phases of $\alpha$ of 90 and 270$\degree$. }\label{fig:FP_all6_combined_90_270}
\end{figure*}

Figure~\ref{fig:FP_all6_combined_90_270} gives $F$ and $P$ as a function of wavelength for different inhomogeneous model atmospheres (i.e. those which vary as a function of longitude $\phi_{\rm long}$  and latitude $\lambda_{\rm latt}$), at orbital phases of 90 and 270$\degree$. The full inhomogeneous atmosphere (orange and turquoise dashed lines) is with a cloud layer that becomes optically thick at 1~$\times$~10$^{-2}$~bar (see discussion at the start of Section~\ref{sec:atmosphere}). We also include models where the cloud layer becomes optically thick higher in the atmosphere, at 1~$\times$~10$^{-4}$~bar. The flux as a function of wavelength is relatively low in both scenarios, but the degree of linear polarisation is markedly different.  We also show completely clear (no cloud) models in Figure~\ref{fig:FP_all6_combined_90_270},  which are nearly identical for both $F$ and $P$ at phase 90 and 270$\degree$. This illustrates that it is the cloud particles (in particular their refractive properties and size distributions) which are causing the differences in reflected flux and degree of polarisation between the different model atmospheres. We explore further why these differences occur, using model homogeneous models and scattering properties of different species, in Sections~\ref{sec:FP_hom} and~\ref{sec:FP_phase_hom}. 

Similar to the models of Jupiter-like exoplanets in \cite{04StHoWa}, the clear atmosphere in Figure~\ref{fig:FP_all6_combined_90_270} has a general trend of decreasing $F$ with $\lambda$, due to a decrease in the molecular scattering optical thickness with $\lambda$. $P$ has a corresponding general increase with $\lambda$, due to less multiple scattering taking place at longer wavelengths because of the lower molecular scattering cross-section. Multiple scattering typically lowers the degree of polarisation $P$ for reflected light. The regions of increased molecular and atomic absorption can clearly be seen in the clear spectra of Figure~\ref{fig:FP_all6_combined_90_270} (left). The most prominent absorption features occur around 0.6~$\mu$m and just under 0.8~$\mu$m, due to strong resonance transition doublets of Na and K~\citep{16AlSpKi.broad,19AlSpLe.broad}. In these regions the strong absorption also causes less multiple scattering to take place. This absorption thus leads to low $F$ and high $P$. 

In the models described so far, it is assumed that all cloud particles are spherical, with scattering properties computed using Mie theory. We also compare to a full inhomogeneous atmosphere with the same properties as in the dashed lines of Figure~\ref{fig:FP_all6_combined_90_270}, but with irregularly shaped instead of spherical particles used for modelling the particles at pressure layers of 0.01~bar and above, for the atmosphere where the clouds become optically thick at 1~$\times$~10$^{-2}$~bar. We use the optool code\footnote{\url{https://github.com/cdominik/optool}}~\citep{21DoMiTa} (see Section~\ref{sec:nonspherical_scattering}) for modelling the scattering properties of the irregularly shaped particles. The value of $f_{\rm max}$ indicates the irregularity of the particle, with 0 a sphere (red) and higher values being more irregular. If the dotted (irregular particles) and dashed (spherical particles) lines of Figure~\ref{fig:FP_all6_combined_90_270} are compared, the difference between inhomogeneous atmospheres with spherical and very irregular particles ($f_{\rm max}$~=~0.8) can be seen, particularly for the degree of linear polarisation $P$. The irregularly shaped particles generally lead to a higher $F$ and $P$ than the spherical particles, although this behaviour does not hold for the higher wavelengths of the morning terminator models for both $F$ and $P$.

\subsection{Inhomogeneous model atmospheres: $F$ and $P$ as a function of orbital phase for selected wavelengths}\label{sec:FP_phase_inhom}

\begin{figure*}
	\centering
	\includegraphics[width=0.49\textwidth]{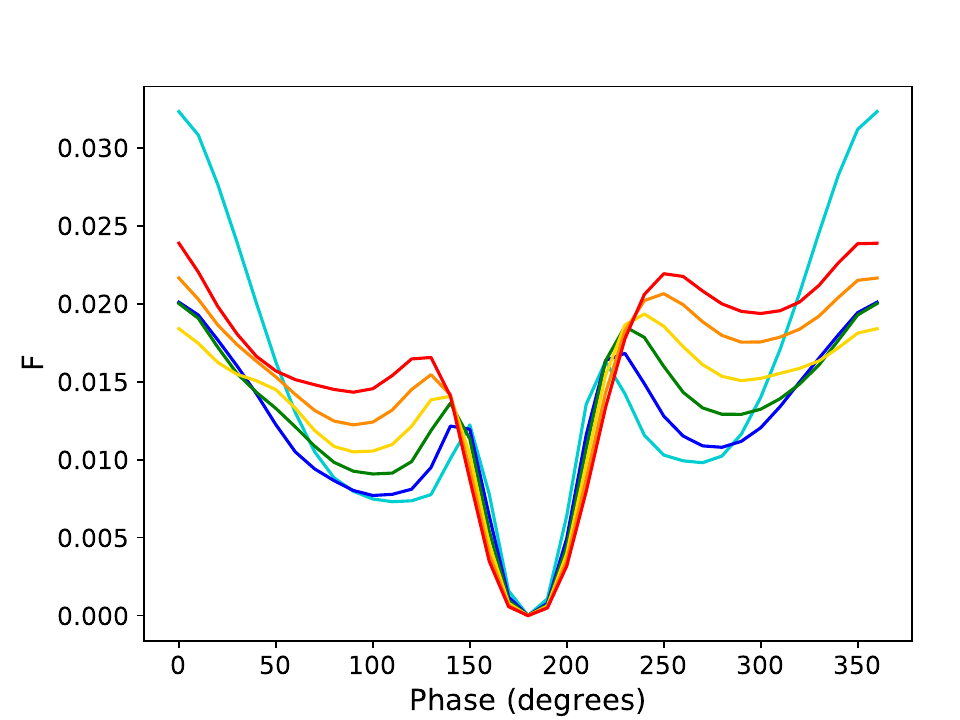}
	\includegraphics[width=0.49\textwidth]{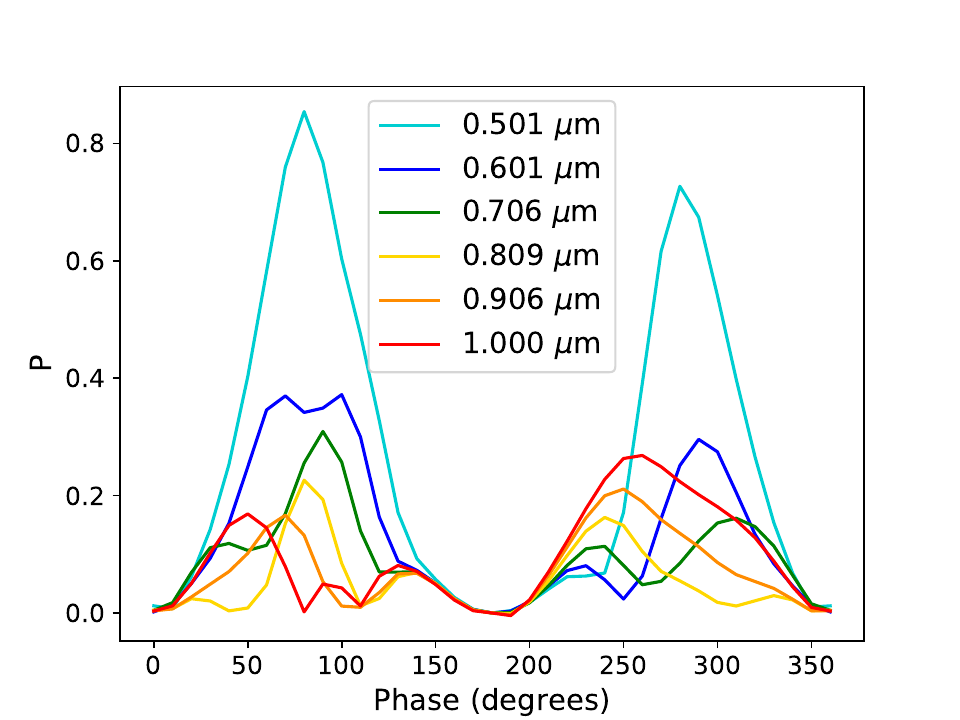}
	\caption{Reflected flux $F$ (left) and degree of linear polarisation $P$ (right) as a function of orbital phase $\alpha$ and at various wavelengths, for the inhomogeneous WASP-96b model atmosphere.}\label{fig:AF_phase_plot}
\end{figure*}

\begin{figure*}
	\centering
	\includegraphics[width=0.49\textwidth]{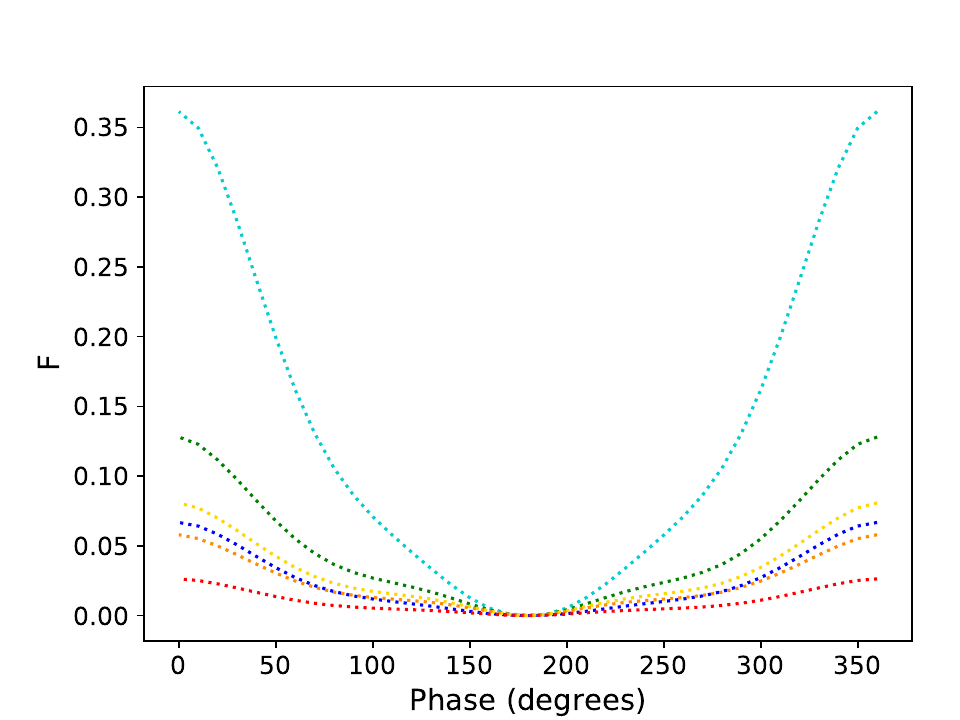}
	\includegraphics[width=0.49\textwidth]{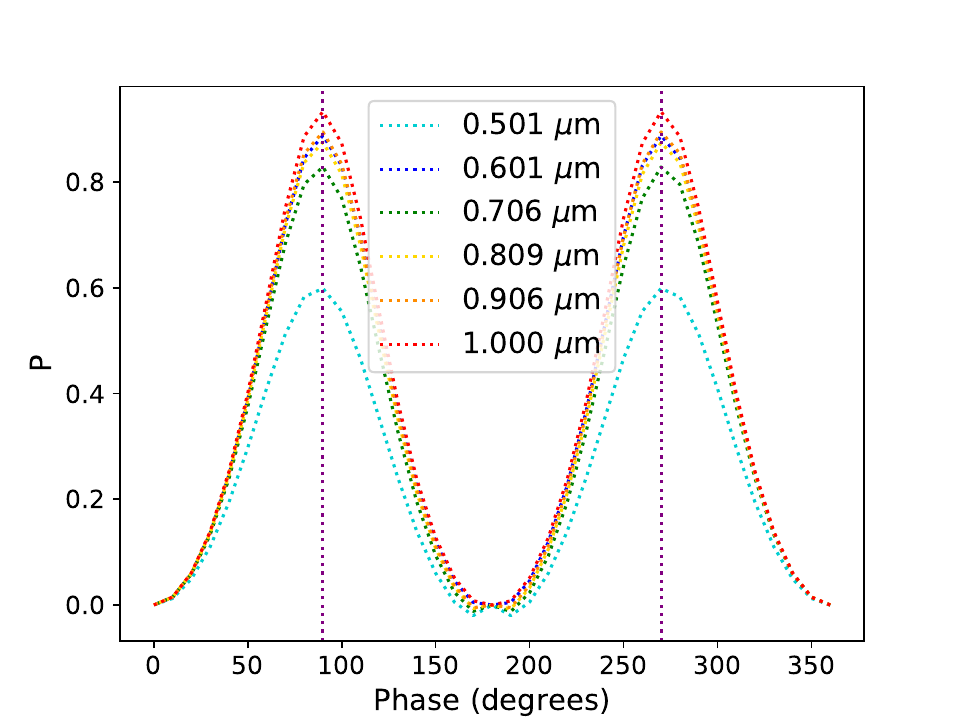}
	\caption{Reflected flux $F$ (left) and degree of linear polarisation $P$ (right) as a function of orbital phase $\alpha$ and at various wavelengths, for the inhomogeneous WASP-96b model atmosphere with a clear (no-cloud) atmosphere. Vertical lines at 90 and 270$\degree$ are shown for reference. }\label{fig:AF_phase_plot_clear}
\end{figure*}

\begin{figure*}
	\centering
	\includegraphics[width=0.49\textwidth]{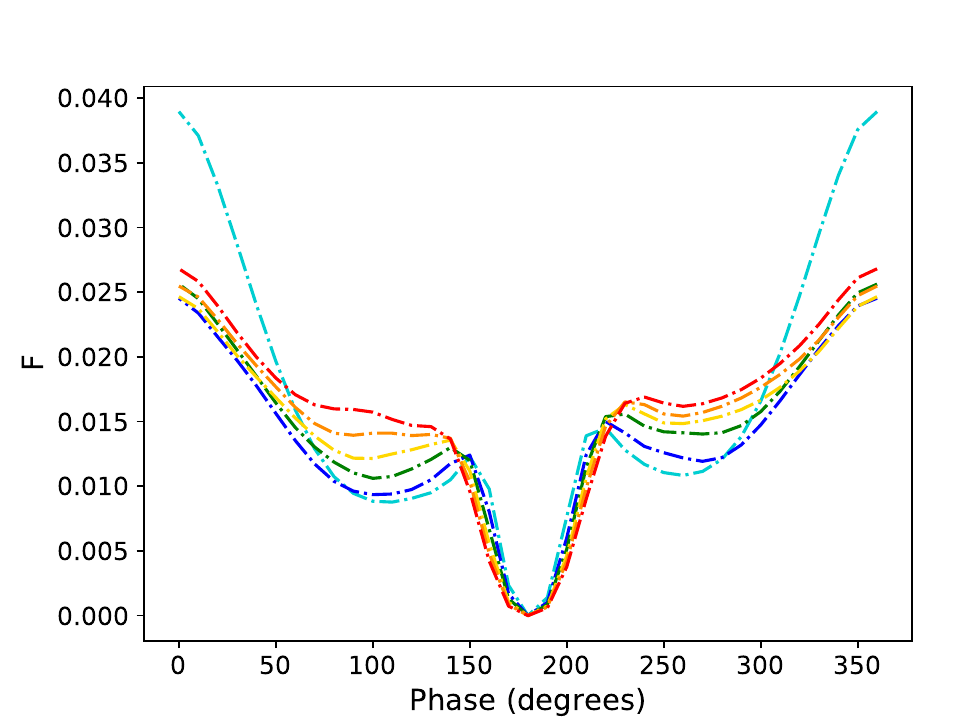}
	\includegraphics[width=0.49\textwidth]{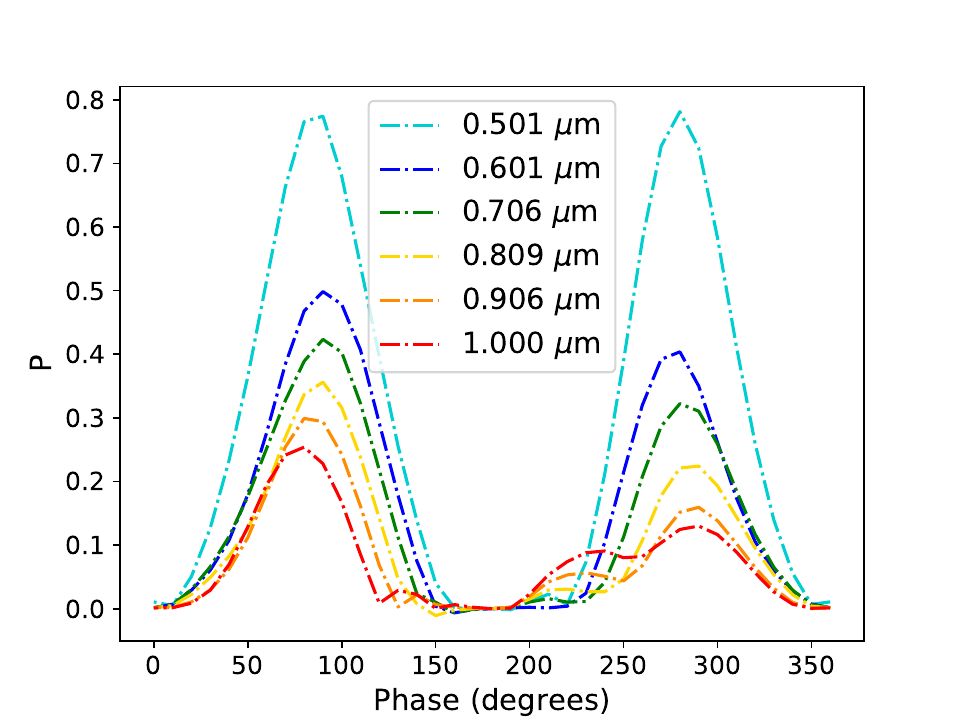}
	\caption{Reflected flux $F$ (left) and degree of linear polarisation $P$ (right) as a function of orbital phase $\alpha$ and at various wavelengths, for the inhomogeneous WASP-96b model atmosphere with irregular instead of spherical particles for pressure layers of 0.01~bar and above.}\label{fig:AF_phase_plot_irregular}
\end{figure*}

In Figure~\ref{fig:AF_phase_plot}, we plot $F$ (left) and $P$ (right) as a function of orbital phase for selected wavelengths between 0.5~-~1~$\mu$m for the full inhomogeneous model atmosphere (with clouds becoming optically thick at 1~$\times$~10$^{-2}$~bar , relating to the dashed lines in Figure~\ref{fig:FP_all6_combined_90_270}).  The inhomogeneity can be clearly seen by the lack of symmetry either side of 180$\degree$ for both $F$ and $P$.  Figure~\ref{fig:AF_phase_plot_clear} gives the same output but for a clear atmosphere, which appears symmetric about 180$\degree$. It can be seen that $P$ peaks at 90 and 270$\degree$, due to Rayleigh scattering by the atoms and molecules in the atmosphere. Figure~\ref{fig:AF_phase_plot_irregular} gives the inhomogeneous model atmospheres with irregular rather than spherical particles at pressure layers of 0.01~bar and above. 

\subsection{Homogeneous model atmospheres: $F$ and $P$ as a function of wavelength for $\alpha$~=~90$\degree$}\label{sec:FP_hom}

In order to explore why the inhomogeneous atmosphere looks as it does for $F$ and $P$, in Figure~\ref{fig:FP_all6_individual} we show a series of models of homogeneous atmospheres (i.e. atmospheres which do not vary as a function of longitude and latitude, but only by altitude) of types A$\dots$F (where A$\dots$F are the six atmospheric regions as defined in Figure~\ref{fig:wasp96b_atm_grid}). The models of  Figure~\ref{fig:FP_all6_individual} can be compared to the dashed lines in Figure~\ref{fig:FP_all6_combined_90_270}, which gives the equivalent inhomogeneous atmosphere. There, the $\alpha$~=~90$\degree$ model has contributions from atmospheric regions D, E, A, and F (i.e. the morning side of the planet), while the $\alpha$~=~90$\degree$ model  has contributions from regions B, C, A and F (the evening side). We note from Figure~\ref{fig:FP_all6_individual} that atmospheric region B and E are very similar to one another, as are A and C. This is due to the hot spot shift away from the sub-stellar point and towards the warmer evening terminator, which is a result of strong equatorial winds driven by the tidally-locked nature of the planet.

In order to now focus on one atmospheric type at a time, we focus on region B (around the evening terminator) and D (around the morning terminator). 
Figures~\ref{fig:B_species} and~\ref{fig:D_species} show $F$ and $P$ for some homogeneous atmospheres, using the atmospheric setup for region B (Figure~\ref{fig:B_species}) and D (Figure~\ref{fig:D_species}). The model with mixed composition clouds are shown (labelled as all species), along with the same setup but with single materials only used to form the clouds. {We note that we use the same particle size distribution for all single-species scenarios here in order to offer a more direct comparison to the mixed-material scenario, but the single-particle models are purely theoretical and not based on physically motivated model atmospheres. See Section~\ref{sec:size_dist} for more discussion on particle size distributions.
It can be seen from Figures~\ref{fig:B_species} and~\ref{fig:D_species} that the model WASP-96b atmosphere is largely dominated by the optical properties of the Fe-bearing species which are used to form the mixed-composition clouds. For Figure~\ref{fig:B_species} in particular, the atmosphere with mixed-composition clouds  and FeO-only clouds are nearly identical in $F$, but differ in $P$. The flux as a function of wavelength for a homogeneous atmosphere of type B is almost identical to the same model atmosphere but with FeO used to form all cloud particles, instead of the mixed cloud particles (see Section~\ref{sec:cloud}).  The examples shown in Figures~\ref{fig:B_species} and~\ref{fig:D_species} are for clouds made purely of Al$_2$O$3$, Fe$_2$O$_3$, FeO, Mg$_2$SiO$_4$, or MgO. It can be seen from Figures~\ref{fig:single_species_FP}~and~\ref{fig:single_species_FP_01} that the optical properties of Fe$_2$SiO$_4$ lead it to share more similarities with the silicate and oxide species than the iron species. Such atmospheres have a high single scattering albedo across all wavelengths (see Figure~\ref{fig:sing_scat_11species_plus90}), so a significant fraction of the incoming stellar light would be reflected out, some of it towards the observer, before it can be absorbed. 

As for the inhomogeneous atmospheres, we also explore homogeneous atmospheres using irregularly shaped instead of spherical cloud particles. Figure~\ref{fig:FP_B_irregular} shows $F$ (left) and degree of linear polarisation $P$ (right) for a model homogeneous planet based on WASP-96b at $\alpha$~=~90$\degree$ assuming atmospheric type B only, but with varying irregularity of the cloud particles. The value of $f_{\rm max}$ indicates the irregularity of the particle, with 0 a sphere (red) and higher values being more irregular. The effect of using irregular instead of spherical particles can be clearly seen. 

Figures~\ref{fig:single_species_FP}~-~\ref{fig:Mg2_size_dist_FP} give an indication of how the particle size distribution affects $F$ and $P$ as a function of wavelength. All three figures give $F$ and $P$ as a function of wavelength at $\alpha$~=~90$\degree$ for model homogeneous atmospheres of type B. Figure~\ref{fig:single_species_FP} uses a Gaussian particle size distribution with average particle size 0.25~$\mu$m and standard deviation 2.5 ~$\times$~10$^{-4}$ for all models, while Figure~\ref{fig:single_species_FP_01} uses a Gaussian particle size distribution with average particle size 0.1~$\mu$m and standard deviation 0.01 for all models. Both present 11 different models, each with a single species used to form the clouds, as labelled. Figure~\ref{fig:Mg2_size_dist_FP} focuses on model atmospheres each with only Mg$_2$SiO$_4$ used to form the clouds, but this time with varying parameters used for the Gaussian size distributions. 

\begin{figure*}
	\centering
	\includegraphics[width=0.49\textwidth]{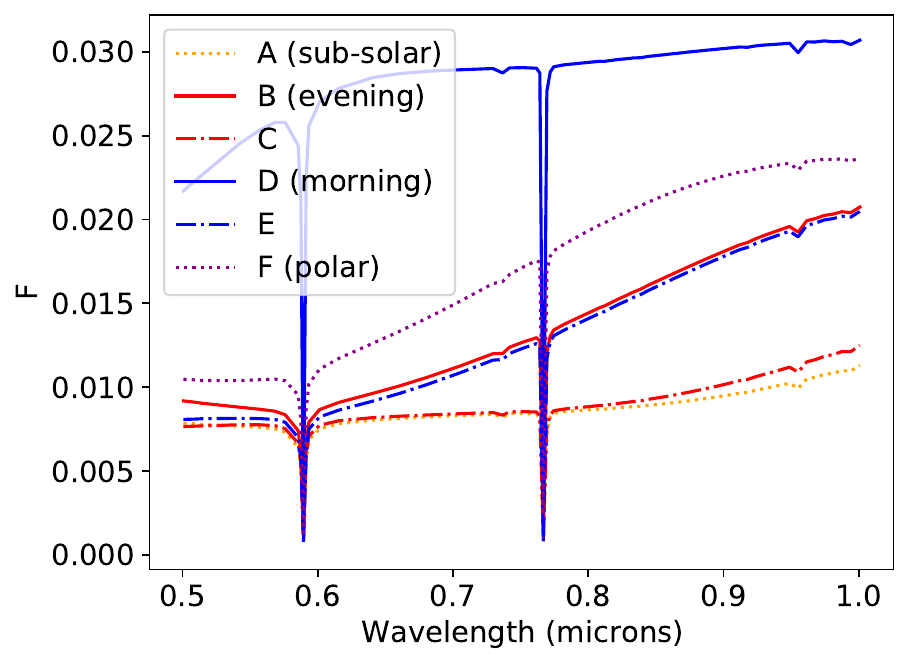}
	\includegraphics[width=0.49\textwidth]{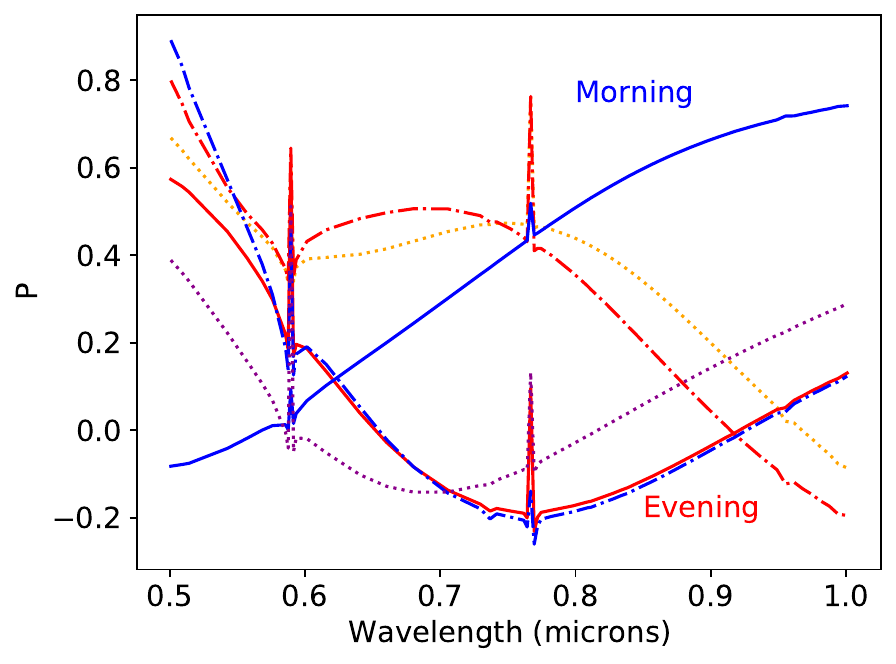}
	\caption{Reflected flux $F$ (left) and degree of linear polarisation $P$ (right) for 6 model homogeneous planets, each assuming a full atmosphere covered by one atmospheric region (A to F), as labelled. Region A covers the region around the sub-stellar point, B the hotter evening terminator, and D the cooler morning terminator. }\label{fig:FP_all6_individual}
\end{figure*}

\begin{figure*}
	\centering
	\includegraphics[width=0.50\textwidth]{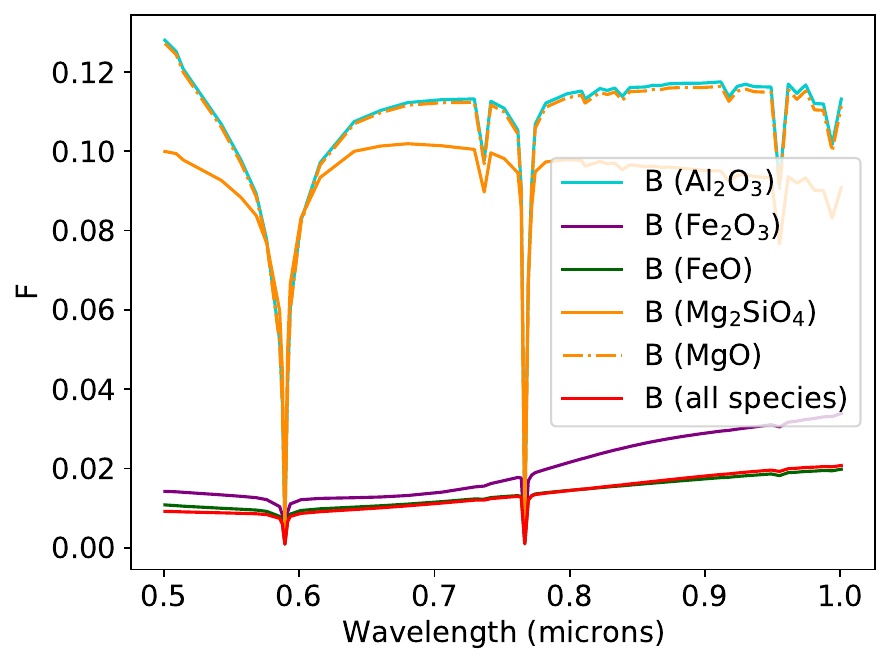}
	\includegraphics[width=0.48\textwidth]{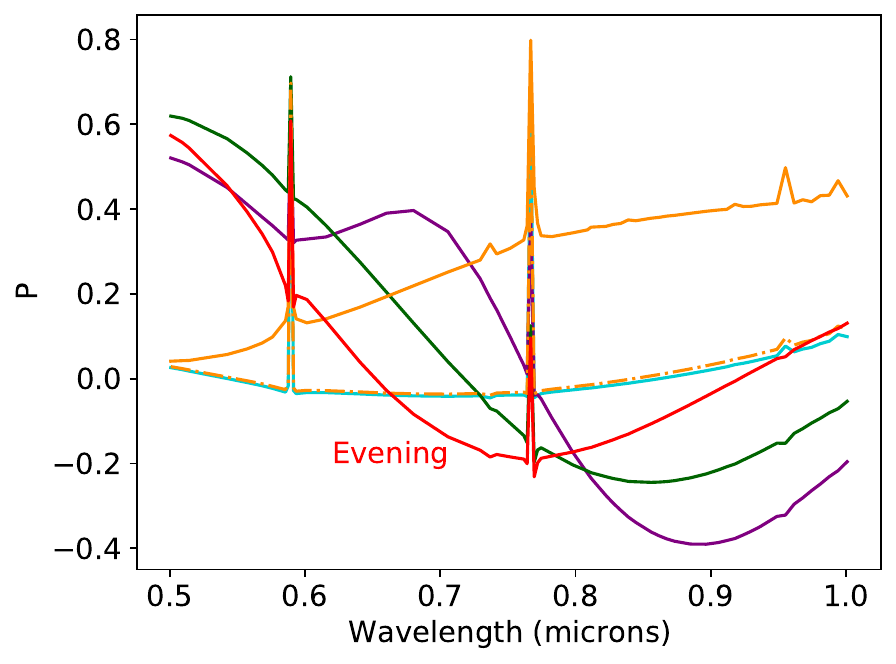}
	\caption{Reflected flux $F$ (left) and degree of linear polarisation $P$ (right) for model homogeneous planets based on WASP-96b at $\alpha$~=~90$\degree$ assuming atmospheric type B (around the evening terminator) only. Some examples of clouds made of single materials only are shown. $F$ and $P$ for a homogeneous atmosphere with the full atmospheric setup for region B are shown for comparison (i.e. with clouds formed from mixed materials). In this case all size distributions of different layers for the single species cases are the same as in the mixed composition atmosphere.}\label{fig:B_species}
\end{figure*}

\begin{figure*}
	\centering
	\includegraphics[width=0.49\textwidth]{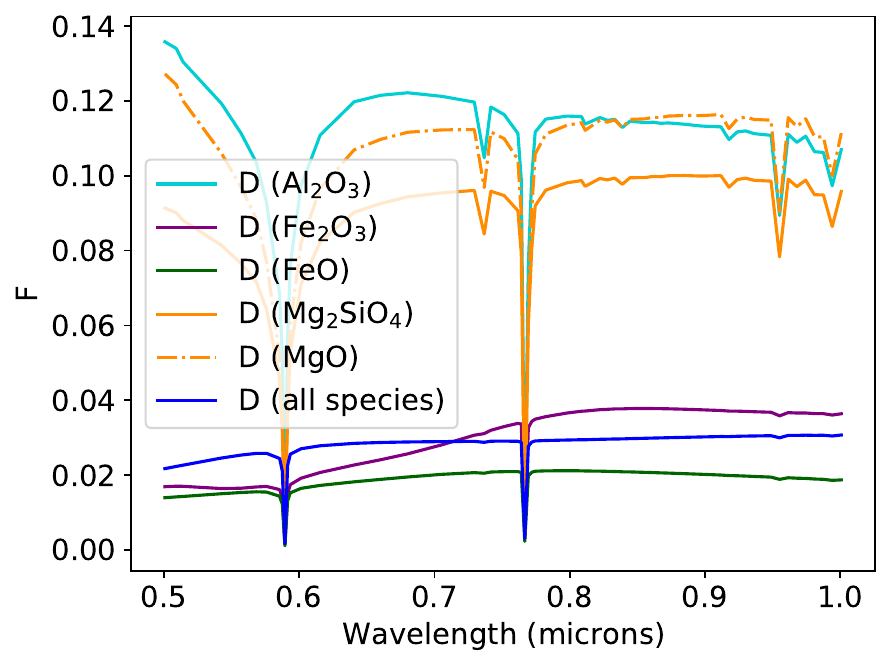}
	\includegraphics[width=0.49\textwidth]{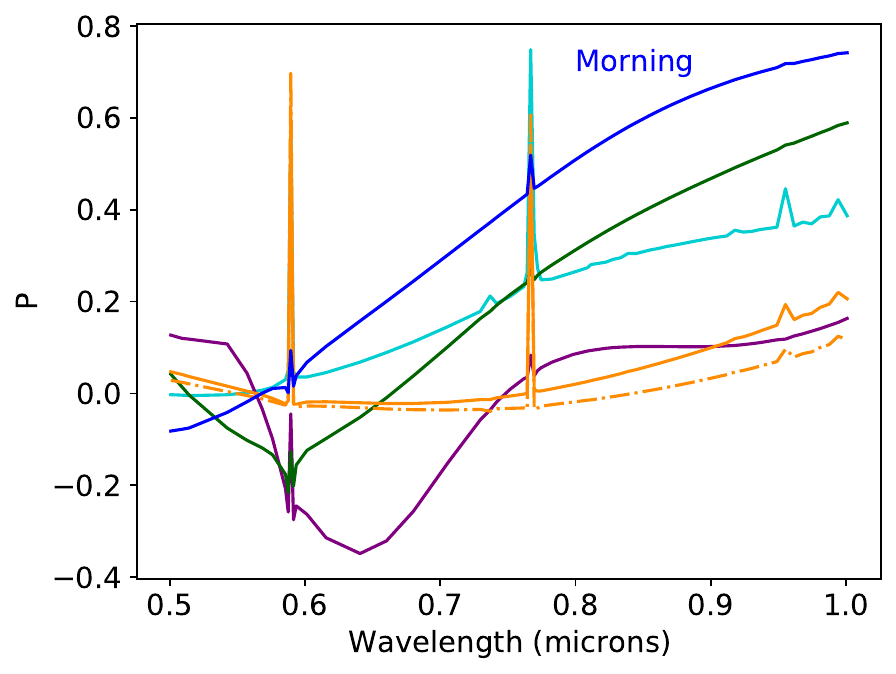}
	\caption{Reflected flux $F$ (left) and degree of linear polarisation $P$ (right) for model homogeneous planets based on WASP-96b at $\alpha$~=~90$\degree$ assuming atmospheric type D (around the morning terminator) only. Some examples of clouds made of single materials only are shown. $F$ and $P$ for a homogeneous atmosphere with the full atmospheric setup for region D are shown for comparison (i.e. with clouds formed from mixed materials). In this case all size distributions of different layers for the single species cases are the same as in the mixed composition atmosphere.}\label{fig:D_species}
\end{figure*}

\begin{figure*}
	\centering
	\includegraphics[width=0.49\textwidth]{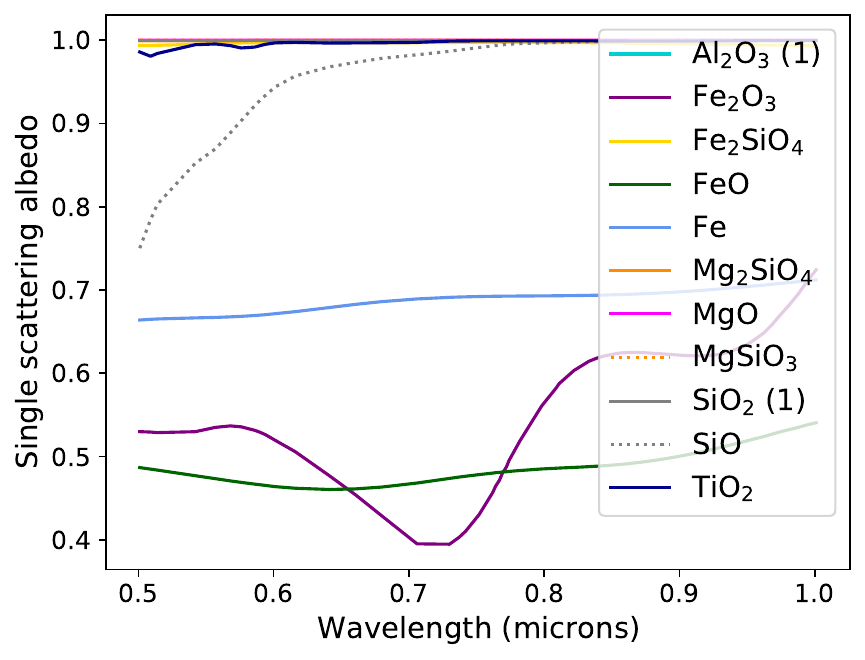}
	\includegraphics[width=0.49\textwidth]{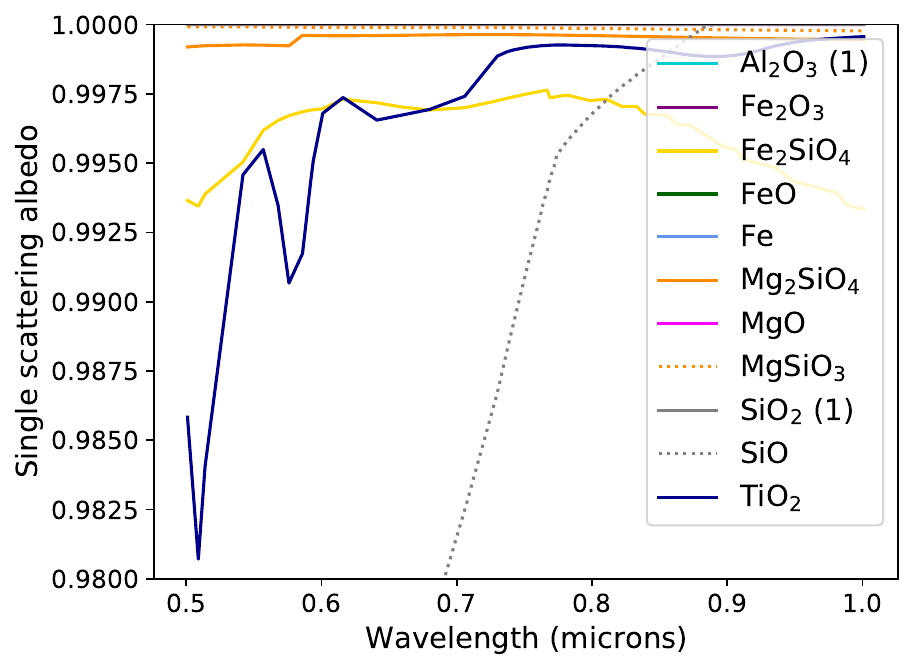}
	\caption{Single scattering albedo (${\rm ssa} = \frac{k_{\rm scat}}{k_{\rm ext}}$) of various species used in this study. The species with high values of the imaginary part of the refractive index $k$ (Fe, Fe$_2$O$_3$, FeO) have lower values of single scattering albedo due to them being highly absorbing in comparison to the other species. Atmospheres with large mixing ratios of these iron-bearing have lower flux (due to high absorption) and relatively low degree of linear polarisation. The species in the legend with a (1) after their name are those with $k$~=~0 and thus  ${\rm ssa}$~=~1 for all wavelengths shown. The panel on the right is the same as the left but zoomed in for clarity. 
 }\label{fig:sing_scat_11species_plus90}
\end{figure*}

\begin{figure*}
	\centering
	\includegraphics[width=0.49\textwidth]{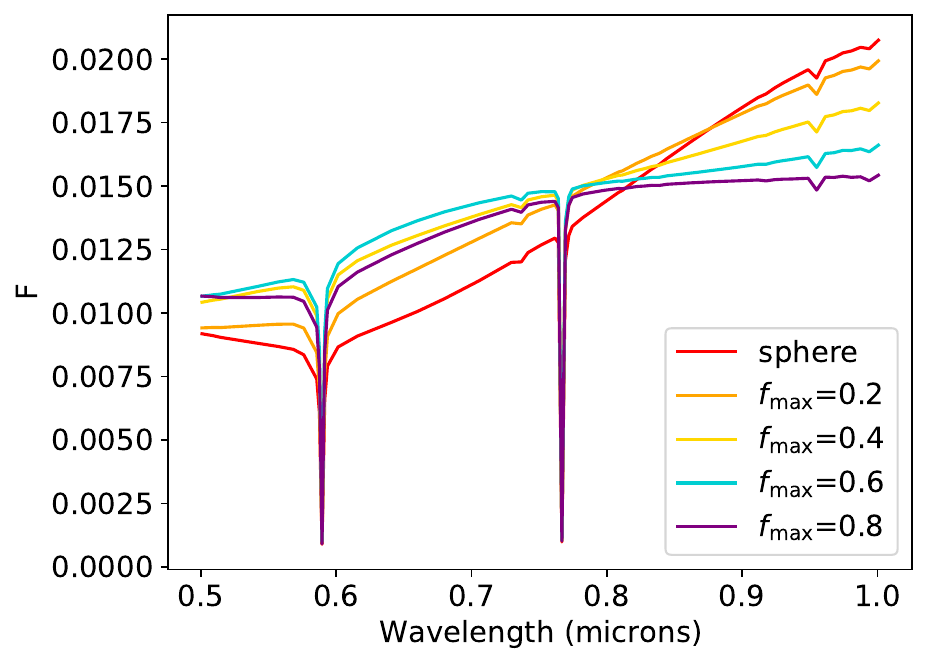}
	\includegraphics[width=0.49\textwidth]{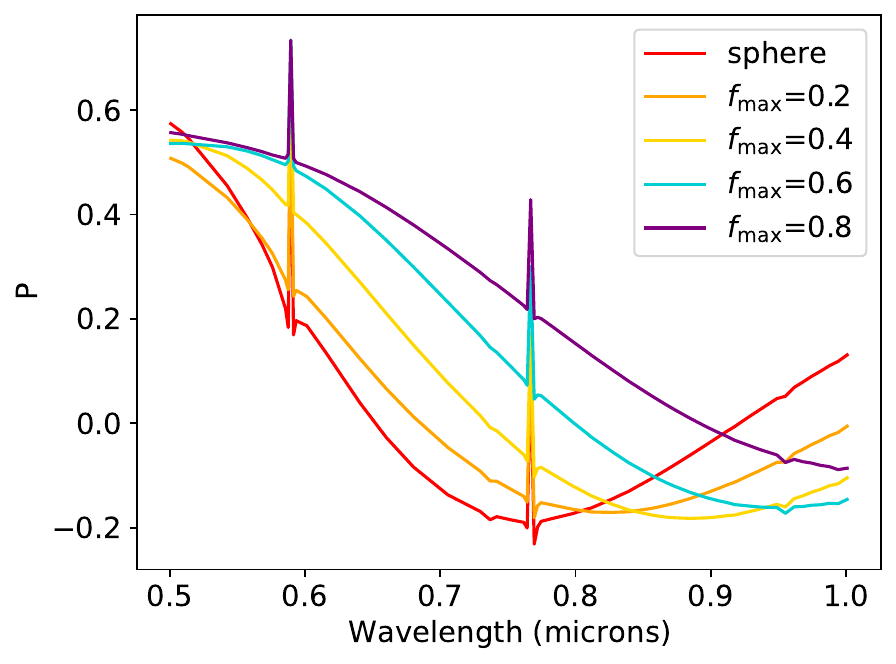}
	\caption{Reflected flux $F$ (left) and degree of linear polarisation $P$ (right) for a model homogeneous planet based on WASP-96b at $\alpha$~=~90$\degree$ assuming atmospheric type B only, but with varying irregularity of the cloud particles. The value of $f_{\rm max}$ indicates the irregularity of the particle, with 0 a sphere (red) and higher values being more irregular.}\label{fig:FP_B_irregular}
\end{figure*}

\begin{figure*}
	\centering
	\includegraphics[width=0.43\textwidth]{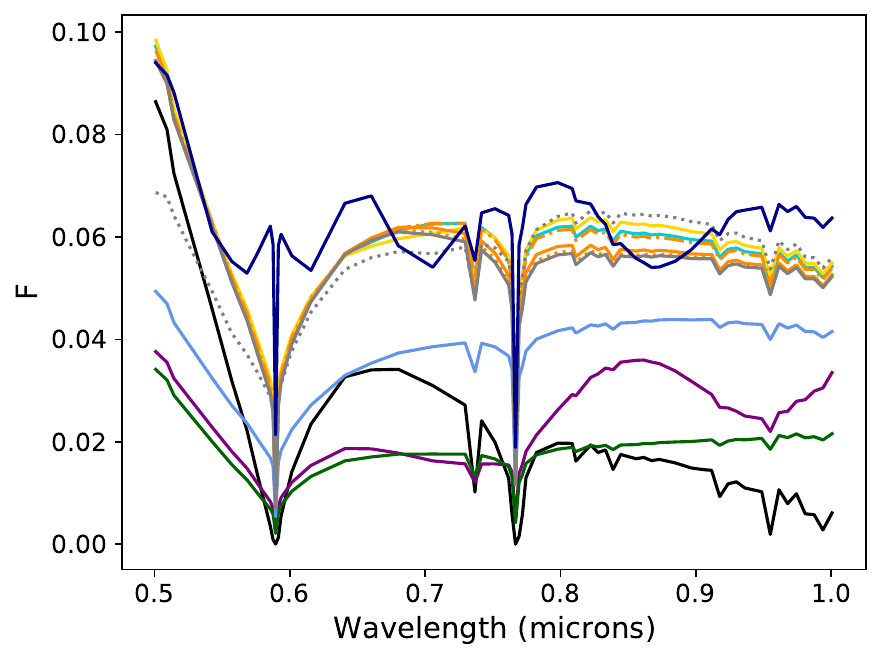}
	\includegraphics[width=0.55\textwidth]{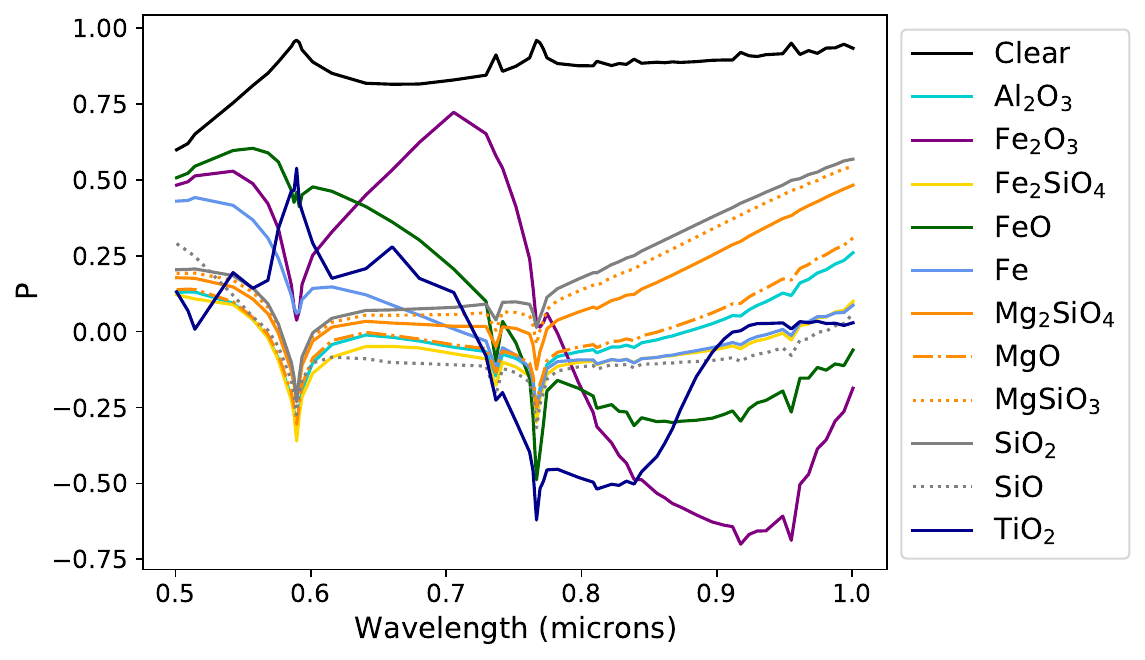}
	\caption{Reflected flux $F$ (left) and degree of linear polarisation $P$ (right) for a model homogeneous atmosphere based on WASP-96b at $\alpha$~=~90$\degree$, using a different single material to form the clouds for each. A Gaussian size distribution is used, with 2.5~$\times$~10$^{-4}$$\sigma$ standard deviation around an average particle size of 0.25~$\mu$m, for atmosphere type B.}
	\label{fig:single_species_FP}
\end{figure*}

\begin{figure*}
	\centering
	\includegraphics[width=0.43\textwidth]{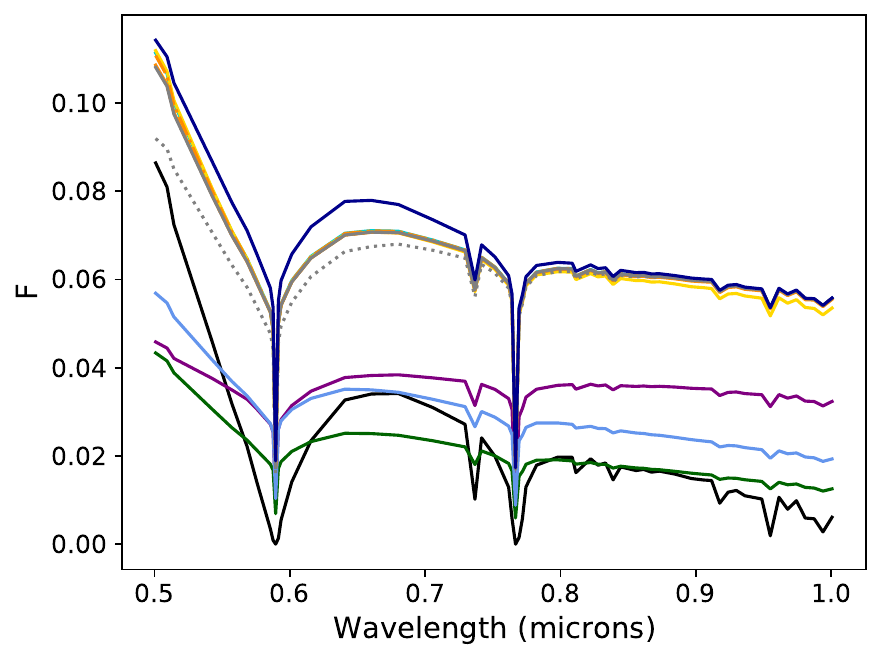}
	\includegraphics[width=0.55\textwidth]{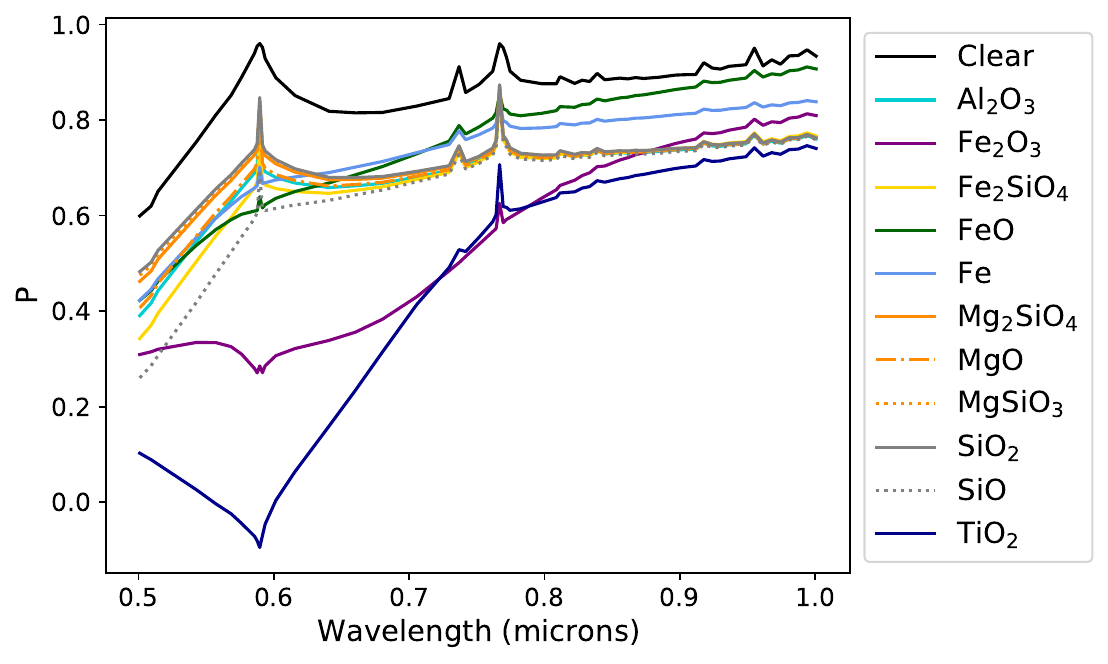}
	\caption{Reflected flux $F$ (left) and degree of linear polarisation $P$ (right) for a model homogeneous atmosphere based on WASP-96b at $\alpha$~=~90$\degree$, using a different single material to form the clouds for each. A Gaussian size distribution is used, with 0.01$\sigma$ standard deviation around an average particle size of 0.1~$\mu$m, for atmosphere type B. }
	\label{fig:single_species_FP_01}
\end{figure*}

\begin{figure*}
	\centering
	\includegraphics[width=0.41\textwidth]{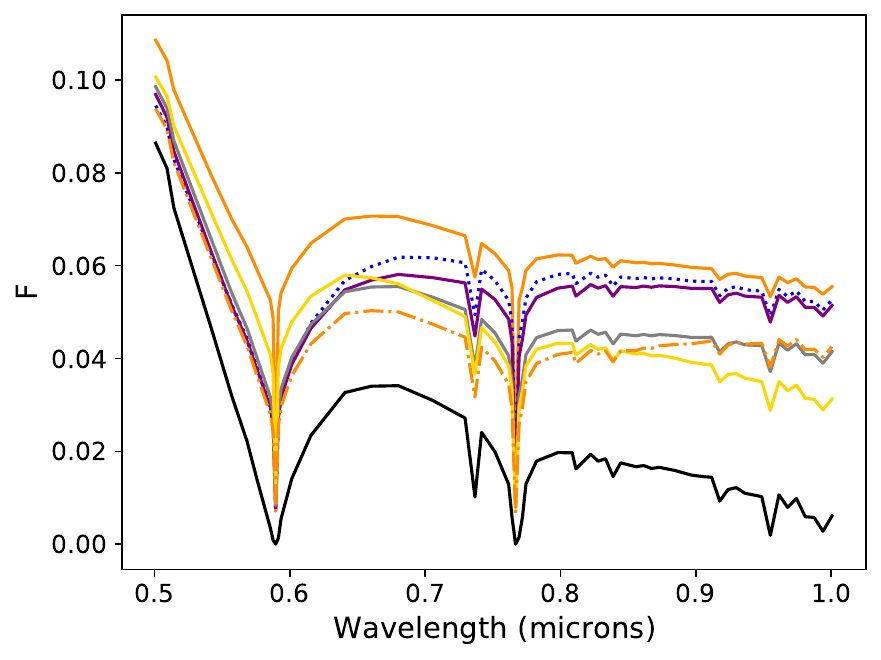}
	\includegraphics[width=0.57\textwidth]{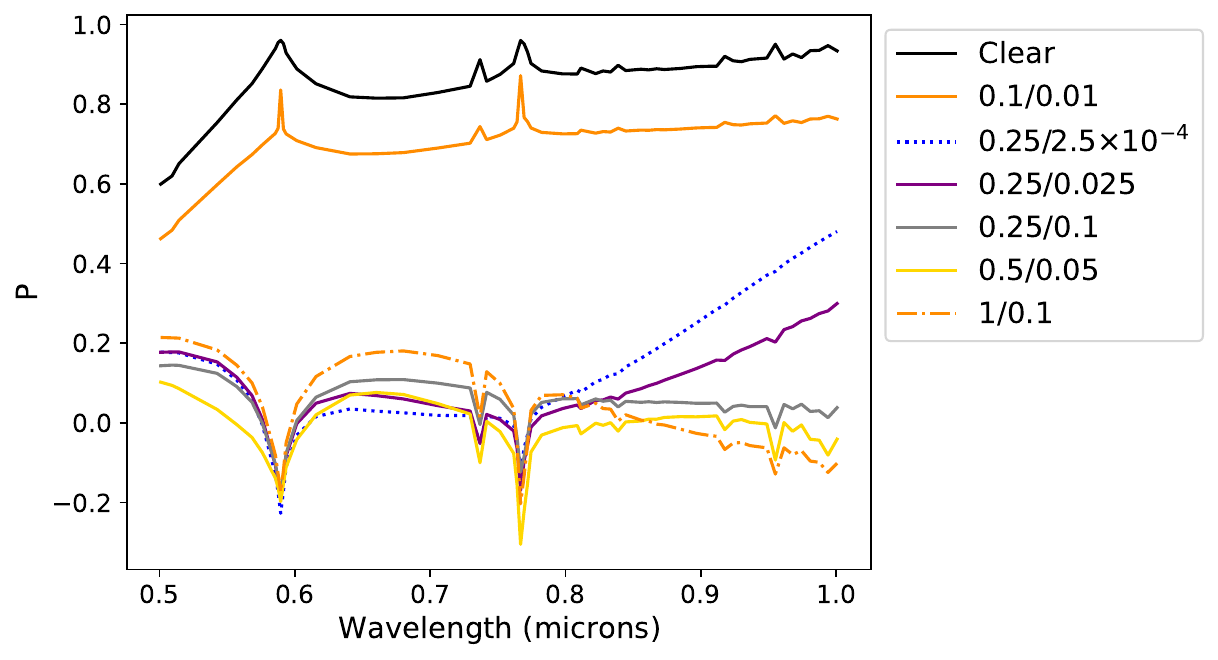}
	\caption{Reflected flux $F$ (left) and degree of linear polarisation $P$ (right) for a model homogeneous atmosphere based on WASP-96b at $\alpha$~=~90$\degree$, using a different single material Mg$_2$SiO$_4$ to form the clouds. Gaussian size distribution are used for each with the average particle size (in units of $\mu$m) and standard deviation $\sigma$ as labelled.}
	\label{fig:Mg2_size_dist_FP}
\end{figure*}

\subsection{Homogeneous model atmospheres: $F$ and $P$ as a function of orbital phase for selected wavelengths}\label{sec:FP_phase_hom}

Figures~\ref{fig:phase_curves_single_species_F} and~\ref{fig:phase_curves_single_species_P} give the phase curves (i.e. variation of $F$ or $P$ with orbital phase) of different homogeneous atmospheres for selected wavelengths. All panels are assuming atmosphere type B, with the majority of the panels showing atmospheres with clouds made up of a single species (Al$_2$O$_3$, Fe$_2$O$_3$, FeO, Mg$_2$SiO$_4$, MgO) only. These are the same models as in Figure~\ref{fig:B_species}. Phase curves for a homogeneous planet with atmosphere type B but containing clouds made up of mixed species (as described in Section~\ref{sec:eff_med_theory}) are shown for comparison. It can be seen that different materials give different signatures, particularly when looking at the degree of linear polarisation $P$. Note the different scales on the y-axis; those which are less reflective such as FeO and Fe$_2$O$_3$ are also generally more highly polarising than  the other more reflective materials.

\subsection{Geometric albedo}

As introduced in Section~\ref{sec:geo_albedo}, the geometric albedo $A_G$ as a function of wavelength can be found by looking at the reflected flux at $\alpha$~=~0$\degree$. $A_G$ is plotted in Figure~\ref{fig:geo_albedo_B} for the full inhomogeneous atmosphere setup, and for the homogeneous atmosphere setup of type B. The latter either includes mixed-species used to form clouds, or only a single species used to form the clouds (A$_2$O$_3$, Fe$_2$O$_3$, FeO, Mg$_2$SiO$_4$, MgO). 
The geometric albedo of  a population of around 20 hot gaseous exoplanets have been measured by studies such as \cite{15AnDeMo} and \cite{15EsDeJa}, with the finding that the majority have albedos typically less than 0.15 in the Kepler bandpass (0.42~-~0.91~$\mu$m).  Two notable exceptions are HAT-P-7b with a measured  geometric albedo of  0.23~\citep{13HeDe} and Kepler-7b with 0.25~\citep{21HeMoKi}. 
In our models, it  can be seen that the highly absorbing Fe-bearing species FeO causes the geometric albedo to be very low for both the full inhomogeneous atmosphere and the homogeneous mixed species atmosphere of type B. If only species with similar properties to silicates and oxides like Al$_2$O$_3$, Fe$_2$O$_3$, Mg$_2$SiO$_4$, MgO  were included in the atmosphere then the geometric albedo (which can be measured from observations) would be much higher. 
We note that it is known that there can be errors in calculated wavelength-dependent planetary phase functions  and albedos due to treating light as a scalar and not as a vector, by neglecting polarization~\citep{05StHo}. An investigation on the impact of cloud materials on measured geometric albedo warrants further investigation. 

\begin{figure*}
	\centering
	\includegraphics[width=0.7\textwidth]{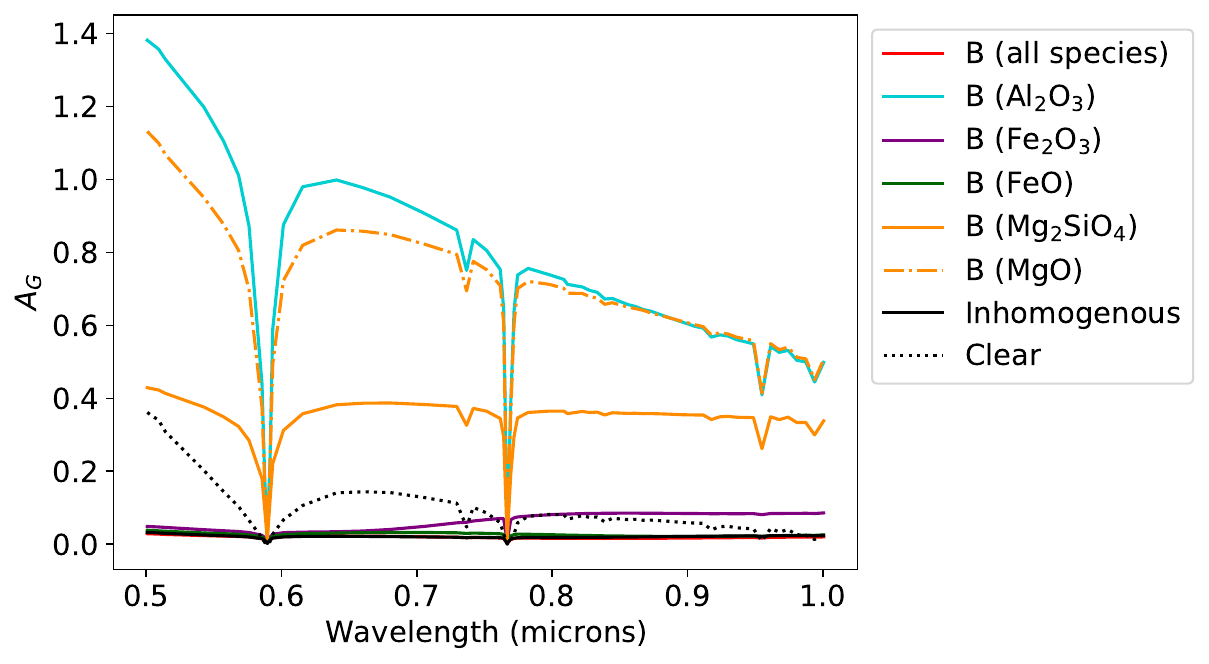}
	\caption{Geometric albedo $A_G$ as a function of wavelength for: the full inhomogeneous setup including mixed-material clouds, the clear inhomogeneous setup, and model homogeneous planets assuming atmospheric type B (around the evening terminator) only. Homogeneous model atmosphere of  atmospheric type B are also shown: either with all cloud species (i.e. with clouds formed from mixed materials), or  with some examples of single materials only.}\label{fig:geo_albedo_B}
\end{figure*}

\section{Discussion of results}\label{sec:discussion}

\subsection{Impact of effective refractive index of materials used to form clouds}\label{sec:ref_index_impact}

As previously mentioned, the imaginary part of the refractive index $k$ of cloud particles relates to absorption, while the real part $n$ relates to scattering. Materials considered in this study which have high values of $k$ and relatively lower values of $n$ (see Figure~\ref{fig:n_k}) and thus low single scattering albedos (Figure~\ref{fig:sing_scat_11species_plus90}) are all Fe-bearing species (Fe, FeO, Fe$_2$O$_3$). Theoretical atmospheres composed of such species, as illustrated by Figure~\ref{fig:single_species_FP},
have relatively lower $F$ across all wavelengths. Figure~\ref{fig:sing_scat_11species_plus90} gives some insights into the scattering behaviour of particles formed from different materials as a function of wavelength. Fe$_2$O$_3$ for example has a single scattering albedo which varies significantly as a function of wavelength. The impact of Fe-bearing materials forming clouds in our model exoplanet atmospheres can be seen in Figures~\ref{fig:B_species} and~\ref{fig:D_species}. The left panel of each shows $F$ as a function of wavelength for single species compared to the full setups (including mixed cloud species) for homogeneous atmosphere setups of type B or D, respectively. It can be seen that the models which include either the mixed cloud species or FeO only have much lower $F$ across all wavelengths than those which include only Al$_2$O$_3$ or Mg$_2$SiO$_4$. This highlights the effect that Fe-bearing materials can have on the reflected flux, and therefore observed albedo, of hot transiting gas giants. Fe and Fe$_2$O$_3$ behave in a similar way to FeO, but interestingly Fe$_2$SiO$_4$ behaves in a similar way to the silicates or oxides, due to it's lower imaginary part of the refractive index $k$. 

It can be seen that the refractive indices of the morning and evening terminators in Figure~\ref{fig:nk_combined} (top and middle) are similar for some pressure layers, but differ around 1~$\times$~10$^{-3}$~bar in particular. The refractive indices at the sub-stellar point are very similar to the evening terminator for all pressure layers shown. The imaginary component of the refractive index $k$ is higher in the hotter evening region than the cooler morning region. From Figure~\ref{fig:cloud_abundance_plus_minus90} it can be seen that this is likely due to the large proportion of clouds formed from Fe extending higher in to the atmosphere.
Figure~\ref{fig:FP_all6_individual} (left), demonstrates that for the hotter regions of the atmosphere which generally have a higher imaginary component of the refractive index $k$ (from Figure~\ref{fig:nk_combined}) also have lower $F$ in comparison to the cooler regions. The imaginary component is slightly higher at lower wavelengths, which leads to a general trend of increasing flux with wavelength, as shown by Figure~\ref{fig:FP_all6_individual} (left).  The shape of $F$ and $P$ as a function of wavelength for homogeneous atmospheres composed of atmospheric region B, D, or A only (using the clouds formed from mixed materials for each region) can be seen from the single scattering properties of the materials used to form the clouds in these regions at various pressure layers,  as shown in  Figure~\ref{fig:scattering_angle_90_layers_B_D}. Here the single scattering matrix elements $F_{11}$ and $P$~=~-$\frac{F_{12}}{F_{11}}$ are plotted as a function of wavelength (see details in Section~\ref{sec:impact}). 

\subsection{Impact of clouds on the degree of linear polarisation}\label{sec:impact}

Different atmospheric layers are probed within (higher in the atmosphere) and outside (lower in the atmosphere) the atomic and molecular absorption features. If vertically inhomogeneous clouds are present in the atmosphere then different cloud layers are thus probed within and outside the absorption features. 
Gas particles scatter strongly at an angle of 90$\degree$, as can be seen in Figure~\ref{fig:AF_phase_plot_clear}. This scattering angle is largely relevant for orbital phases of $\alpha$~=~90$\degree$ or $\alpha$~=~270$\degree$. 
We therefore plot single scattering matrix elements $F_{11}$ and $P$~=~-$\frac{F_{12}}{F_{11}}$ at 90$\degree$ as a function of  wavelength $\lambda$ in order to gain some insight into why the models with mid-altitude cloud layers in Figure~\ref{fig:FP_all6_combined_90_270} of $P$ as a function of $\lambda$ for $\alpha$~=~90$\degree$ or 270$\degree$ look like they do. This is demonstrated by Figure~\ref{fig:FP_all6_individual}, showing the contribution to $P$ from different atmospheric regions A$\dots$F, and Figure~\ref{fig:scattering_angle_90_layers_B_D}, showing the single scattering properties of the mixed-material cloud particles used to form each layer in regions B (evening), D (morning), and A (sub-stellar). 
Figure~\ref{fig:scattering_angle_90_B_D} shows the single scattering properties of selected different materials, and so does not depend on the atmospheric setup of WASP-96b such as the temperature pressure profile. The only difference between atmosphere regions $B$ (upper panels) and $D$ (lower panels) in this case are the size distributions of the particles, which are based on the size distributions at 0.01~bar for these regions. Figure~\ref{fig:scattering_angle_90_B_D} therefore really highlights the impact that size distribution alone can have on the scattering properties of a cloud material. 

The atmosphere containing mixed-composition clouds in Figure~\ref{fig:B_species} has a low value of $P$ between 0.7 and 0.8~$\mu$m. The behaviour around the absorption feature, where $P$ dips just before and after the peak, is indicative of different atmospheric levels being probed here. Particles with different material composition and size distribution are in the atmosphere at different levels.

\subsection{Particle size distributions}\label{sec:size_dist_discussion}

We use a Gaussian distribution to  describe the sizes of the local particles in this work. Other particle size distributions are discussed in various works such as \cite{SamraThesis,74HaTr,84RoSt}. Studies which use the cloud-model of \cite{01AcMa} usually use a log-normal distribution~\citep[e.g.][]{20LaBu,21LuMo}, which was initially employed by \cite{01AcMa} based on measurements of terrestrial cloud particles, while other works such as \cite{18PoZhGa,19PoLoKr} typically compute particle size distributions from first principles. A modified-gamma distribution, based on terrestrial water clouds~\citep{64Deirmendjian}, is also sometimes used in atmospheric models~\citep[e.g.][]{17PiMa}. Although particle distributions based on measured Earth cloud particles can be considered a reasonable assumption, it is tricky to know the exact nature of size distributions in exoplanetary atmospheres such as the one we are modelling here. One benefit of using a Gaussian size distribution is the ease of determining a broad vs narrow distribution, as well as computational requirements. For example, some distributions have very long tails and a few very large particles can lead to very lengthy radiative transfer computations while having only a small influence on the reflected light spectra~\citep{74HaTr}, although we note some mentioned here such as the log-normal distribution can be considered relatively efficient. The impact of the particle size distribution on the reflected flux and polarisation warrants further exploration.

\subsection{Close-in planets}

Our focus is on characterising close-in transiting exoplanets. Although we do not explicitly take it into account here, we are aware of expected deviations of $F$ and $P$ for extreme orbital phases $\alpha$. This has been explored before by, for example, 
\cite{17PalmerEPSC}. They show the variation of $F$ and $P$ as a function of  distance from the host star for hot Jupiter exoplanets around solar-type stars.  Situations where the angular size of the host star in an exoplanet's sky is non-negligible are also investigated in \cite{19Palmer}. They define close-in planets which start to be affected, largely at the extreme orbital phase angles, as those closer than 0.05~AU, although the exact distance depends on the planet and star sizes also. With a semi-major axis of 0.045~AU~\citep{14HeAnCa}, WASP-96b is very close to this cutoff and thus has potential to be affected by the geometry, although noticeably less than even closer-in planets at 0.005 or 0.01~AU, as demonstrated by \cite{17PalmerEPSC}. \cite{19Palmer} find the flux to be particularly affected at extremes of orbital phase angle. \cite{17KoYaBe} investigate the difference in flux and polarization curves for transiting exoplanets in the cases of either plane parallel or a spherical stellar atmosphere used in models. They find that for most cases of known transiting systems the plane-parallel approximation can be safely used due to only a very small difference between the results using the two approaches. We therefore do not expect our resulting spectra as a function of wavelength which we typically take at orbital phases of $\alpha$~=~90 or 270$\degree$ to be significantly affected by such affects, but for our results which show the variation of $F$ or $P$ with orbital phase at set values of wavelength, some caution should be exercised at the extreme values of phase (close to 0$\degree$ and 180$\degree$). This also applies to our figures of geometric albedo as a function of wavelength, although we do expect the difference to be small and for the general trends to hold. In all our models the geometry causes $F$ to go to 0 at 180$\degree$, but in reality there could be expected to be some scattered flux at such a phase angle for very close-in exoplanets. 

There are other aspects of an exoplanet's orbit which we do not consider in the present study. For example, there is an expectation that close-in exoplanets are more impacted by tidal deformation than those further out, and the rotation rate of these planets affect their oblateness.  \cite{19Palmer} also investigate such effects and show, for example, that increasing the oblateness of a planet 
increases the amount of scattering at high atmospheric altitudes, which is typically expected to lead to an increase in the maximum degree of polarisation. 
We assume the orbital inclination angles of our model planets are 90$\degree$. For reference, the inclination of WASP-96b has been measured very close to 90$\degree$, at 85.6$\degree$~\citep{14HeAnCa}.

\subsection{Temperature dependence of optical properties}

Some studies have been done regarding the temperature dependence of the optical properties of various species, such as olivine, enstatite~\citep{15ZeHuPo}, and corundum, spinel, and alpha-quartz~\citep{13ZePoMu}, although these studies are generally focused on larger wavelengths than in our study. \cite{20YaZh} found that the optical properties of forsterite (Mg$_2$SiO$_4$) undergo a blue shift with increasing pressure. More studies exploring the temperature dependence of materials used to form clouds would be beneficial to future work. 

\subsection{Porosity of cloud particles}

As predicted by \cite{22SaHeCh}, the atmosphere of WASP-96b will also be impacted by the porosity of the material used to form the cloud particles.  We have not investigated this here, but it will be of interest to look into how different degrees of porosity of the cloud particles will affect $F$ and $P$ of a model WASP-96b planet.

\section{Conclusion}\label{sec:conclusion}

WASP-96b is a relatively homogeneous planet, in terms of temperature contrast and differences in composition between morning and evening terminators, when compared to some other hotter gas giant planets which have been modelled using global climate models coupled with kinetic cloud modelling~\citep{21HeLeSa,22SaHeCh}. The planet is expected to be relatively warm throughout, and thus expected to have clouds throughout the majority of the atmosphere. The molecular composition is relatively consistent across the morning and evening sides of the planet, with the exception of CH$_4$. We find that the degree of polarisation of reflected flux in particular is highly dependent on the types of clouds  and the properties of the materials used to form them and therefore highlights even subtle differences between the morning and evening sides of the planet (i.e. when the planet is at 90 or 270$\degree$ phase angle). We have investigated the effect of using irregularly shaped particles as opposed to assuming perfectly spherical particles in our models, and find there is a considerable difference in the modelled polarisation signals, in particular for an orbital phase of 90$\degree$ (face-on to the cooler morning terminator). This highlights the importance of using more physically realistic models of cloud particles, in situations where the particles are expected to be more fluffy or irregular shaped. The exact shape of aerosol particles in such exotic atmospheres is not necessarily known at present, but it is an avenue which warrants future exploration. In either case, knowing the scattering properties of different shapes of particles will be an advantage when fitting models to observed spectra in the future. 

In general we demonstrate in this study using the PolHEx code that we expect measuring the polarisation state of reflected flux of hot close-in transiting planets to give detailed insights into the atmosphere. It is an extremely complimentary tool to analysing transmission and emission spectroscopy observations of the same exoplanets, as it is sensitive to different components of an exoplanet atmosphere, most notably the material composition of clouds and their size distribution. 
	
	\section*{Acknowledgements}
	
	This project has received funding from STFC, under project number  ST/V000861/1. Ch.H. further acknowledges funding from the European Union H2020-MSCA-ITN-2019 under grant agreement no. 860470 (CHAMELEON). D.S. acknowledge financial support from the Austrian Academy of Sciences.

	\section*{Data Availability}
	
	The data underlying this article will be shared on reasonable request to the corresponding author.
	

	
	
	\bibliographystyle{mnras}
	\bibliography{polarimetry} 

	
	
	
	\appendix
	\section{Detectability calculations}\label{sec:app1}
	
Here we demonstrate the detectability of the models presented in this paper for hot gaseous exoplanet WASP-96b. For this we first use Eq~\ref{eq:ref_flux} to compute the total reflected Flux arriving at the observer on (or near) Earth $F_{\rm p}(\lambda,\alpha)$. We use c.g.s units for all parameters. To compute the stellar surface flux, we assume that the luminosity of a G8 star such as WASP-96~\citep{14HeAnCa} is 0.68~$L_{\odot}$, which translates to a surface flux of 3.88~$\times$~10$^{10}$~erg~s$^{-1}$~cm$^{2}$. This corresponds to a maximum reflected planetary flux arriving to the observer $F_p$ of 6.57~$\times$~10$^{-15}$~erg~s$^{-1}$~cm$^{2}$. We can scale this against the total stellar flux arriving at the observer $F_s$ using the relation:
\begin{equation}\label{eq:lum_flux_eq}
F_{s} = \frac{L}{4 \pi d^2},
\end{equation}
where $L$ is the stellar luminosity (erg s$^{-1}$), and $d$ is the stellar-Earth distance (cm). This gives $F_s$~=~1.72~$\times$~10$^{-10}$~erg~s$^{-1}$~cm$^{2}$. If we scale the maximum value of $F_p$ with this we get $\frac{F_p}{F_s}$~=~38~ppm. Using typical values of $F$ and $P$ from this figures shown in this paper gives us an approximate range of typical values for $\frac{P}{F_s}$ of 0.1~-~30~ppm. This is called the ``observed polarisation'' in \cite{15BaKeCo} (i.e. polarisation as a fraction of light from the star). The HIPPI-2 instrument can measure a polarised signal with a precision of around 3.5~ppm~\citep{20BaCoKe}, with higher precisions expected from potential future instruments.

\newpage 
 \onecolumn
\section{Additional figures}

\begin{figure*}
	\centering
	\includegraphics[width=0.45\textwidth]{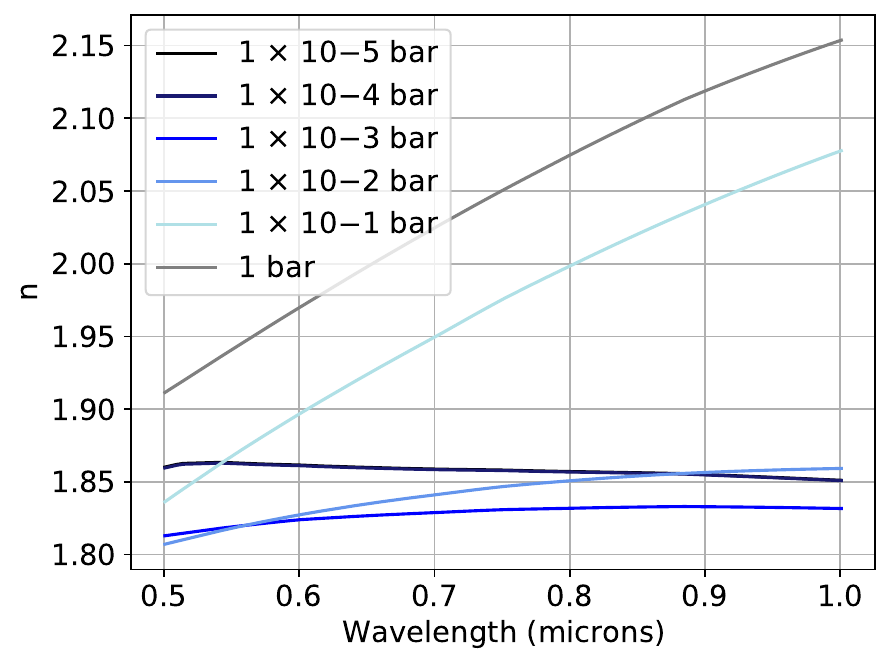}
	\includegraphics[width=0.45\textwidth]{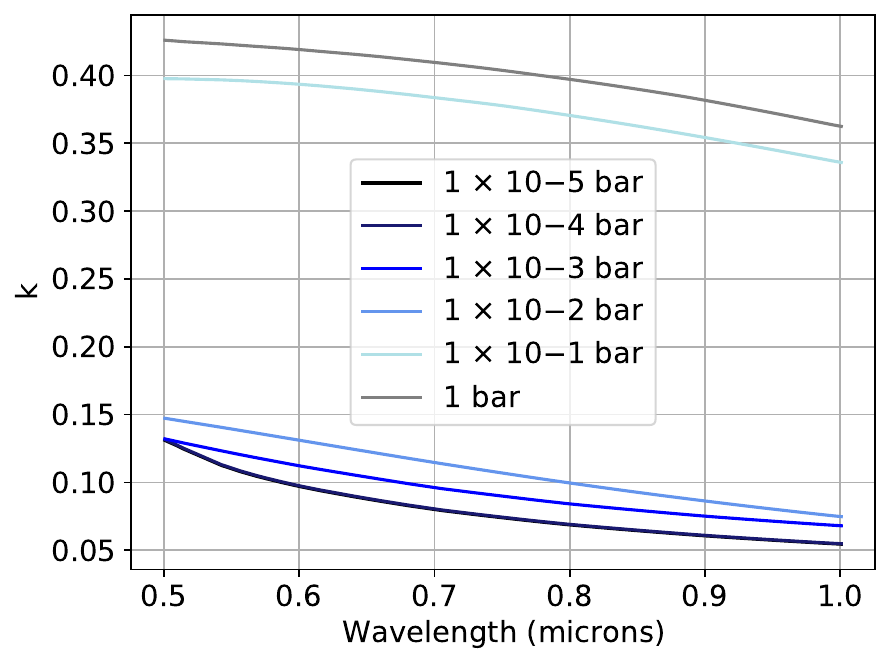}
	\includegraphics[width=0.45\textwidth]{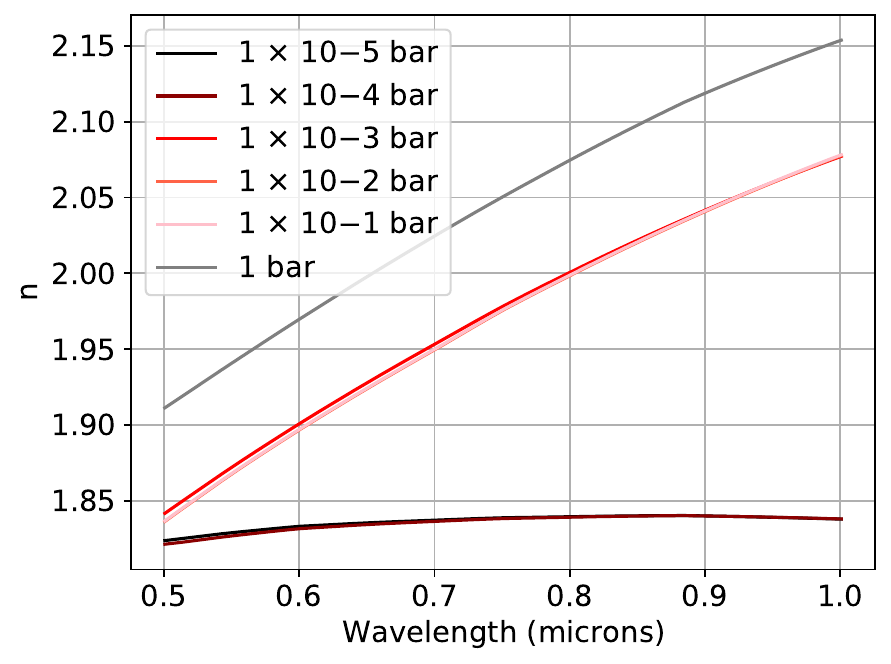}
	\includegraphics[width=0.45\textwidth]{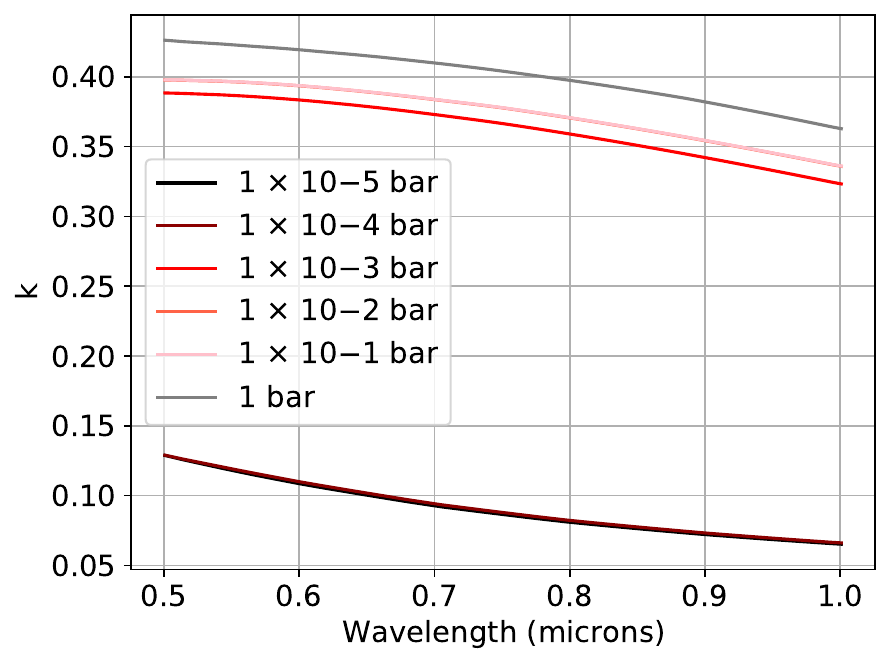}
	\includegraphics[width=0.45\textwidth]{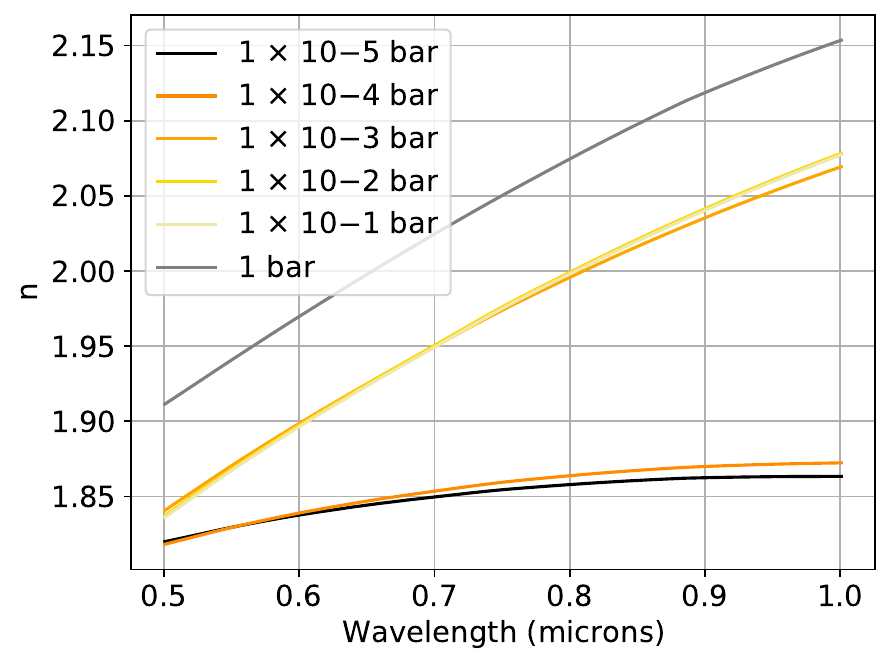}
	\includegraphics[width=0.45\textwidth]{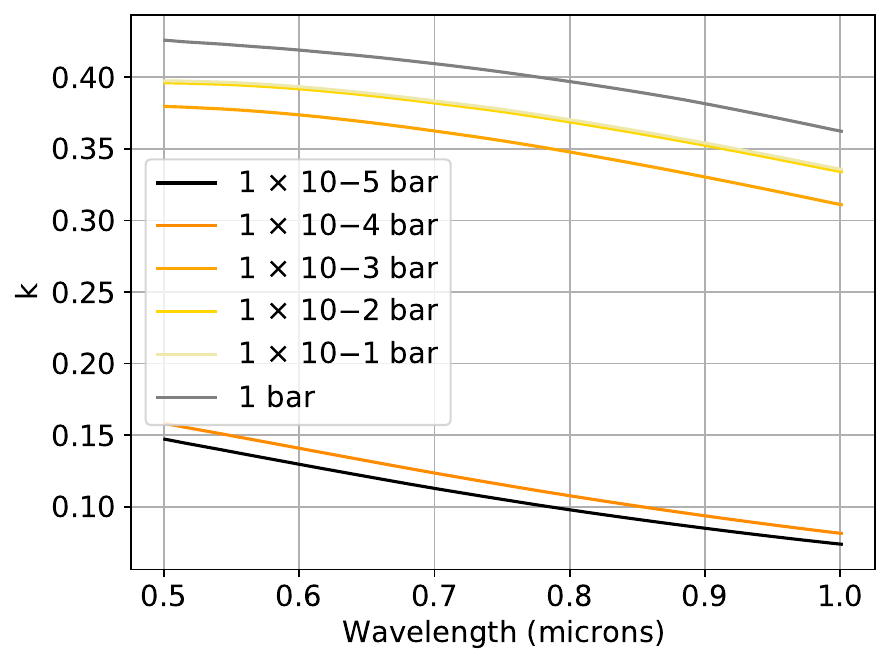}
	\caption{Complex refractive indices (real $n$ and imaginary $k$ component) of combined materials to form clouds as a function of wavelength for  different atmospheric regions and pressure layers. Top: region D, -90/0$\degree$ (morning). Middle: region B, 90/0$\degree$ (evening), Bottom: region A: 0/0$\degree$ (sub-stellar point).}\label{fig:nk_combined}
\end{figure*}

 \begin{figure*}
	\centering
	\includegraphics[width=0.49\textwidth]{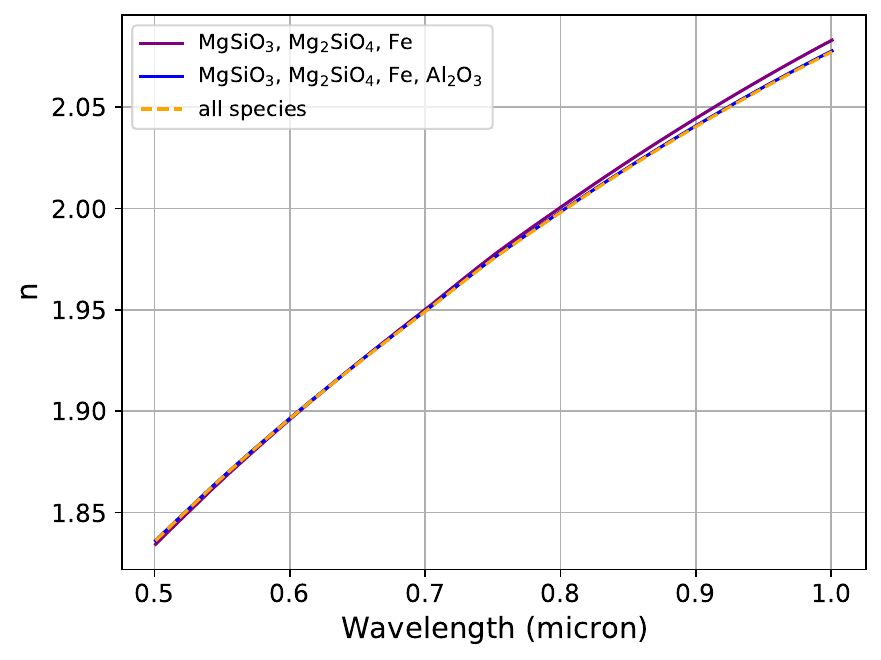}
	\includegraphics[width=0.49\textwidth]{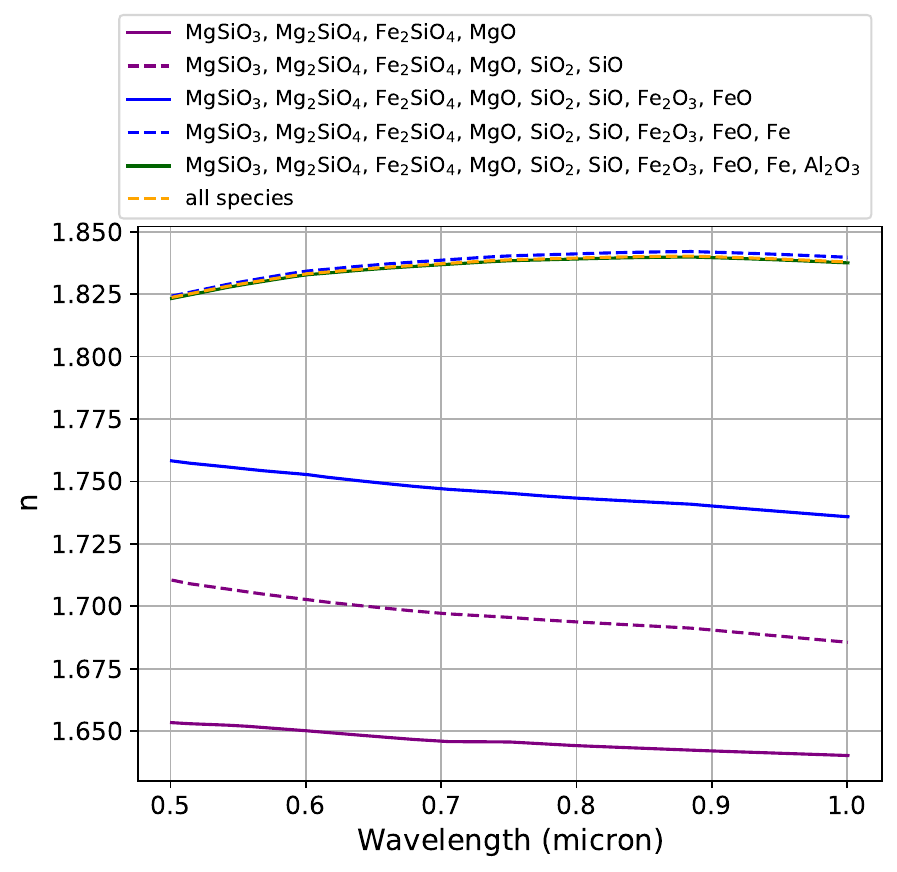}
	\caption{The real $n$ components of the complex refractive indices of various amounts of combined materials to form clouds as a function of wavelength for region B 90/0$\degree$ (evening) and pressure layers 1$\times$10$^{-1}$~bar (left) and 1$\times$10$^{-4}$~bar (right).}\label{fig:nk_combined_1e-1_1e-4}
\end{figure*}

\begin{figure*}
	\centering
	\includegraphics[width=0.49\textwidth]{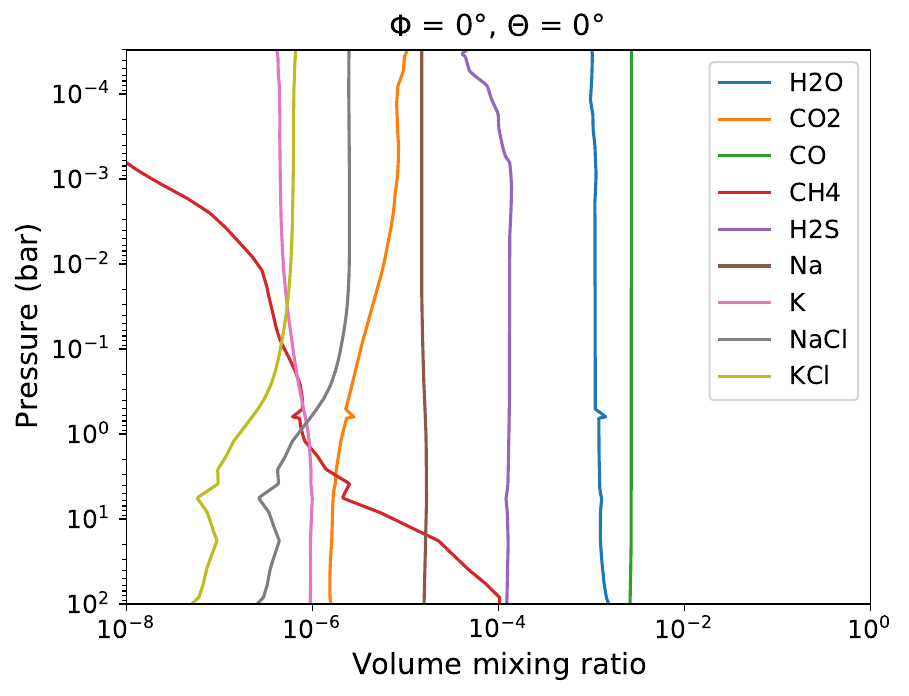}
	\includegraphics[width=0.49\textwidth]{mixingratios_WASP96b_DRIFT_all_pasted_Phi90-0Theta0-0.pdf}
	\includegraphics[width=0.49\textwidth]{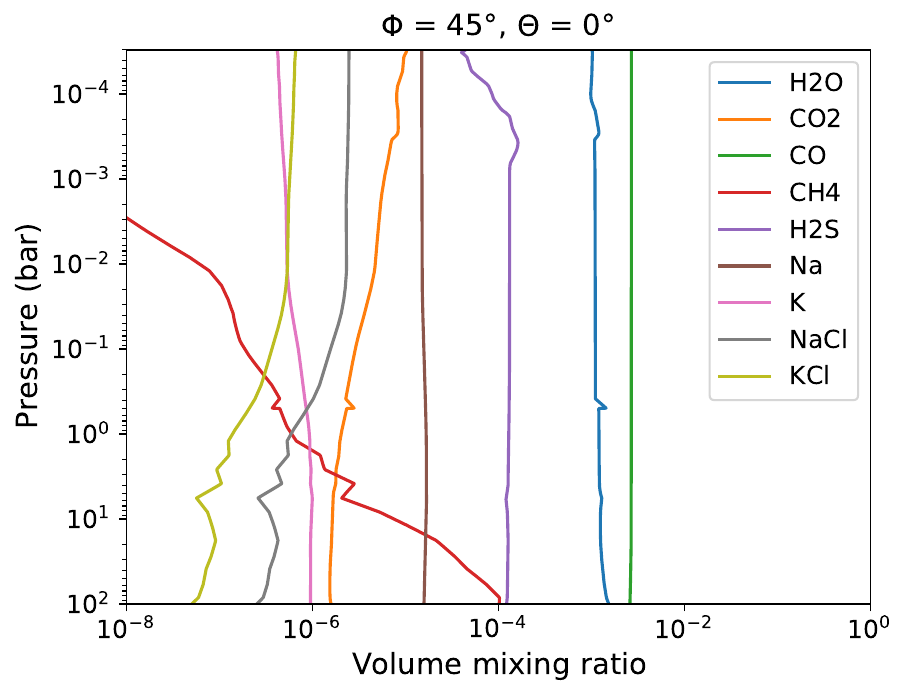}
	\includegraphics[width=0.49\textwidth]{mixingratios_WASP96b_DRIFT_all_pasted_Phi-90-0Theta0-0.pdf}
	\includegraphics[width=0.49\textwidth]{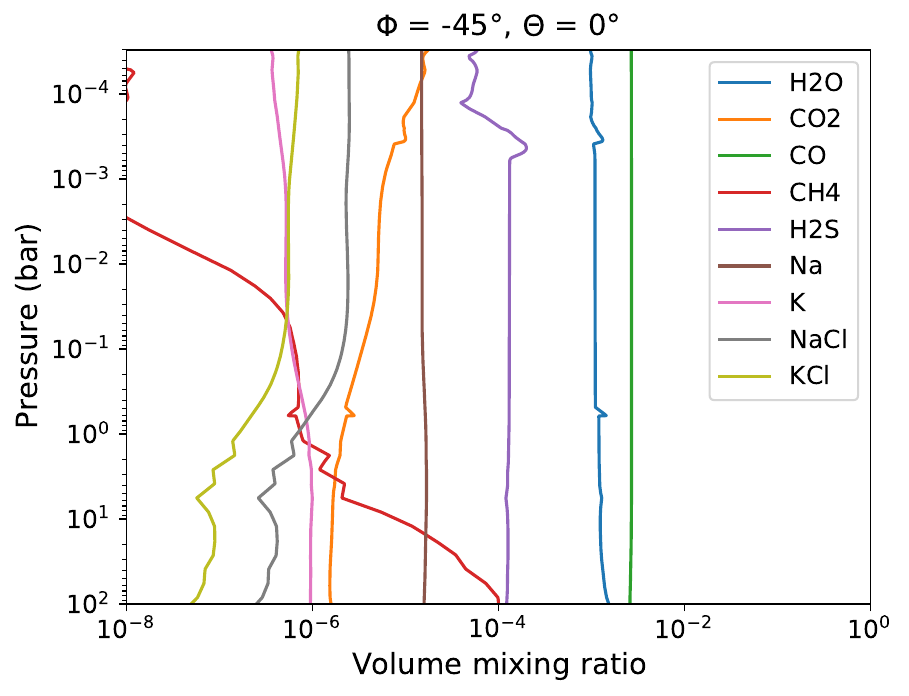}
	\includegraphics[width=0.49\textwidth]{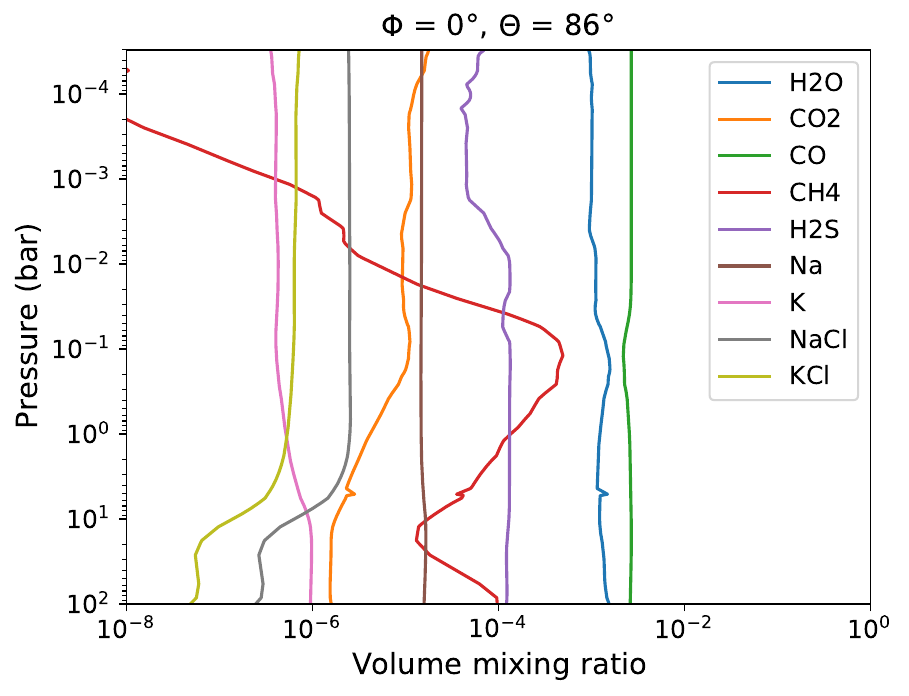}
	\caption{Molecular and atomic concentrations $\frac{n_i}{n_{\rm tot}}$ for models  
		of WASP-96b used in this work as a function of pressure, and at various longitude and latitude points, as labelled in each sub-plot.}\label{fig:mixing_part1}
\end{figure*}

\begin{figure*}
	\centering
	\includegraphics[width=0.45\textwidth]{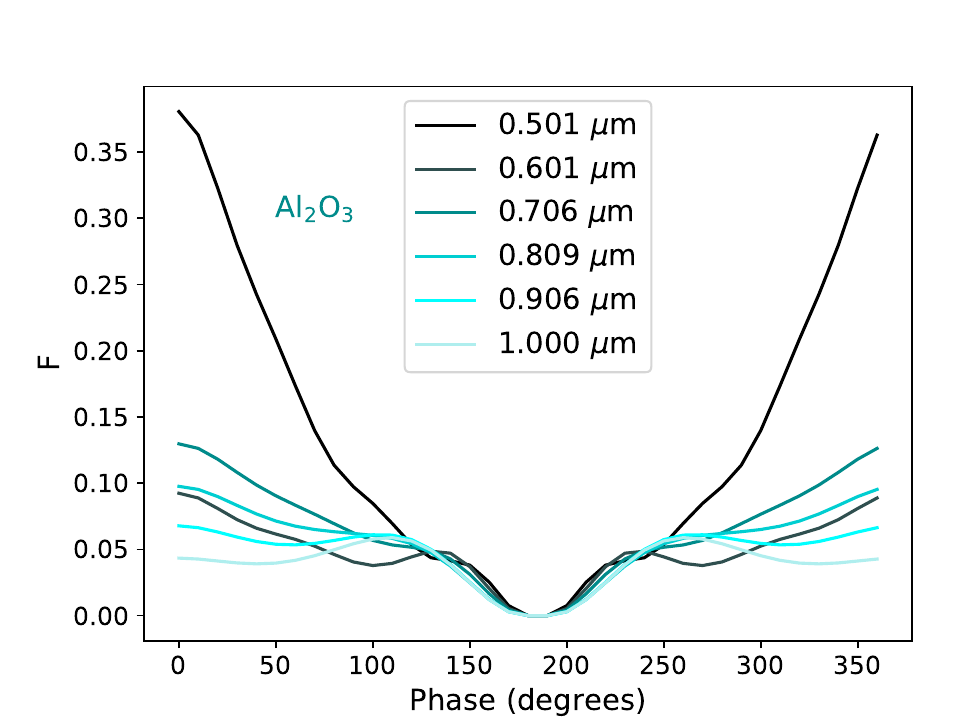}
	\includegraphics[width=0.45\textwidth]{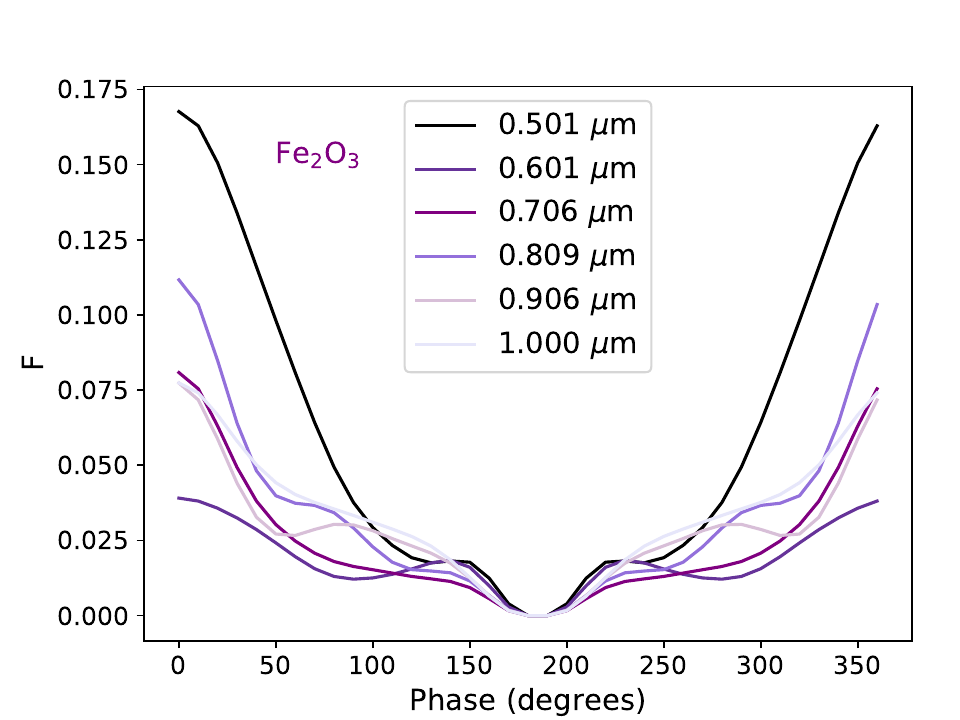}
	\includegraphics[width=0.45\textwidth]{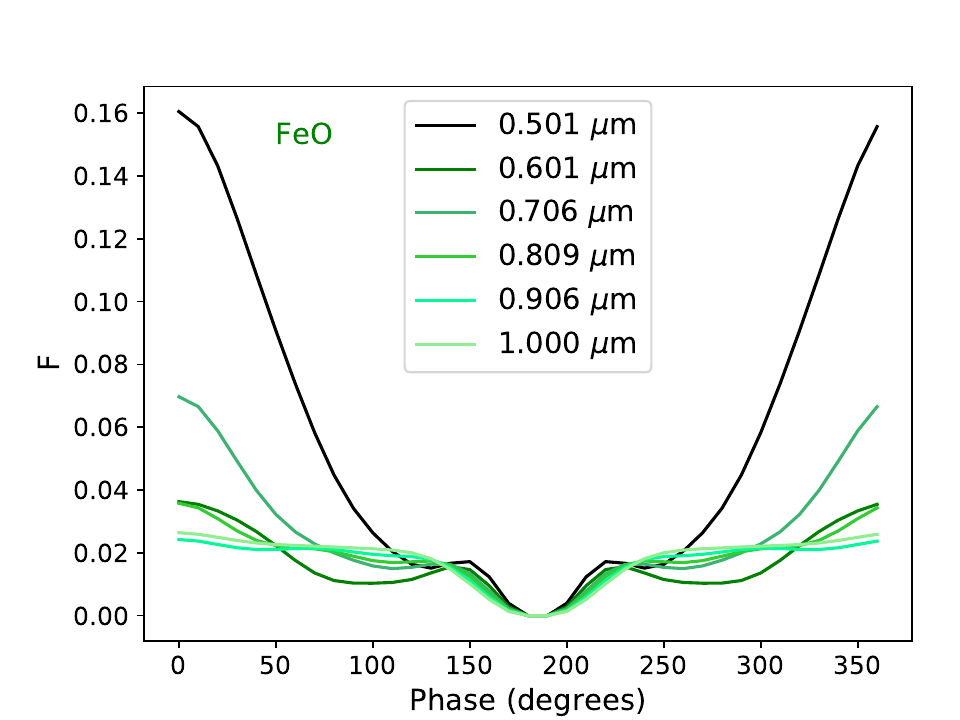}
	\includegraphics[width=0.45\textwidth]{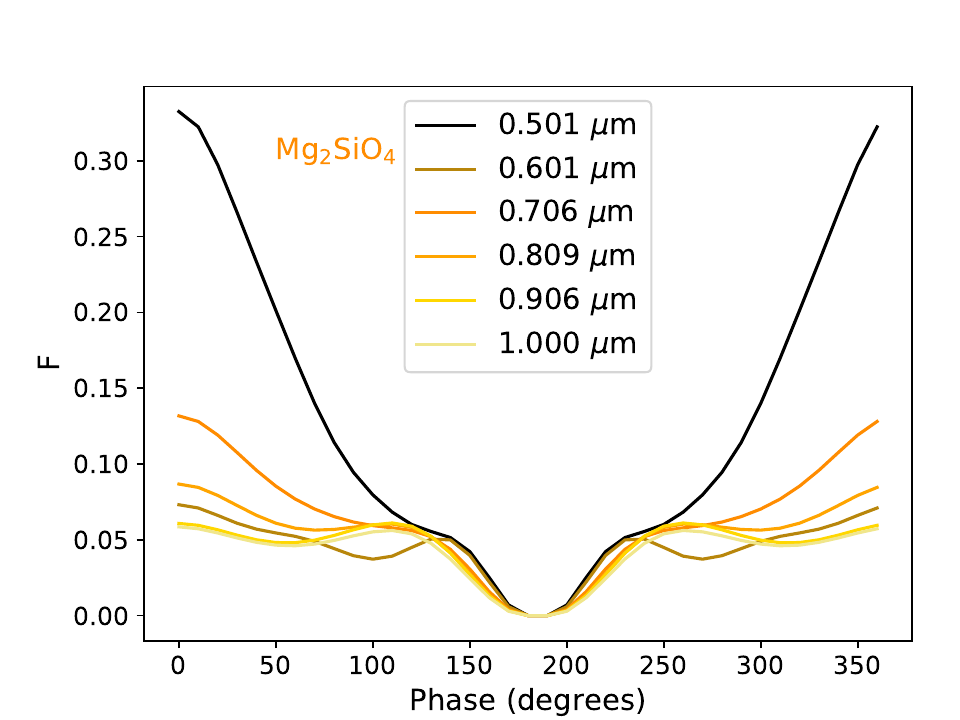}
	\includegraphics[width=0.45\textwidth]{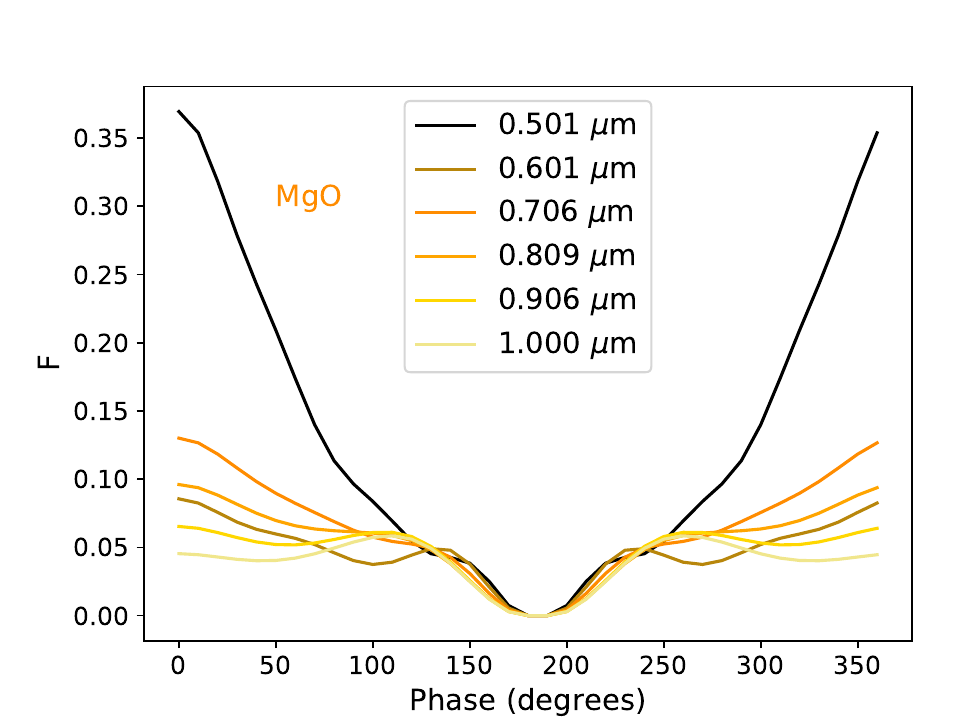}
	\includegraphics[width=0.45\textwidth]{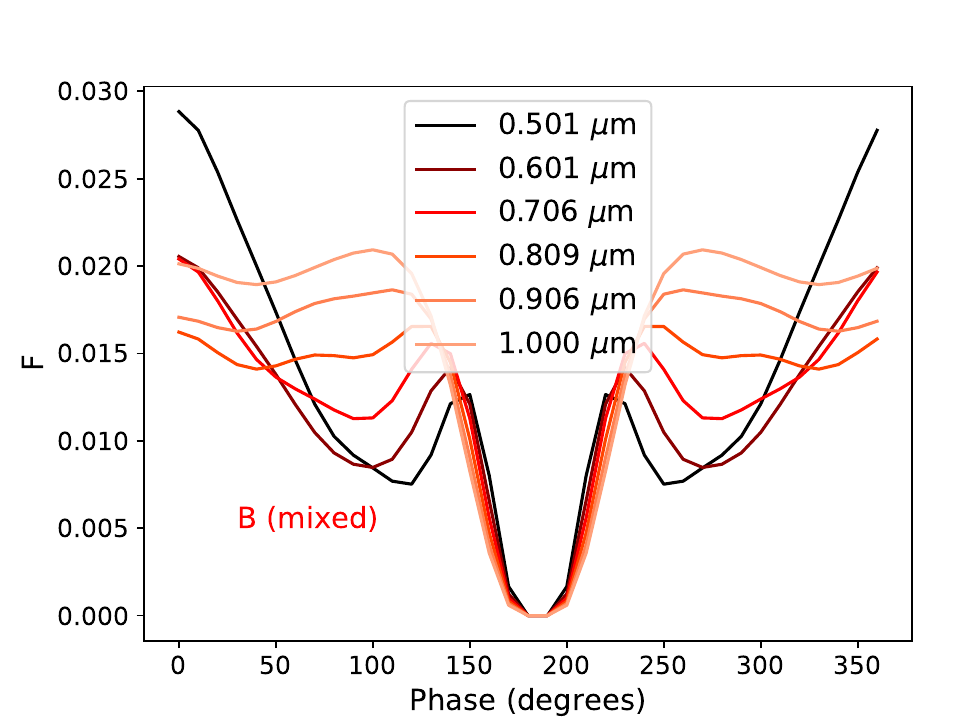}
	\caption{Phase curves for  reflected flux $F$ for homogeneous planetary atmospheres composed of clouds formed from a single species (as labelled in each panel), and atmosphere type B. The phase curve for the homogeneous planetary atmosphere type B but with mixed species for the clouds is shown in the lower right panel for comparison.}\label{fig:phase_curves_single_species_F}
\end{figure*}

\begin{figure*}
	\centering
	\includegraphics[width=0.45\textwidth]{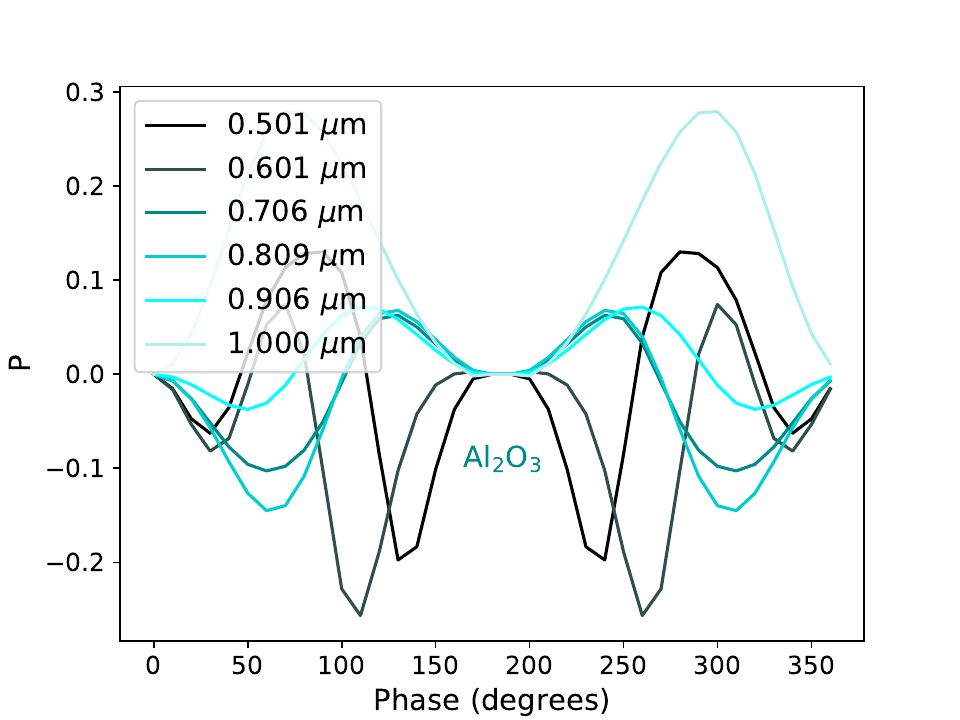}
	\includegraphics[width=0.45\textwidth]{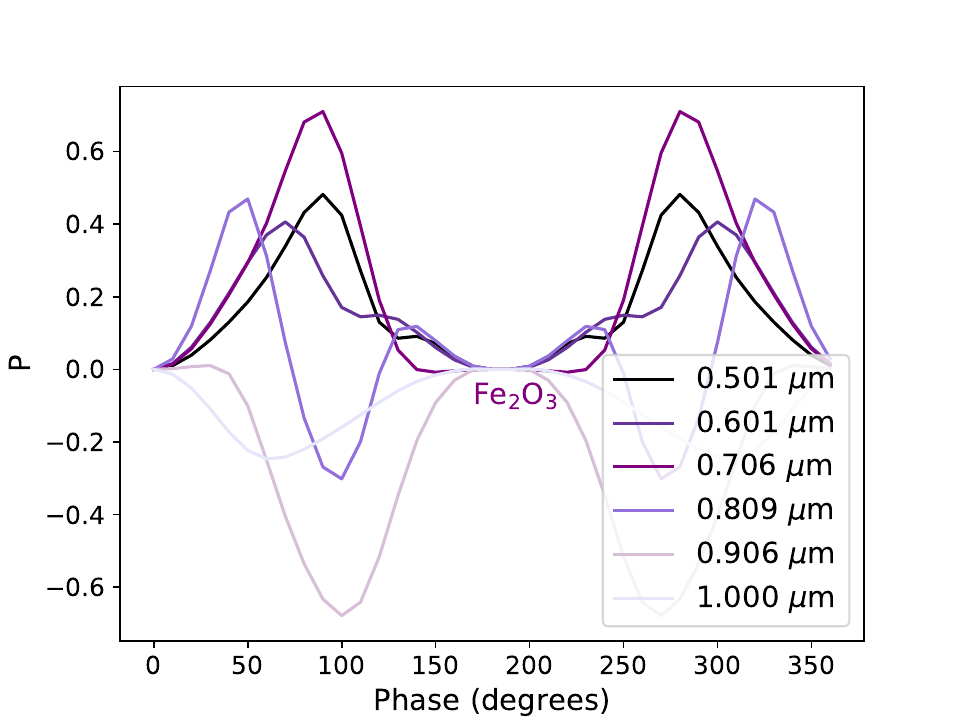}
	\includegraphics[width=0.45\textwidth]{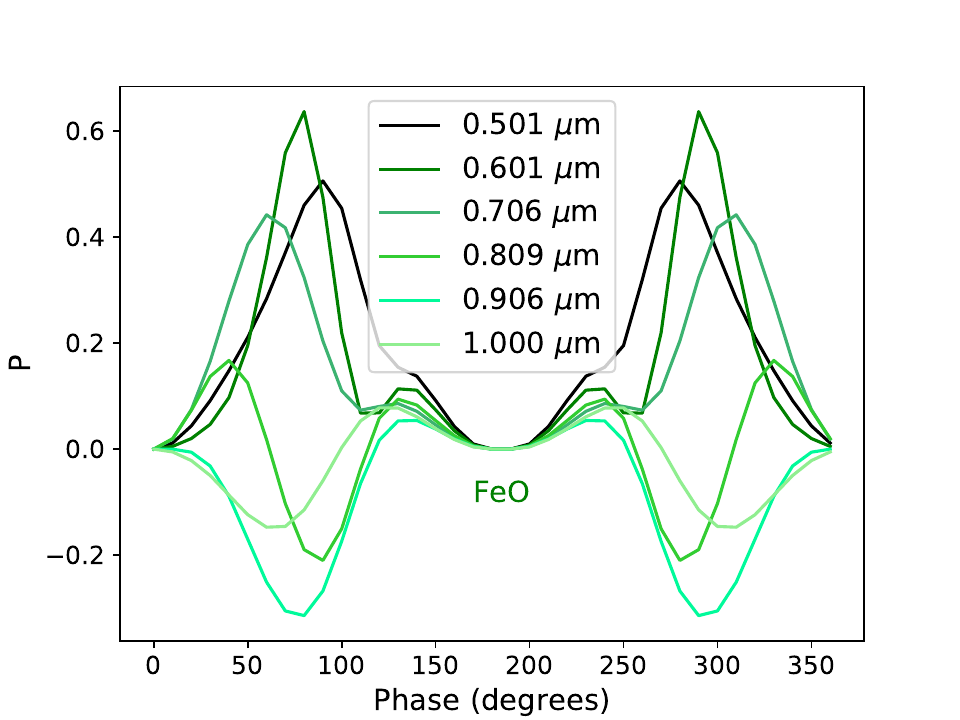}
	\includegraphics[width=0.45\textwidth]{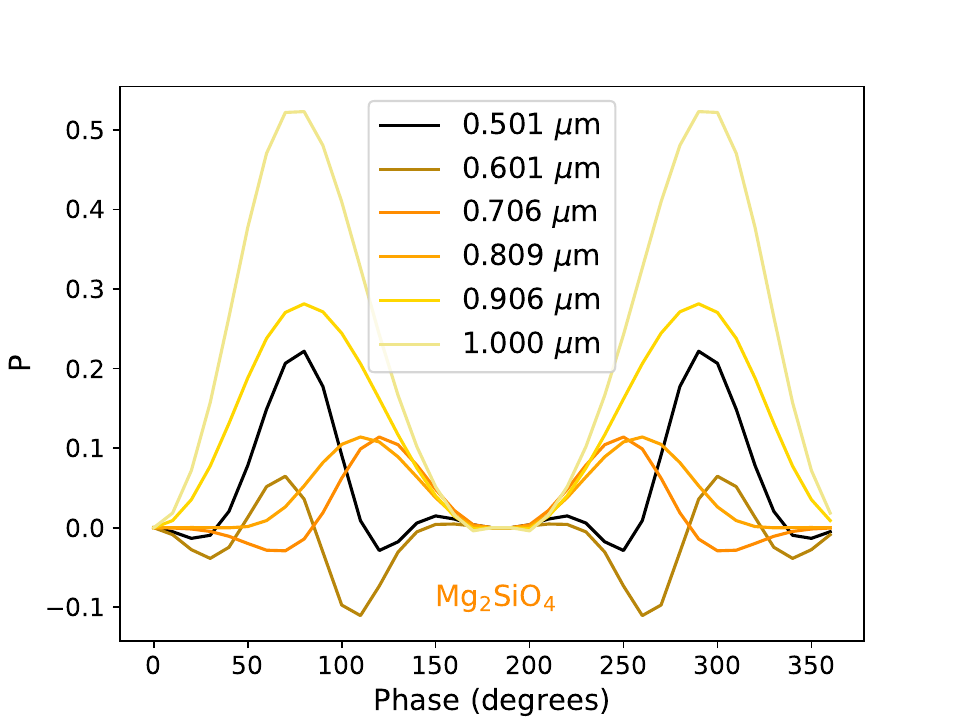}
	\includegraphics[width=0.45\textwidth]{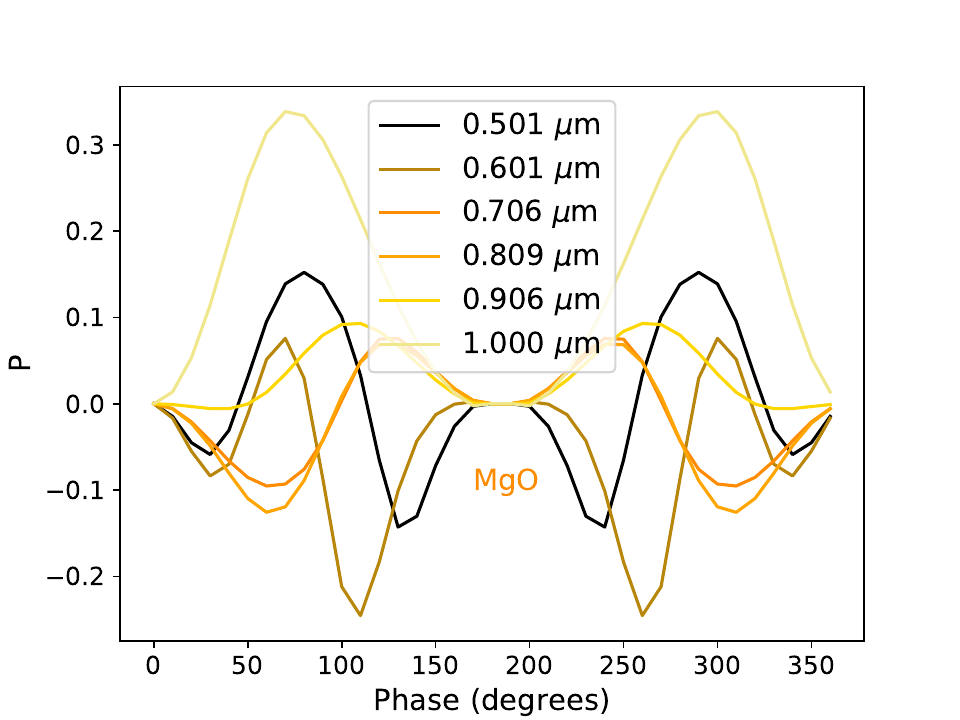}
	\includegraphics[width=0.45\textwidth]{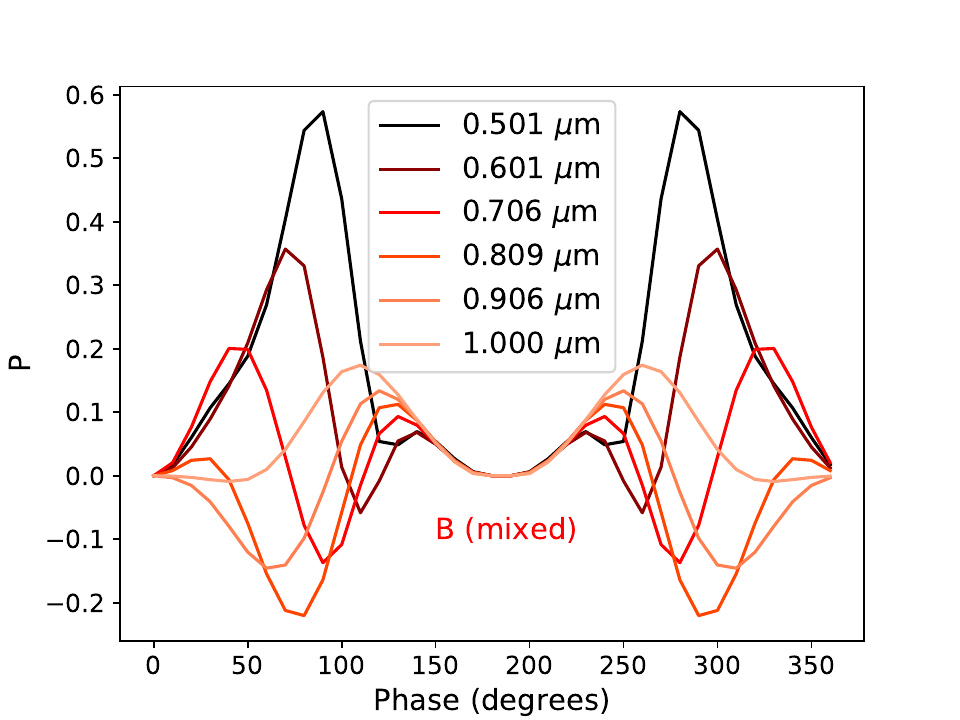}
	\caption{Phase curves for  degree of linear polarisation $P$  for homogeneous planetary atmospheres composed of clouds formed from a single species (as labelled in each panel), and atmosphere type B. The phase curve for the homogeneous planet composed of atmosphere type B but with mixed species for the clouds is shown in the lower right panel for comparison.}\label{fig:phase_curves_single_species_P}
\end{figure*}

\begin{figure*}
	\includegraphics[width=0.49\textwidth]{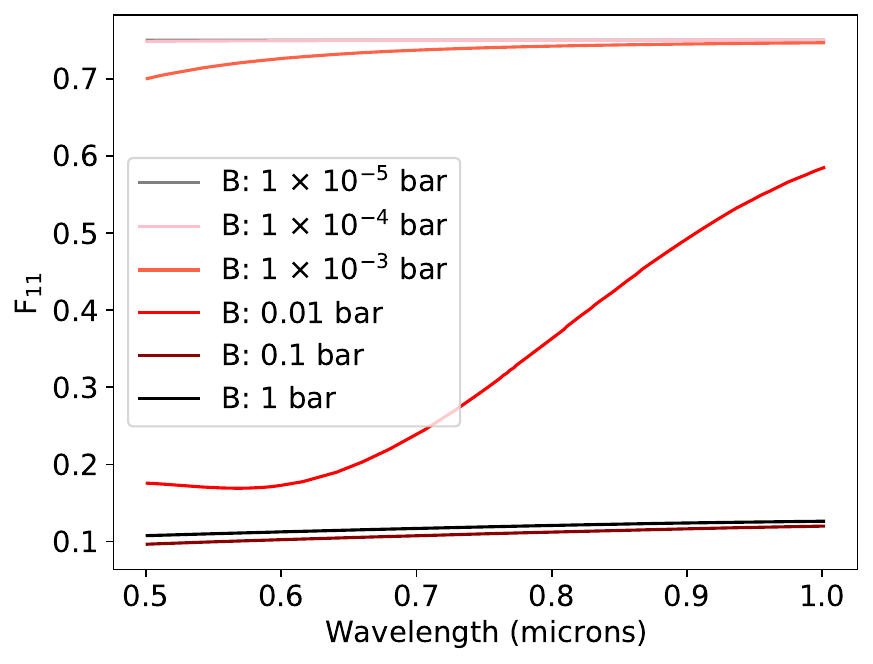}
	\includegraphics[width=0.49\textwidth]{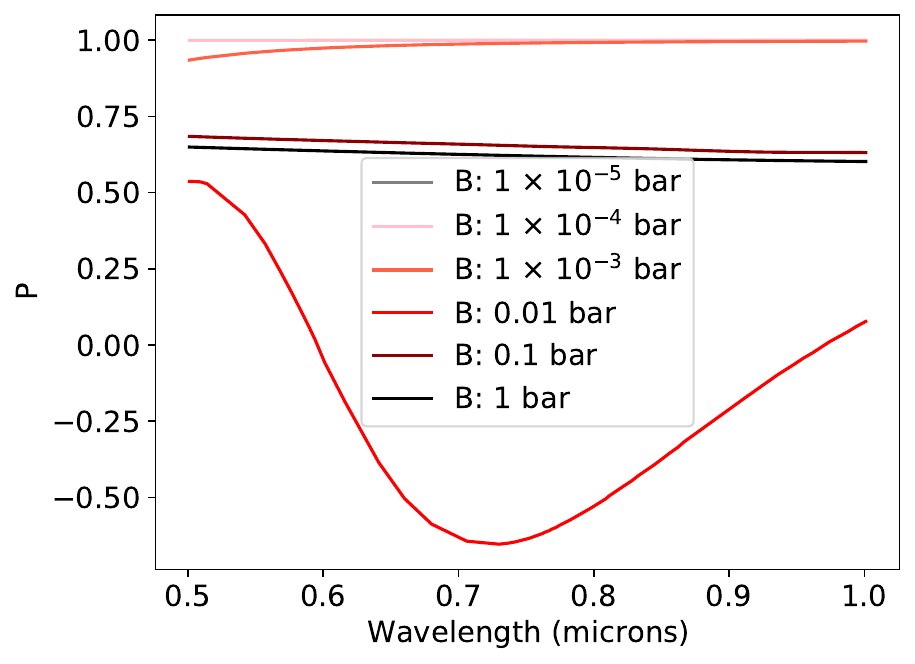}
	\includegraphics[width=0.49\textwidth]{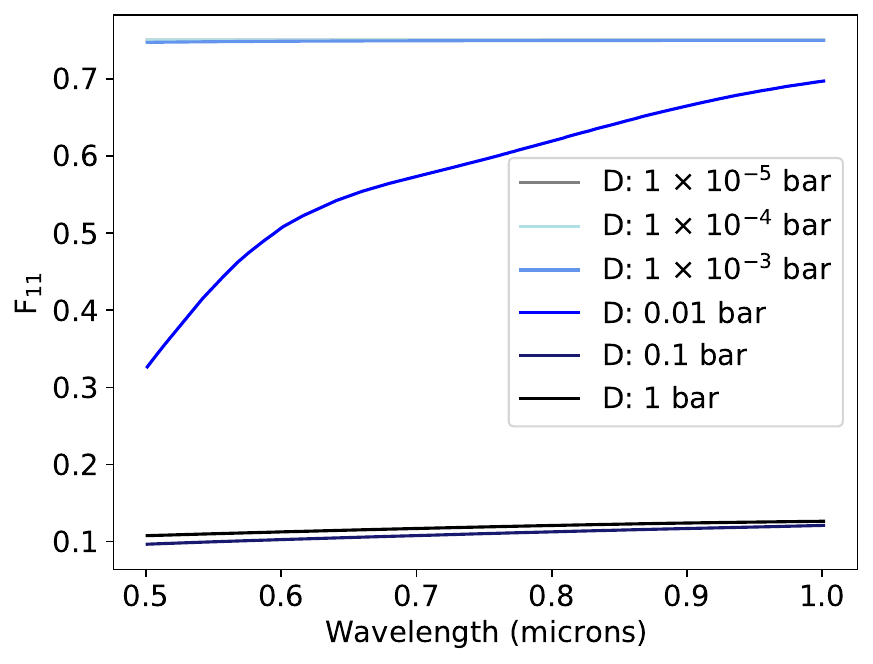}
	\includegraphics[width=0.49\textwidth]{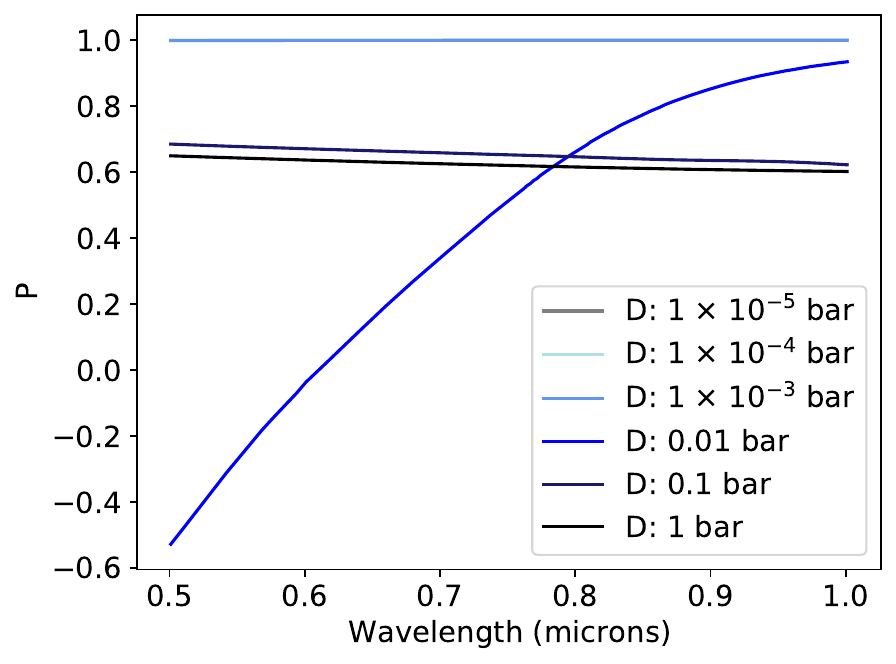}
	\includegraphics[width=0.49\textwidth]{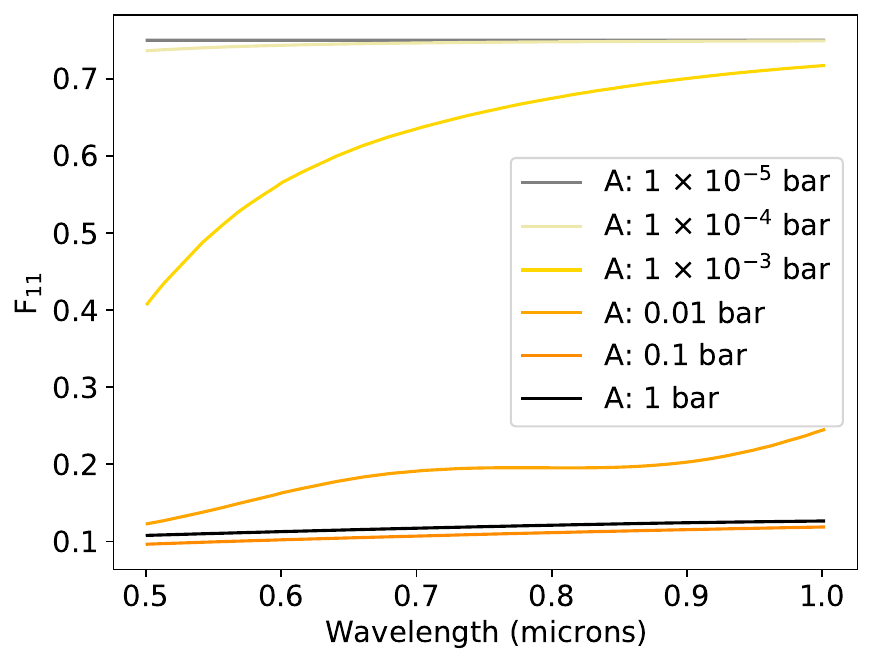}
	\includegraphics[width=0.49\textwidth]{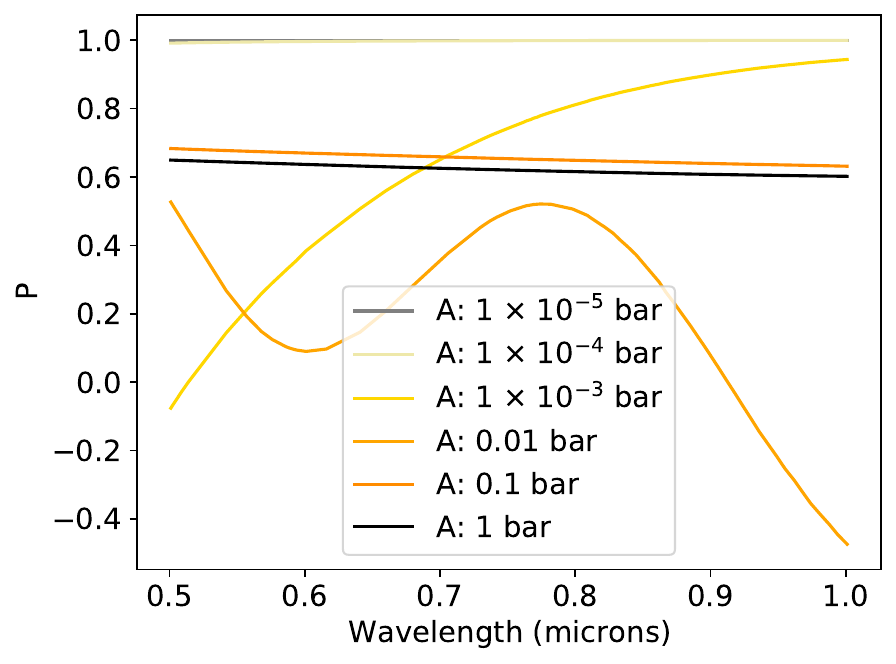}
	\caption{Single scattering matrix elements F$_{11}$ (left panels) and $P$ (right panels) for a scattering angle of 90$\degree$ as a function of wavelength, for the cloud setups used for different pressure layers (as labelled) of atmosphere regions B (upper panels), D (middle panels), and A (lower panels). }
	\label{fig:scattering_angle_90_layers_B_D}
\end{figure*}

\begin{figure*}
	\includegraphics[width=0.49\textwidth]{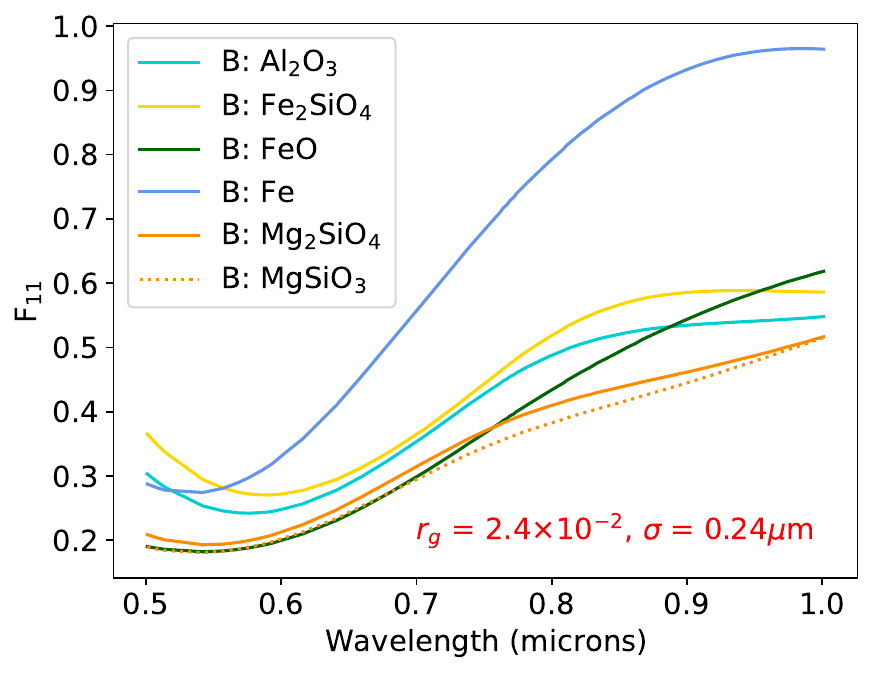}
	\includegraphics[width=0.49\textwidth]{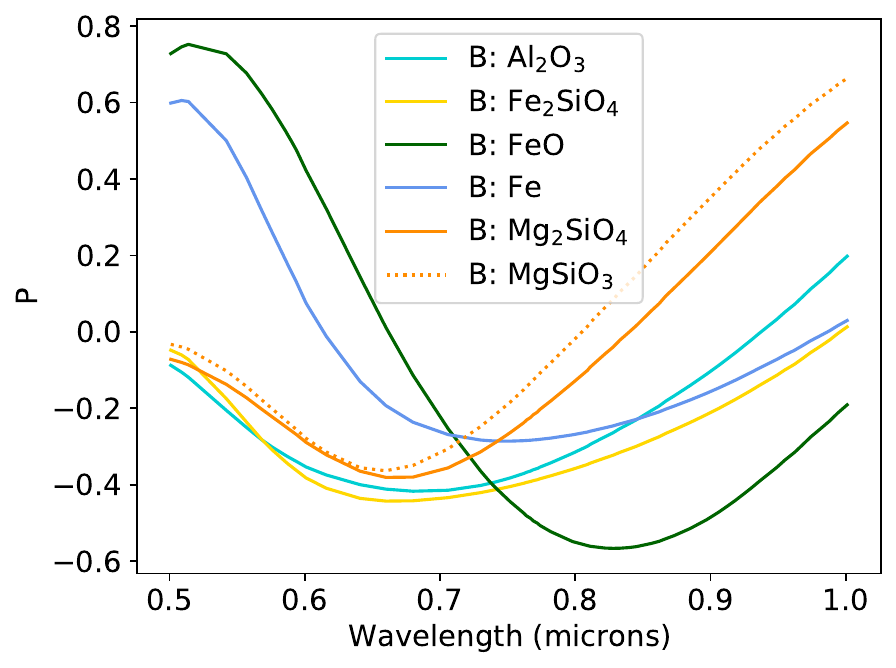}
	\includegraphics[width=0.49\textwidth]{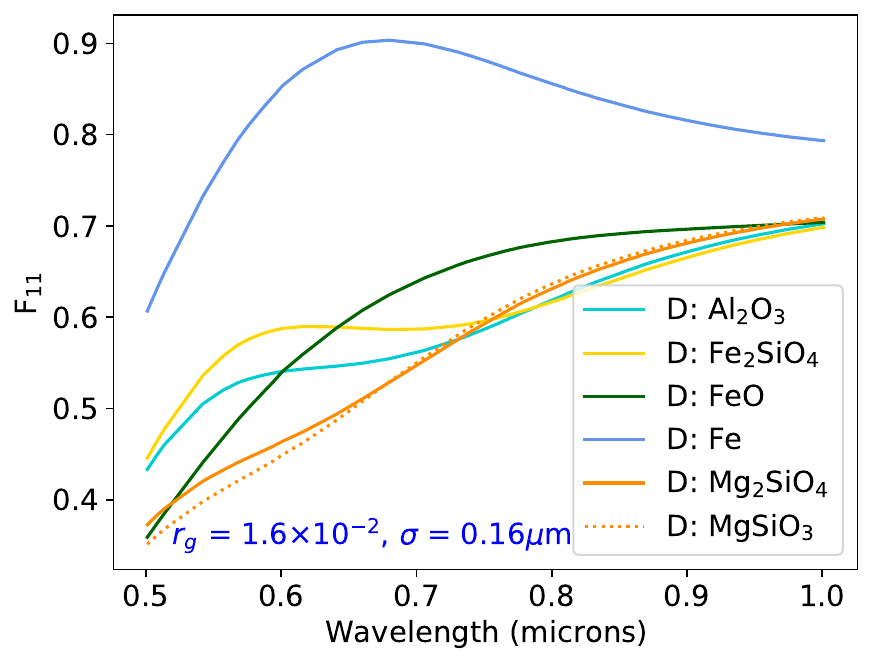}
	\includegraphics[width=0.49\textwidth]{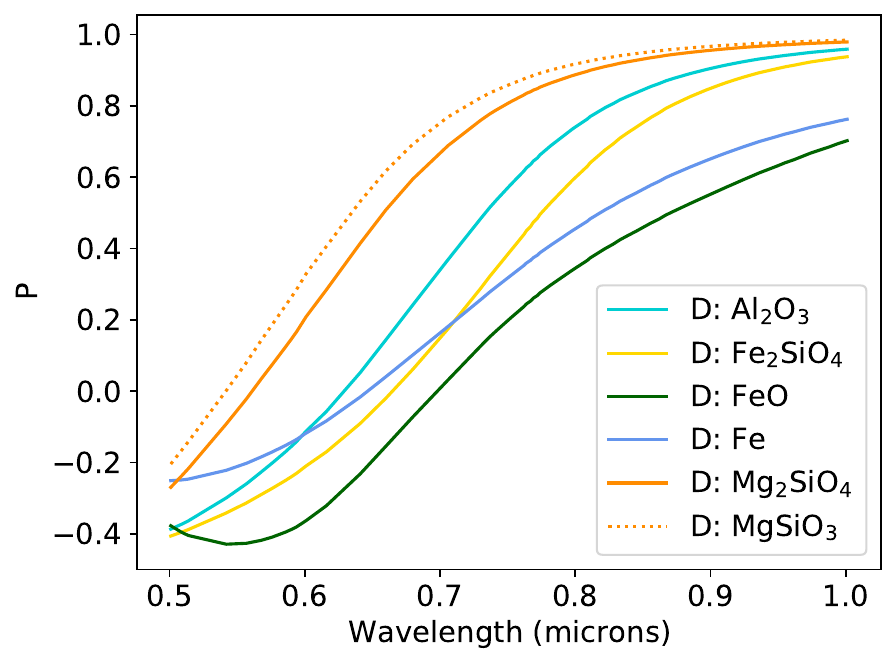}
	\caption{Single scattering matrix elements F$_{11}$ (left panels) and P (right panels) for a scattering angle of 90$\degree$ as a function of wavelength, for various species as labelled. The size distribution from atmosphere type B (upper) at 0.01~bar (2.4~$\times$~10$^{-2}$ around 0.24~$\mu$m), and for atmosphere type D (lower) at 0.01~bar (1.6~$\times$~10$^{-2}$ around 0.16~$\mu$m) are compared.}
	\label{fig:scattering_angle_90_B_D}
\end{figure*}

\newpage


	\label{lastpage}
\end{document}